\documentclass[a4paper]{article}

\usepackage{jheppub} 
                     
\usepackage{graphicx,color}
 \usepackage{bm}
   \usepackage{amsmath}
    \usepackage{amssymb}    
     \usepackage{pifont}

\usepackage{standalone}
\usepackage{slashed}
\usepackage{multirow}
\usepackage{hhline}
\usepackage{colortbl}
\usepackage{soul}

\newcommand{\nn}{\nonumber}

\newcommand{\const}{\text{const.}}

\newcommand{\ot}{\leftarrow}

\renewcommand{\(}{\left(}
\renewcommand{\)}{\right)}
\renewcommand{\[}{\left[}
\renewcommand{\]}{\right]}

\renewcommand{\vec}[1]{\bm{#1}}

\newcommand{\specialcellcenter}[2][c]{\begin{tabular}[#1]{@{}c@{}}#2\end{tabular}}

\title{Extraction of unpolarized transverse momentum distributions from the fit of Drell-Yan data at N$^4$LL}

\author[a]{Valentin Moos,}
\author[b]{Ignazio Scimemi,}
\author[b]{Alexey Vladimirov,}
\author[a,b]{Pia Zurita}

\affiliation[a]{Institut f\"ur Theoretische Physik, Universit\"at Regensburg, Universit{\"a}tsstra{\ss}e  31, D-93040 Regensburg, Germany}
\affiliation[b]{Departamento de F\'isica Te\'orica \& IPARCOS, Universidad Complutense de Madrid, Plaza de las Ciencias 1, E-28040 Madrid, Spain}

\emailAdd{valentin.moos@physik.uni-regensburg.de}
\emailAdd{ignazios@ucm.es}
\emailAdd{alexeyvl@ucm.es}
\emailAdd{marzurit@ucm.es}

\abstract{
We present an extraction of unpolarized transverse momentum dependent parton distributions functions (TMDPDFs) and Collins-Soper kernel from the fit of Drell-Yan and weak-vector boson production data. The TMDPDF are parameterized, as commonly done, using their (large transverse momentum) asymptotic  matching to PDF. The analysis is done at the next-to-next-to-next-to-next-to leading logarithmic accuracy (N$^4$LL) (performed only approximately  because   PDF  evolution is known so far at next-to-next-to leading order (NNLO)). The non-perturbative model used for TMDPDF is flavor dependent to reduce the colllinear PDF bias. The estimation of uncertainties is done with the replica method and, for the first time, it includes the propagation of uncertainties due to the collinear distributions.
}

\preprint{IPARCOS-UCM-035}

\begin{document} 
\allowdisplaybreaks

\maketitle 
\section{Introduction}
\label{sec:introduction}

The formulation of the transverse-momentum dependent (TMD) factorization theorems for Drell-Yan (DY) and semi-inclusive deep inelastic scattering (SIDIS) about a decade ago~\cite{Becher:2010tm, Collins:2011zzd, Echevarria:2011epo, Echevarria:2012js} has driven an increasing effort towards the understanding the TMD distribution of partons within the nucleon. The cornerstone of this program is made up by the unpolarized transverse momentum-dependent parton distribution functions (TMDPDFs) because
they have an impact  on the determination of all the other TMD distributions.
For this reason, the precise knowledge of unpolarized TMDPDFs is of utmost importance. In this work, we perform a global analysis of the vector boson production data and determine the unpolarized TMDPDFs. 
We include an extended dataset and we incorporate consistently the perturbative QCD information from the highest available orders.


The previous unpolarized TMDPDFs extractions~\cite{Scimemi:2017etj, Bacchetta:2019sam, Bertone:2019nxa, Scimemi:2019cmh, Bacchetta:2022awv} were based on the next-to-next-to-leading logarithm (N$^2$LL) resummation, and  next-to-next-to-next-to-leading order (N$^3$LO) of perturbative accuracy. The perturbative ingredients for this order were computed in refs.~\cite{Gehrmann:2012ze, Gehrmann:2014yya,Echevarria:2015byo, Echevarria:2016scs, Vladimirov:2016dll, Li:2016ctv}. High perturbative orders are needed to meet the precision of modern experiments, mainly ATLAS and CMS, at the LHC. The latest measurements of differential Z-boson-production cross-section at ATLAS~\cite{ATLAS:2019zci} reach an extraordinary precision of $\sim 0.1\%$. The progress on the experimental side has stimulated the calculation of even higher perturbative orders in Quantum Chromodynamics (QCD), reaching very recently the three-loop accuracy  for all terms of the factorized cross-section, and even higher for the anomalous dimensions~\cite{Das:2019btv, Luo:2019szz, vonManteuffel:2020vjv, Duhr:2020seh, Ebert:2020yqt, Moch:2021qrk, Chen:2021rft, Duhr:2022yyp, Moult:2022xzt}. All this has prompted the explicit calculation of cross-sections, including next-to-leading logarithmic effects up to power four\footnote{
The evolution of collinear PDFs is taken at NNLO since an extraction with the N$^3$LO evolution is not yet available. In this respect  the N$^4$LL order  of this work is only approximate.} (N$^4$LL).

The perturbative calculations, however, are just the tip of the iceberg. The TMD distributions, the heart of the TMD factorization theorem, are essentially non-perturbative functions. As such, they cannot be computed in perturbative QCD, but must be determined from the data. 
For this task we  take advantage of 
 a well-known  relation between TMDPDFs and PDFs when the TMDPDFs are written in the transverse momentum conjugate variable $b$. Then, going to the asymptotic small-$b$ limit, one has 
\begin{equation}\label{def:F1=C.f1}
\lim_{b\rightarrow 0}
f_{1,i\ot h}(x,b;\mu,\zeta)\simeq\sum_{j}\int_{x}^1 \frac{dy}{y}C_{i\ot j}\(y,b;\mu,\zeta\)q_{j}\(\frac{x}{y},\mu\) \, ,
\end{equation}
where $x$ is Bjorken collinear momentum fraction,  $q_j(x,\mu)$ and $f_i(x,b;\mu,\zeta)$ are the collinear PDFs and TMDPDFs respectively. These distributions  depend on the factorization  and rapidity scales, $\mu$ and $\zeta$. In the eq.~(\ref{def:F1=C.f1}), $C_{i\leftarrow j}$'s are the Wilson coefficient functions,  the subscripts $i$ and $j$ label the flavors of partons and we have omitted  power-suppressed corrections for simplicity. 
Implementing eq.~(\ref{def:F1=C.f1}) in a fit ensures a correct behavior of TMDPDFs in the collinear limit. Simultaneously, its usage propagates all the biases of collinear PDFs into the TMDPDFs, and   extractions done using  different collinear PDFs (keeping fixed all the rest) might show significant deviations from one another, up to the point of not overlapping the uncertainty bands. This bias is called PDF-bias~\cite{Bury:2022czx}, and it represents one of the largest problems of modern TMD phenomenology. A way to mitigate the PDF-bias consists in  taking into account the theoretical uncertainty of the collinear PDFs along with a flexible non-perturbative ansatz in the TMDPDF determination. The effects of PDF selection were discussed in ref.~\cite{Scimemi:2019cmh} and explicitly shown in ref.~\cite{Bury:2022czx} using four sets of collinear proton PDFs. 

The main goal of this work is to update the  unpolarized TMDPDFs determined within the 
\texttt{artemide}-framework~\cite{Scimemi:2017etj, Bertone:2019nxa, Vladimirov:2019bfa, Scimemi:2019cmh}. For this reason, we name this extraction ``ART23". Compared to the previous extraction~\cite{Scimemi:2019cmh} (SV19), ART23 is based on a larger dataset, which is greater by almost 50\%. It includes the recently released data~\cite{STAR:SX, ATLAS:2019zci, CMS:2019raw, CMS:2021oex, CMS:2022ubq, LHCb:2021huf} on the Z-boson production and, for the first time in the TMD phenomenology, the data on W-boson production~\cite{CDF:1991pgi, D0:1998thd}. We increase the perturbative accuracy to N$^4$LL and include the flavor dependence into the non-perturbative ansatz. The baseline collinear distribution that we use is the MSHT20 PDF set~\cite{Bailey:2020ooq}. We provide a critical and comprehensive treatment of uncertainties. For the first time, we consistently include the PDF uncertainty in the analysis along with the data uncertainty obtaining  a more trustful result.

This article is organised as follows. We devote section \ref{sec:theory} to the theoretical framework used for the computations. It includes the formulas for cross-sections, a brief description of the $\zeta$-prescription for the TMD evolution employed by \texttt{artemide}~\cite{artemide}, and the ans\"atze for the non-perturbative parts of the TMDPDFs and Collins-Soper (CS) kernel. In section \ref{sec:data} we discuss the experimental data and the selection criteria employed, while section \ref{sec:fit} contains the details pertaining to the fitting procedure. The outcome of the fit can be found in section \ref{sec:finalresults}. Finally, we present our conclusions in section \ref{sec:conclusions}. The plots comparing the experimental data and theoretical predictions are presented in Appendix \ref{app:plots}. In Appendix \ref{app:NNPDF}, we present the outcome of the same analysis using the NNPDF3.1 collinear PDFs~\cite{NNPDF:2017mvq} as  baseline.

\section{Theory overview}
\label{sec:theory}

There are several implementations of the TMD factorization framework. All of them are based on the same evolution equations, but differ in the realization of the solution, which is not unique. Conceptually, all realisations produce the same final result~\cite{Chiu:2012ir, Scimemi:2018xaf}. Practically, there are differences due to the truncation of the perturbative series. Also, the correlations between non-perturbative (NP) elements are different, which could affect the results of the extractions. In our fit we use the $\zeta$-prescription~\cite{Scimemi:2017etj, Scimemi:2018xaf}, which eliminates (theoretically) the correlation between the CS kernel and the TMD distributions. 

In this section, we present the relevant expressions for TMD factorization used in the current fit, and point the reader to the original works in refs.~\cite{Collins:1989gx, Becher:2010tm, Collins:2011zzd, Echevarria:2011epo, Echevarria:2012js, Chiu:2012ir, Vladimirov:2017ksc, Scimemi:2018xaf, Vladimirov:2021hdn} for details about their derivation. 

\subsection{Cross-section in the TMD factorization}

The DY lepton pair production is defined by the process
\begin{eqnarray}
h_1(P_1)+h_2(P_2)\to l(l)+l'(l')+X,
\end{eqnarray}
with $h_1$, $h_2$ the colliding hadrons, $l$, $l'$ the final state leptons, and the symbols in parentheses denoting the momentum of each particle. In the following, we neglect both hadron and lepton masses, i.e., 
$P_1^2\simeq P_2^2\simeq
l^2\simeq l'^2\simeq0,
$ 
since the target-mass corrections are negligible at the typical energies of DY data. 

The relevant kinematic variables in DY read
\begin{eqnarray}
s=(P_1+P_2)^2,\qquad q^\mu=(l+l')^\mu,\qquad q^2=Q^2,\qquad y=\frac{1}{2}\ln\frac{q^+}{q^-},
\end{eqnarray}
where $q^+$ and $q^-$ are the components of $q^\mu$ along $P_1^\mu$ and $P_2^\mu$, correspondingly.
The transverse components of vectors are projected by a tensor $g_T^{\mu\nu}$,  orthogonal to $P_1^\mu$ and $P_2^\mu$,
\begin{eqnarray}\label{eq:gmunuDY}
g_T^{\mu\nu}=g^{\mu\nu}-\frac{2}{s}\(P_1^\mu P_2^\nu+P_2^\mu P_1^\nu\).
\end{eqnarray}
The transverse momentum of the exchanged boson is $q_T^2=-\vec q_T^2=g_T^{\mu\nu}q_\mu q_\nu$. In the center-of-mass frame, the components of momenta are $P_1^+=P_2^-=\sqrt{s/2}$, and the variables $x_{1,2}$ are respectively
\begin{eqnarray}
x_1=\sqrt{\frac{Q^2+\vec q_T^2}{s}}e^{+y},\qquad x_2=\sqrt{\frac{Q^2+\vec q_T^2}{s}}e^{-y}.
\end{eqnarray}

At leading power in the TMD factorization, the cross section of the DY process mediated by a neutral boson reads
\begin{eqnarray}\label{DY:xSec}
&&\frac{d\sigma}{dQ^2dy d\vec q_T^2}=\frac{2\pi}{3N_c}\frac{\alpha^2_{\text{em}}}{sQ^2} \(1+\frac{\vec q_T^2}{2Q^2}\)\mathcal{P}_1\sum_{f}W_{f_1f_1}^{ff}(Q,\vec q_T^2)
\\\nn && \qquad
\times \Big[z^{\gamma\gamma}_lz_f^{\gamma\gamma}+z^{\gamma Z}_lz_f^{\gamma Z}\frac{2Q^2(Q^2-M_Z^2)}{(Q^2-M_Z^2)^2+\Gamma_Z^2M_Z^2}+
z^{ZZ}_lz_f^{ZZ}\frac{Q^4}{(Q^2-M_Z^2)^2+\Gamma_Z^2M_Z^2}\Big],
\end{eqnarray}
where $M_Z$ and $\Gamma_Z$ are the mass and decay width of the Z-boson, respectively. The summation index $f$ runs over all active quarks and anti-quarks. The function $W^{ff}_{f_1f_1}$ describes the hadronic part of the process and is defined below in eq.~(\ref{def:Wff}). The function $\mathcal{P}_1$ accounts for the modifications of the lepton phase space (fiducial cuts) and it is given in eq.~(\ref{DY:P1}). The factors 
$z_i^{jk}$ are the combinations of $Z$ and $\gamma$ couplings to quarks and leptons :
\begin{eqnarray}
z_f^{\gamma\gamma}&=&e_f^2;
\qquad
z_f^{\gamma Z}=\frac{T_3-2e_fs_W^2}{2s_W^2c_W^2};
\qquad
z_f^{Z Z}=\frac{(1-2|e_f|s_W^2)^2+4e_f^2s_W^4}{8s_W^2c_W^2},
\end{eqnarray}
with $s_W$ and $c_W$ the sine and cosine of the Weinberg angle, respectively. Here  $z_{f}^{\gamma Z}$ corresponds to the interference term in the product of amplitudes, and $z_f^{\gamma\gamma}$ and $z_f^{ZZ}$ corresponds to the diagonal terms. 

Some experiments do not correct the data for the detector acceptance  of leptonic phase space. In these cases, the collaborations provide the description of the fiducial regions to be accounted for in the theoretical predictions. On the theory side, the fiducial cuts are  part of the leptonic interactions described by the leptonic tensor, and do not interfere with the hadronic tensor. Due to this, the corrections for fiducial cuts can be incorporated exactly, as  part of the integration over the phase-space volume of the lepton pair. In the fits that we present here, these corrections are collected in a factor $\mathcal{P}_1$ defined as
\begin{eqnarray}\label{DY:P1}
\mathcal{P}_1&=&\int \frac{d^3l}{2E}\frac{d^3l'}{2E'}\delta^{(4)}(l+l'-q)((ll')-(ll')_T)\theta(\text{cuts})\Big/\[\frac{\pi}{6}Q^2\(1+\frac{\vec q_T^2}{2Q^2}\)\],
\end{eqnarray}
where $E$ and $E'$ are the energy components of the leptonic momenta $l$ and $l'$, and $\theta(\text{cuts})$ is the 
 Heaviside step function 
defining the volume of integration. Typically, the cuts on the lepton pair are reported as 
\begin{eqnarray}
\eta_{\text{min}}<\eta,\eta'<\eta_{\text{max}},\qquad l_T^2>p_1^2,\qquad {l'}^2_T>p_2^2,
\end{eqnarray}
where $\eta$ and $\eta'$ are the pseudo-rapidity of the leptons. The derivation of eq.~(\ref{DY:P1}) can be found in refs.~\cite{Scimemi:2017etj, Scimemi:2019cmh}. The factor $\mathcal{P}_1$ is normalized such that it is equal to unity in the absence of cuts. The integral on the right hand side of eq.~(\ref{DY:P1}) is computed numerically.

In the case of W-boson production the formula eq.~(\ref{DY:xSec}) changes to
\begin{eqnarray}\label{W:xSec}
&&\frac{d\sigma}{dQ^2dy d\vec q_T^2}=\frac{2\pi}{3N_c}\frac{\alpha^2_{\text{em}}}{s} \frac{Q^2}{(Q^2-M_W^2)^2+\Gamma_W^2M_W^2}\(1+\frac{\vec q_T^2}{2Q^2}\)\mathcal{P}_1\sum_{ff'}z^{WW}_{ll'}z_{ff'}^{WW} W_{f_1f_1}^{ff'}(Q,\vec q_T^2),\nn \\
\end{eqnarray}
where $M_W$ and $\Gamma_W$ are the mass and the decay width of the W-boson, and
\begin{eqnarray}
z_{ff'}^{WW}=\frac{|V_{ff'}|^2}{4s_W^2},
\end{eqnarray}
with $V_{ff'}$ either an element of the Cabibbo-Kobayashi-Maskawa (CKM) matrix (for quarks) or unity (for leptons). Notice that for the data considered in this work it is not necessary to pass from the variable $Q^2$ to the square of the transverse mass $m_T^2$; it is sufficient to know that $Q^2_{min}> m_T^2$ and one can integrate on $Q^2>Q^2_{min}$ (see also ref.~\cite{Gutierrez-Reyes:2020ouu}).

The hadronic function $W_{f_1f_1}^{ff'}$ is given by
\begin{eqnarray}\label{def:Wff}
W^{ff'}_{f_1f_1}(Q,q_T,x_1,x_2)&=&|C_V(-Q^2,\mu_H^2)|^2
\\\nn && \qquad \times
\int_0^\infty db\,bJ_0(bq_T)f_{1,f\ot h_1}(x_1,b;\mu_H,\zeta_1)f_{1,\bar f'\ot h_2}(x_2,b;\mu_H,\zeta_2),
\end{eqnarray}
where $f_1$ is the unpolarized TMD distribution,  $C_V$ is the hard coefficient function (that coincides with the vector form factor of the quark), and $J_0$ is the Bessel function of the first kind. The variable  $b$ is the Fourier conjugate of the transverse momentum $q_T$, and $\mu_H$ is the hard factorization scale. Finally, the argument $\zeta$ is the rapidity evolution scale, which is typical in TMD factorization~\cite{Collins:1989gx, Becher:2010tm, Collins:2011zzd, Echevarria:2011epo, Echevarria:2012js, Chiu:2012ir, Vladimirov:2017ksc, Scimemi:2018xaf, Vladimirov:2021hdn} that prescribes $\zeta_1\zeta_2=(2q^+q^-)^2$.

\begin{table}[]
\centering
\begin{tabular}{||c|c|c|c|c||c|c||}
\hline
$M_Z$ & $\Gamma_Z$& $M_W$ & $\Gamma_W$ & $\sin^2\theta_W$ & $m_c$ & $m_b$\\
 \hline
 91.1876 \,GeV & 2.4942\,GeV & 80.379 \,GeV & 2.089\,GeV& 0.2312 &
 1.40GeV & 4.75GeV\\
 \hline
\end{tabular}
\caption{The masses and electroweak parameters used in the present work. The CKM matrix elements are taken from the Particle Data Group (ed. 2022)~\cite{ParticleDataGroup:2022pth}. The quark masses are taken as in the MSHT20 collinear PDF extraction~\cite{Bailey:2020ooq}.}
\label{tab:ew}
\end{table}

Eq.~(\ref{DY:xSec}) and eq.~(\ref{W:xSec}) are only the leading power terms of the TMD factorization theorem. The power corrections scale as $q_T^2/Q^2$ and $\Lambda_{\text{QCD}}^2/Q^2$. Currently the theory of these corrections is in a developing stage, see e.g. refs.~\cite{Balitsky:2017gis, Vladimirov:2021hdn}. Thus, in what follows, we consider only the data for which the power corrections are (presumably) negligible.

The values of the electroweak parameters and heavy-quark masses used in this work are reported in table~\ref{tab:ew}. For the electroweak parameters we have taken the central values published in the Particle Data Group~\cite{ParticleDataGroup:2022pth}. We do not include their uncertainties, since they are smaller than other uncertainties involved. The strong coupling constant and the quark masses are taken from the PDF set that we use, that is MSHT20~\cite{Bailey:2020ooq} for the main fit.

\subsection{TMD evolution and optimal TMD distributions}
\label{sec:evolution}
The scale evolution of TMDPDFs is essential to include high and low energy data in an unique theoretical frame.  The TMD evolution equations are 
\begin{eqnarray}\label{def:TMD_ev_UV}
\mu^2 \frac{d}{d\mu^2} f_{1,q\ot h}(x,b;\mu,\zeta)&=&\frac{\gamma_F(\mu,\zeta)}{2}f_{1,q\ot h}(x,b;\mu,\zeta),
\\\label{def:TMD_ev_RAP}
\zeta\frac{d}{d\zeta}f_{1,q\ot h}(x,b;\mu,\zeta)&=& -\mathcal{D}(b,\mu)f_{1,q\ot h}(x,b;\mu,\zeta).
\end{eqnarray}
Note that these equations do not depend on the quark's flavor. This system of equations consists of a standard renormalization group equation, eq.~(\ref{def:TMD_ev_UV}), coming from the renormalization of  ultraviolet (UV) divergences, and a rapidity evolution equation, eq.~(\ref{def:TMD_ev_RAP}), specific of the TMD factorization that  comes from the factorization of rapidity divergences. The function $\mathcal{D}(\mu,b)$ is called the CS kernel and it is a non-perturbative (NP) function. The integrability condition (also known as Collins-Soper equation~\cite{Collins:1981va})
\begin{eqnarray}\label{def:integrability}
-\zeta\frac{d\gamma_F(\mu,\zeta)}{d\zeta}=\mu\frac{d\mathcal{D}(b,\mu)}{d\mu}=\Gamma_{\text{cusp}}(\mu),
\end{eqnarray}
holds, where $\Gamma_{\text{cusp}}(\mu)$ is the cusp anomalous dimension. Eq.~(\ref{def:integrability}) expresses the formal path-independent evolution  in the $(\mu$, $\zeta)$-plane. The TMD anomalous dimension is 
\begin{eqnarray}
\gamma_F(\mu,\zeta)=\Gamma_{\text{cusp}}(\mu)\ln\(\frac{\mu^2}{\zeta}\)-\gamma_V(\mu).
\end{eqnarray}
The perturbative expansion  for $\gamma_F$ is 
\begin{eqnarray}\label{def:gamma-pert}
\Gamma_{\text{cusp}}(\mu)=\sum_{n=0}^\infty a_s^{n+1}(\mu)\Gamma_n,\qquad
\gamma_V(\mu)=\sum_{n=1}^\infty a_s^{n}(\mu)\gamma_n,\qquad 
\text{with}\quad
a_s(\mu)=\frac{g^2(\mu)}{(4\pi)^2}.
\end{eqnarray}
In the present work we use the five-loop $\Gamma_{\text{cusp}}$~\cite{Herzog:2018kwj} and four-loop $\gamma_V$~\cite{Agarwal:2021zft}.

The selection of the initial evolution scale (i.e. the scale where the NP functions are extracted) is a key point. In our work, we use the initial scale associated with the $\zeta$-prescription~\cite{Scimemi:2018xaf, Scimemi:2019cmh} so that 
 the boundary conditions for the system in eq.~(\ref{def:TMD_ev_UV},~\ref{def:TMD_ev_RAP}) are given by the optimal TMD distribution~\cite{Scimemi:2018xaf}. This optimal TMDPDF is defined at the scale $(\mu,\zeta_\mu(b))$, which belongs to the special equi-potential (or null-evolution) line (of the evolution potential  introduced in ref.~\cite{Scimemi:2018xaf}) defined by the equation
\begin{eqnarray}\label{def:zeta-line}
\Gamma_{\text{cusp}}(\mu)\ln\(\frac{\mu^2}{\zeta_\mu(b)}\)-\gamma_V(\mu)=2\mathcal{D}(b,\mu)\frac{d \ln\zeta_\mu(b)}{d\ln\mu^2}.
\end{eqnarray}
with boundary conditions
\begin{eqnarray}\label{th:saddle-point}
\mathcal{D}(\mu_0,b)=0,\qquad \gamma_F(\mu_0,\zeta_0)=0.
\end{eqnarray}
For that reason, the optimal TMDPDF is exactly scale-independent (for any $\mu$ and $b$) and it is denoted without scales,
\begin{eqnarray}
f_{1,q\ot h}(x,b)\equiv f_{1,q\ot h}(x,b,\mu,\zeta_\mu(b)).
\end{eqnarray}
Eq.~(\ref{th:saddle-point}) defines the (unique) saddle point $(\mu_0,\zeta_0)$ of the evolution potential. Due to it, the value of $\zeta_\mu$ is finite for any $\mu$ (bigger than $\Lambda_\text{QCD}$) and $b$. A TMDPDF at any other scale can be obtained evolving the optimal TMDPDF along the path $\mu=\const$,
\begin{eqnarray}\label{def:TMD-evolved}
F(x,b;\mu,Q^2)=\(\frac{Q^2}{\zeta_\mu(b)}\)^{-\mathcal{D}(b,\mu)}F(x,b).
\end{eqnarray}

The hard factorization scale can be arbitrary since the dependence on it cancels between factors in the factorized expression of eq.~(\ref{def:Wff}). Practically, we set it to
\begin{eqnarray}\label{def:muH}
\mu_H=Q,
\end{eqnarray}
in order to minimize logarithms in the hard coefficient function.

Substituting eq.~(\ref{def:TMD-evolved},~\ref{def:muH}) into the definition of the structure functions $W_{f_1f_1}^{f}$ we obtain,
\begin{eqnarray}\label{def:Wff-final}
W_{f_1f_1}^f(Q,q_T;x_1,x_2)&=&|C_V(-Q^2,Q^2)|^2
\\\nn &&\qquad\times \int_0^\infty db\, bJ_0(bq_T)f_{1,f\ot h}(x_1,b)f_{1,\bar f\ot h}(x_2,b)\(\frac{Q^2}{\zeta_Q(b)}\)^{-2\mathcal{D}(b,Q)}.
\end{eqnarray}
These are the final expressions used to extract the NP functions.

The most important feature of the $\zeta$-prescription is that it exactly removes the correlation between the CS kernel and TMDPDF due to renormalization scales. The optimal TMDPDF does not depend on the CS kernel because it is determined exactly at the saddle point $\mathcal{D}=0$. Therefore, the CS kernel is treated as an independent NP function. Thus, the solution of eq.~(\ref{def:zeta-line}) with boundary conditions eq.~(\ref{th:saddle-point}) must be found for a generic $\mathcal{D}$ since it will change during the fitting procedure. This problem is solved in ref.~\cite{Vladimirov:2019bfa}. The corresponding solution for $\zeta_\mu(b)$ as a functional of $\mathcal{D}$ is denoted as $\zeta_\mu^{\text{exact}}[\mathcal{D}]$. The expression is rather lengthy, and can be found in ref.~\cite{Scimemi:2019cmh} at N$^3$LO, and directly in the code of \texttt{artemide}~\cite{artemide} at N$^4$LO. In contrast to SV19, in this work we use $\zeta_\mu^{\text{exact}}$ without any modification at small values of $b$.

\subsection{TMD distributions at small-$b$}
\label{sec:matching}

As mentioned in sec. \ref{sec:introduction}, in the regime of small-$b$ the TMDPDF can be expressed via the collinear PDFs with the help of the operator product expansion (OPE). The relation between TMDPDF and PDFs reads
\begin{eqnarray}\label{def:F-match}
\lim_{b\to0}f_{1,f\ot h}(x,b)&=&\sum_{f'}\int_x^1 \frac{dy}{y} \sum_{n=0}^\infty a_s^n(\mu_{\text{\tiny OPE}})C_{f\ot f'}^{[n]}\(\frac{x}{y},\mathbf{L}_{\mu_{\text{\tiny OPE}}}\)q_{f'\ot h}(y,\mu_{\text{\tiny OPE}}),
\end{eqnarray}
where $q(x,\mu)$ is the unpolarized PDF, the label $f'$ runs over all active quarks, anti-quarks and gluon, and
\begin{eqnarray}
\mathbf{L}_\mu=\ln\(\frac{\vec b^2 \mu^2 e^{2\gamma_E}}{4}\),
\end{eqnarray}
with $\gamma_E$ being the Euler constant. At LO the coefficient function reads $C_{f\ot f'}^{[0]}=\delta(1-x)\delta_{ff'}$ and 
the higher order terms are known 
up to  N$^3$LO~\cite{Echevarria:2016scs,Gehrmann:2014yya, Luo:2020epw, Ebert:2020yqt}. In the $\zeta$-prescription, the expressions of the coefficient functions are different from those presented in refs.~\cite{Echevarria:2016scs,Gehrmann:2014yya, Luo:2020epw, Ebert:2020yqt}, e.g. all double-logarithm contributions disappear. Up to N$^3$LO the expressions are
\begin{eqnarray}\label{th:match-coeff}
C^{[1]}_{f\ot f'}(x,\mathbf{L}_{\mu})&=&-\mathbf{L}_\mu P^{(1)}_{f\ot f'}+C_{f\ot f'}^{(1,0)}
\\
C^{[2]}_{f\ot f'}(x,\mathbf{L}_{\mu})&=&
\frac{\mathbf{L}^2_\mu}{2}\(P^{(1)}_{f\ot k}\otimes P^{(1)}_{k\ot f'}-\beta_0 P_{f\ot f'}^{(1)}\)
\\\nn && 
-\mathbf{L}_\mu \(P^{(2)}_{f\ot f'}+C_{f\ot k}^{(1,0)}\otimes P^{(1)}_{k\ot f'}-\beta_0 C_{f\ot k}^{(1,0)}\)
+C_{f\ot f'}^{(2,0)}+\frac{d^{(2,0)}\gamma_1}{\Gamma_0}
\\
C^{[3]}_{f\ot f'}(x,\mathbf{L}_{\mu})&=& 
\frac{\mathbf{L}_\mu^3}{6}\(-P^{(1)}_{f\ot k}\otimes P^{(1)}_{k\ot k'}\otimes P^{(1)}_{k'\ot f'}
+3P^{(1)}_{f\ot k}\otimes P^{(1)}_{k\ot f'}\beta_0
-2 P^{(1)}_{f\ot f'}\beta_0^2\)
\\\nn &&+\frac{\mathbf{L}_\mu^2}{2}\Big(
P^{(1)}_{f\ot k}\otimes P^{(2)}_{k\ot f'}
+
P^{(2)}_{f\ot k}\otimes P^{(1)}_{k\ot f'}
+
C^{(1,0)}_{f\ot k}\otimes P^{(1)}_{k\ot k'}\otimes P^{(1)}_{k'\ot f'}
\\\nn &&\qquad
-3C^{(1,0)}_{f\ot k}\otimes P^{(1)}_{k\ot f'}\beta_0
-2P^{(2)}_{f\ot f'}\beta_0
+2 C^{(1,0)}_{f\ot k}\beta_0^2
-P^{(1)}_{f\ot k}\beta_1\Big)
\\\nn &&
-\mathbf{L}_\mu\Big(
P^{(3)}_{f\ot f'}
+C^{(1,0)}_{f\ot k}\otimes P^{(2)}_{k\ot f'}
+C^{(2,0)}_{f\ot k}\otimes P^{(1)}_{k\ot f'}
-2C^{(2,0)}_{f\ot f'}\beta_0
-C^{(1,0)}_{f\ot f'}\beta_1
\\\nn && \qquad
+\frac{d^{(2,0)}\gamma_1}{\Gamma_0}P^{(1)}_{f\ot f'}
-2\frac{d^{(2,0)}\gamma_1}{\Gamma_0}\beta_0\Big)
+C_{f\ot f'}^{(3,0)}+C_{f\ot f'}^{(1,0)}\frac{d^{(2,0)}\gamma_1}{\Gamma_0}
\\\nn && +\frac{(d^{(2,0)})^2+d^{(3,0)}\gamma_1+d^{(2,0)}\gamma_2}{\Gamma_0}-\frac{d^{(2,0)}\gamma_1\Gamma_1}{\Gamma_0^2}
.
\end{eqnarray}
where the symbol $\otimes$ denotes the Mellin convolution, a summation over the intermediate flavour index $k$ is to be understood, and we have omitted the argument $x$ of the $C_{f\ot f'}$ on the r.h.s. for brevity. The functions $P^{(n)}(x)$ are the coefficients of the DGLAP kernel, $P(x)=\sum_n a_s^n P^{(n)}(x)$ and, up to three-loops, they can be found in ref.~\cite{Moch:2004pa}. The functions $C_{f\ot f'}^{(n,0)}(x)$ are the finite parts of the coefficient functions given in refs.~\cite{Gehrmann:2014yya, Echevarria:2016scs, Luo:2020epw, Ebert:2020yqt}. In particular, the NLO terms are
\begin{eqnarray}
C_{q\ot q}^{(1,0)}(x)=C_F\(2\bar x-\delta(\bar x)\frac{\pi^2}{6}\),\qquad C_{q\ot g}^{(1,0)}(x)=2x\bar x,
\end{eqnarray}
with  $\bar x=1-x$.

The OPE has an internal renormalization scale, $\mu_\text{OPE}$, which is not connected to the scales of the TMD evolution, as it happens f.i. in the case of the $b^*$-prescription~\cite{Collins:2011zzd, Bacchetta:2022awv}. Therefore, the expansion in eq.~(\ref{def:F-match}) is independent of $\mu_\text{OPE}$, and its value can be conveniently chosen such that it minimizes the logarithmic contributions at $b\to 0$ and, at the same time, it avoids the Landau pole at large-$b$. We have decided to use the same value for $\mu_{\text{OPE}}$ as in the SV19 extraction, i.e,
\begin{eqnarray}
\mu_{\text{OPE}}=\frac{2e^{-\gamma_E}}{b}+2\;\text{GeV}.
\end{eqnarray} 
The choice of the large-$b$ offset of $\mu_{\text{OPE}}$ as 2 GeV is motivated by a typical reference scale for PDFs (and lattice calculations). We remark that the factorization of the cross-section with TMD distributions is superior to a particular realization of the TMD distributions in terms of PDFs. Therefore, the actual choice of $\mu_{\text{OPE}}$ is a part of a TMDPDF modeling which (in the present case) includes the asymptotic collinear limit. Any modifications of $\mu_{\text{OPE}}$ would be absorbed by the NP parameters.

The CS kernel is an independent NP function, defined by the vacuum matrix element of a certain Wilson loop~\cite{Vladimirov:2020umg}. Analogously to TMDPDFs, the CS kernel can be computed at small values of $b$ using the OPE. The leading power expression has the form
\begin{eqnarray}\label{CS:small-b}
\lim_{b\to 0}\mathcal{D}(b,\mu)=\sum_{n=0}^\infty a_s^n(\mu) \sum_{k=0}^n \mathbf{L}_\mu^k d^{(n,k)},
\end{eqnarray}
where the explicit expressions for $d^{(n,k)}$ are given in~\cite{Vladimirov:2016dll, Li:2016ctv, Vladimirov:2017ksc} at N$^3$LO, and in~\cite{Duhr:2022yyp,Moult:2022xzt} at N$^4$LO. The power corrections to eq.~(\ref{CS:small-b}) have been computed in ref.~\cite{Vladimirov:2020umg} and they are proportional to the gluon condensate.

The coefficients of OPE and the values of the anomalous dimensions depend on the number of active quark flavors $N_f$. To treat this number we use the (naive) variable flavor number scheme, which sets $N_f=3$ for $\mu<m_c$, $N_f=4$ for $m_c<\mu<m_b$, and $N_f=5$ for $\mu>m_b$. In this scheme, the evolution integrals are smooth, whereas the coefficient functions have discontinuities (steps) at some $b$ values, those corresponding to the thresholds of $\mu_{\text{OPE}}(b)$. These discontinuities produce tiny oscillations in $W_{f_1f_1}$ after the Fourier transform. The input PDF distribution that we choose is the MSHT20 extraction ref.~\cite{Bailey:2020ooq}, which uses a similar scheme. We noticed that it is important to set our threshold parameters identical to those used in MSHT20, so that the oscillations are reduced to  a negligible $\sim 0.01\%$. The values of the threshold masses are reported in tab.~\ref{tab:ew}.

\subsection{Models for TMD distributions and CS kernel}
\label{sec:modelTMD}

We use the following phenomenological ansatz for our optimal unpolarized TMDPDFs:
\begin{eqnarray}\label{def:phen-f1}
f_{1,f\ot h}(x,b)&=&\int_x^1 \frac{dy}{y}\sum_{f'}C_{f\ot f'}\(y,\mathbf{L}_{\mu_{\text{OPE}}},a_s(\mu_{\text{\tiny OPE}})\)q_{f'\ot h}\(\frac{x}{y},\mu_{\text{\tiny OPE}}\)f^f_{\text{NP}}(x,b),
\end{eqnarray}
where the functions $f^f_{\text{NP}}$ accumulate the effect of power corrections to the small-$b$ matching. To satisfy the general structure of OPE~\cite{Moos:2020wvd}, $f^f_{\text{NP}}$ must be a function of $\vec b^2$ and behave as $f^f_{\text{NP}}(x,b)\sim 1+\mathcal{O}(\vec b^2)$ at small $b$. Additionally, $f^f_{\text{NP}}$ must decay at large $b$ to ensure the convergence of the Hankel transformation. Note that, in the ansatz of eq.~(\ref{def:phen-f1}), the logarithm of $b$ in the coefficient function grows unrestricted at large-$b$ (the so-called ``local $\ln(b)$''-setup). 

There is a large freedom in the definition of the functions $f^f_{\text{NP}}$. The main criterion for their construction is to have the maximum flexibility with the smallest number of free parameters. From our experience in  previous extractions we  deduce that  the optimal $b$-profile is the one with an exponential decay at $b\to\infty$ and Gaussian behavior at intermediate $b$~\cite{Scimemi:2017etj, Bertone:2019nxa, Scimemi:2019cmh}. The $x$-profile should distinguish large and small-$x$ contributions~\cite{Bertone:2019nxa, Scimemi:2019cmh, Bacchetta:2019sam, Bacchetta:2022awv}. After several attempts we opted for the following functional form
\begin{eqnarray}\label{def:fNP}
f_{NP}^{f}(x,b)=\frac{1}{\cosh\(\(\lambda^f_1(1-x)+\lambda^f_2 x\)b\)},
\end{eqnarray}
where $\lambda^f_{1,2}$ are free parameters. In the present fit, we distinguish $\{u,d,\bar u, \bar d, sea\}$ flavors, where $sea$ stands for $\{s, \bar s, c, \bar c, b, \bar b\}$-quarks. This decomposition is suggested by  the data as they  do not allow for the flavor separation of $sea$ quarks yet.
In total we have 10 free parameters, $\{\lambda_1^u, \lambda_2^{u}, \lambda_1^d, \lambda_2^{d}, $ $\lambda_1^{\bar u}, \lambda_2^{\bar u}, \lambda_1^{\bar d}, \lambda_2^{\bar d}, \lambda_1^{sea}, \lambda_2^{sea}\}$. 

The novel feature of the present model is the flavor dependence. In previous determinations of unpolarized TMD distributions $f_{\text{NP}}$ was chosen to be flavor-independent, which led to a number of undesirable effects, see ref.~\cite{Bury:2022czx} for a discussion. First of all, the extraction of the TMD distribution appeared to be strongly dependent on the choice of the collinear PDF, and often a choice of $f_{\text{NP}}$ valid for one PDF set was not successful for another (we call this effect ``PDF bias''). Secondly, the uncertainties of $f_{\text{NP}}$ were essentially underestimated. The inclusion of flavor dependence significantly reduces these problems. Additionally, the functional form used for each flavor $f$ in eq.~(\ref{def:fNP}) is much simpler in comparison to SV19~\cite{Scimemi:2019cmh} or MAP22~\cite{Bacchetta:2022awv}. Note that the TMD flavor dependence were also studied in ref.~\cite{Signori:2013gra, Bacchetta:2018lna}.

The CS kernel reads
\begin{eqnarray}\label{CS:ansatz}
\mathcal{D}(b,\mu)=\mathcal{D}_{\text{small-b}}(b^*,\mu^*)+\int^{\mu}_{\mu^*}\frac{d\mu'}{\mu'}\Gamma_{\text{cusp}}(\mu') +\mathcal{D}_{\text{NP}}(b),
\end{eqnarray}
where $\mathcal{D}_{\text{small-b}}$ is  given in eq.~(\ref{CS:small-b}), and $\mathcal{D}_{\text{NP}}$ provides the rest of the NP terms. The term with the integral in eq.~(\ref{CS:ansatz}) performs the evolution of the CS kernel\footnote{
In SV19 the evolution was taken into account by using the resummed version of $\mathcal{D}_{\text{small-b}}$~\cite{Echevarria:2012pw}. Formally, the resummed expression is the solution of the evolution equation eq.~(\ref{def:integrability}). However, for $b=b^*$, the resummed solution deviates from the exact solution at large $b$. For that reason, we prefer to use the explicit integral in the present fit. On top of this, the current implementation allows us to introduce the control scale $\mu^*$, often discussed by other groups\cite{Bacchetta:2019sam,Neumann:2022lft,Ebert:2020dfc}.
} from the scale $\mu^*$ to the scale $\mu$. Therefore, generally, eq.~(\ref{CS:ansatz}) does not depend on $\mu^*$, apart from the truncation of the perturbative series. The functions $b^*$ and $\mu^*$ are
\begin{eqnarray}
b^*(b)=\frac{b}{\sqrt{1+\frac{\vec b^2}{B^2_{\text{NP}}}}}
,\qquad \mu^*(b)=\frac{2e^{-\gamma_E}}{b^*(b)},
\end{eqnarray}
with a free parameter $B_{\text{NP}}$. This definition implies that $\mathbf{L}_{\mu^*}(b^*)=0$. 

Analogously to $f_{\text{NP}}$, the NP part of the CS kernel must be a function of $b^2$ to support the structure of the OPE. At large $b$ the CS kernel must be positive (to guarantee the convergence of the Hankel transform in eq.~(\ref{def:Wff-final})), and not grow faster than $(\vec b^2)^{1/2-\delta}$ with $\delta \geqslant 0$~\cite{Vladimirov:2020umg}. The expression for $\mathcal{D}_{\text{NP}}$ generalizes the one used in SV19 including logarithmic corrections,
\begin{eqnarray}\label{CS:NP}
\mathcal{D}_{\text{NP}}(b)=bb^*
\left[c_0 +c_1\ln\(\frac{b^*}{B_\text{NP}}\)
\right],
\end{eqnarray}
where $c_{0,1}>0$. One can easily identify three free parameters in our ansatz for the CS kernel, namely, $\{B_{\text{NP}}, c_0, c_1\}$. At large-$b$, the logarithmic term vanishes and the expression for the CS kernel becomes linear in $b$: $\mathcal{D}_{\text{NP}}\sim c_0B_{\text{NP}} b$. The term proportional to $c_1$ simulates the logarithmic dependence of the power corrections, and gives an extra flexibility to the ansatz at $b\sim B_{\text{NP}}$. In preliminary studies, we have found that such a correction provides a better agreement with  data in comparison to other models. This fact conveys the important message that both theory and experiment have achieved a degree of precision at which these effects become measurable.

\subsection{Definition of perturbative order and scale variation uncertainties}

In the factorized cross section defined above, we encounter three perturbative inputs and associated scales: 
\begin{itemize}
\item The perturbative hard coefficient function $C_V$, and associated hard factorization scale $\mu$, that separates $C_V$ and the TMD distributions in eq.~(\ref{def:Wff-final}).
\item The coefficient function of the small-$b$ operator product expansion for TMDPDF $C^{[n]}_{f\ot f'}$ and the associated scale $\mu_{\text{OPE}}$ in eq.~(\ref{def:F-match}).
\item The small-$b$ expansion for the CS kernel $\mathcal{D}_{\text{small-b}}$ and the associated scale $\mu^*$ in eq.~(\ref{CS:ansatz}).
\end{itemize}
Thanks to the $\zeta$-prescription, each perturbative  series can be truncated irrespectively of the perturbative orders included in the others. 

In this work, we use the highest known orders for all perturbative ingredients: the N$^4$LO (four-loop, $\sim a_s^4$) hard coefficient function $C_V$~\cite{Lee:2022nhh}, the N$^3$LO (four-loop, $\sim a_s^4$) light-like-quark anomalous dimension $\gamma_V$~\cite{Agarwal:2021zft}, the N$^3$LO (four-loop, $\sim a_s^4$) expression for the CS kernel $\mathcal{D}_{\text{small-b}}$~\cite{Duhr:2022yyp,Moult:2022xzt}, and the N$^3$LO (three-loop, $\sim a_s^3$) expression for the matching coefficient functions $C_{f\to f'}$~\cite{Luo:2020epw, Ebert:2020yqt}. The cusp anomalous dimension is taken at order $(\sim a_s^5)$~\cite{Herzog:2018kwj} (the expression in ref.~\cite{Herzog:2018kwj} is approximate, we consider the central value).

The input collinear PDFs, MSHT20~\cite{Bailey:2020ooq} (and NNPDF3.1~\cite{NNPDF:2017mvq} discussed in appendix \ref{app:NNPDF}), were obtained at NNLO, which implies the usage of the NNLO evolution kernel $P^{(3)}$. As a result, the logarithms included in the N$^3$LO small-$b$ coefficient functions are entirely compensated by the PDF evolution. The value and evolution of $a_s$ is provided together with the collinear PDF. The orders of the anomalous dimensions and coefficients functions are adjusted to each other, such that the scale-dependence is canceled at a given perturbative order. In the resummation nomenclature this combination of orders is referred to as N$^4$LL~\cite{Bacchetta:2019sam, Neumann:2022lft} (or N$^4$LL$^-$ in~\cite{Bacchetta:2022awv}, or, here, approximate N$^4$LL). The summary of the perturbative orders is also given in tab.~\ref{tab:pert}. 

\begin{table}[b]
\begin{center}
\begin{tabular}{||c|c||c|c|c|| c|c||}\hline
 $\Gamma_{\text{cusp}}$ &  $\gamma_V$   & $\mathcal{D}_{\text{small-b}}$ &  $C_{f\ot f'}$ & $C_V$ & PDF
\\\hline
 $a_s^5$ ($\Gamma_4$) & $a_s^4$ ($\gamma_4$) & $a_s^4$ ($d^{(4,0)}$) & $a_s^3$ ($C^{[3]}_{f\ot f'}$) & $a_s^{4}$ & NNLO
\\\hline
\end{tabular}
\end{center}
\caption{\label{tab:pert} Summary of the perturbative orders used for each part of the factorized cross section. The evolution of $\alpha_s$ is provided by the LHAPDF library and comes together with the PDF set (uniformly NNLO). In parentheses we write the last included term of the corresponding perturbative expansion (eq.~(\ref{def:gamma-pert}), (\ref{def:F-match}) and (\ref{CS:small-b})).}
\end{table}

\begin{figure}[t]
\centering
\includegraphics[width=0.45\textwidth]{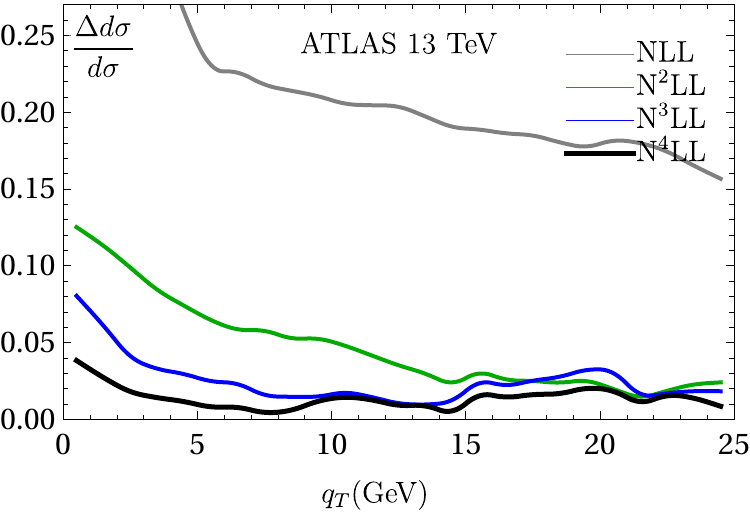}
\includegraphics[width=0.45\textwidth]{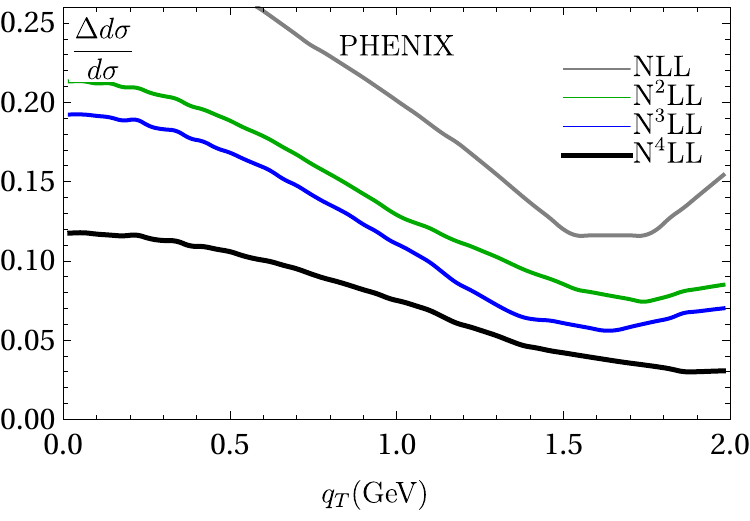}
\caption{\label{fig:scale} 
Ratio of scale variation band over theoretical cross section at different perturbative orders for Z/$\gamma$-boson production at ATLAS at $\sqrt{s}=13$ TeV (left), and for the DY process at PHENIX (right). The NP parameters and the PDF set are kept fixed. The definition of the variation band is given in eq.~(\ref{def:scale-vary}).}
\end{figure}
\begin{figure}[t]
\centering
\includegraphics[width=0.45\textwidth]{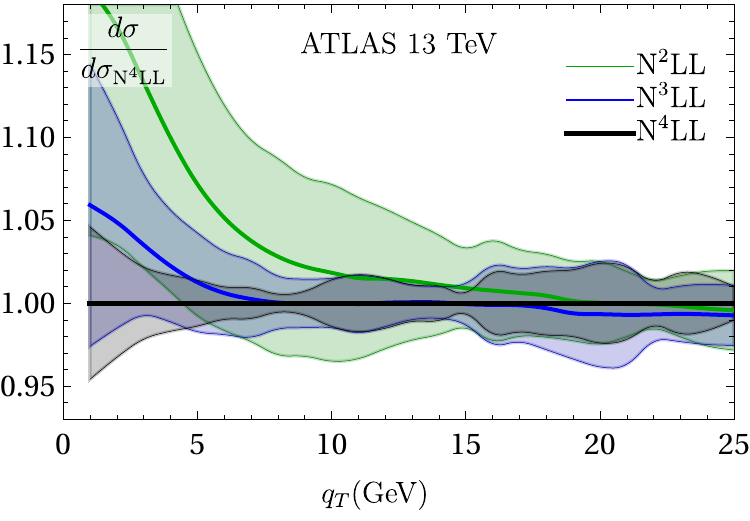}
\includegraphics[width=0.45\textwidth]{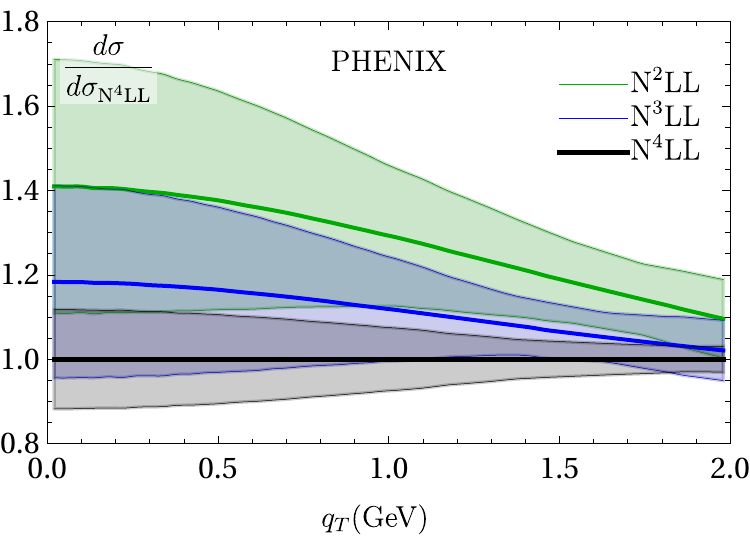}
\caption{\label{fig:scale_converge}
Ratio of cross sections at different orders over the one at N$^4$LL with the corresponding scale-variation band for the kinematics of Z/$\gamma$-boson production at ATLAS at $\sqrt{s}=13$ TeV, and for DY process at PHENIX.}
\end{figure}

To define the scale-variation band we multiply each scale with an independent factor $s_i$ ($i=2,3,4$), defined by the rule
\begin{eqnarray}
\Big\{\mu_H \to s_2 \mu_H, \mu^* \to s_3 \mu^*, \mu_{\text{OPE}}\to  s_4\frac{2e^{-\gamma_E}}{b}+2\text{GeV}\Big\}.
\end{eqnarray} 
The labels of parameters $s_i$ follow the enumeration used in ref.~\cite{Landry:2002ix}. The rule for $\mu_{\text{OPE}}$ is designed such that the variation of scale does not impact the NP large-$b$ part of the TMD distribution. As customary, each $s_i$ is allowed to vary by a factor 2, i.e. they can take the values $\{0.5, 1, 2\}$. In total, there are 27 combinations of $\{s_2,s_3,s_4\}$. For each one of them we compute the cross-section $d \sigma_i$. The size of the variation band is defined as the maximum (symmetric) deviation from the central value, i.e.
\begin{eqnarray}\label{def:scale-vary}
\Delta d\sigma=\max\(|d\sigma_i-d\sigma|_{i=1...27}\).
\end{eqnarray}

In fig.~\ref{fig:scale} we compare the sizes of variation bands for a representative sample  of high and low energy experiments for four consecutive orders, while using the same values of the NP parameter and the same PDFs. In fig.~\ref{fig:scale_converge} the same comparison is done for the cross-sections. As expected, we observe that the size of the bands reduces with the  increasing of the perturbative order. Each next-order curve is inside the variation band of the previous one. We also notice that for $q_T>5$ GeV the curves are close to each other. It indicates that convergence of the perturbative series is better than what one could estimate from the variation bands and  that the rule in eq.~(\ref{def:scale-vary}) is too conservative.

For small values of transverse momentum ($q_T<5$ GeV), the dominant contribution to the variation band arises from the factor $s_4$. However, in this regime the scale variation band is still not significant, as the effects of non-perturbative parameters override those of the perturbation theory. For $q_T>5$ GeV, the variation band is largely determined by the variation of $s_3$, and in this range, it remains nearly constant. For the ATLAS experiment at $\sqrt{s}=13$ TeV, the mean value of the variation band for $5$ GeV $<q_T<25$ GeV is 1.3$\%$. At low energies, the variation band remains large (around 10$\%$) even at the N$^4$LL level, but this is not a problem because in this range the theory prediction is largely non-perturbative.

We also note that the oscillations happen in the variation band at $q_T>10$ GeV. Studying  this effect in detail, we have  concluded that it is connected to our implementation of the flavor variable number mass scheme in a complex way. Basically, the discontinuities in the shapes of the distributions (due to the quark mass thresholds) change positions with the  variation of the  parameters $s_i$. These discontinuities generate tiny oscillations in the cross-section. For a natural selection of scales (that minimizes logarithms), the discontinuities (and hence the oscillations) are negligible, but some combinations of $s_i$'s are especially oscillatory since they generate many discontinuities. Therefore, the final shapes in fig.~\ref{fig:scale} are generated by the overlapping of the 27 curves with small oscillations in each one. We will address this problem in future works.

\subsection{Summary of inputs from theory}
\label{sec:summary-theory}

The cross-section of the vector boson production is computed with eq.~(\ref{DY:xSec}) and eq.~(\ref{W:xSec}), for neutral and charged EW boson, respectively. The structure function $W_{f_1f_1}$ is evaluated in the TMD factorization framework and given in eq.~(\ref{def:Wff-final}). The evolution of TMDPDFs is computed using the $\zeta$-prescription, and the phenomenological ans\"atze for the optimal unpolarized TMDPDFs and CS kernel are defined in eq.~(\ref{def:phen-f1}) and eq.~(\ref{CS:NP}). At small-$b$ the TMDPDFs are matched to the unpolarized PDFs, for which we use the MSHT20 extraction at NNLO ~\cite{Bailey:2020ooq}.  In tab. \ref{tab:pert} we list the perturbative orders used in each factor of the approximated N$^4$LL cross section. In total, there are 13 free parameters to be determined by the fitting procedure: 3 of these describe the CS kernel, while the remaining 10 are for the unpolarized TMDPDFs.

\section{Overview of data}
\label{sec:data}

The factorization formulas, eq.~(\ref{DY:xSec}) and eq.~(\ref{W:xSec}), are valid at small values of $q_T^2/Q^2$. This restriction has been studied from the phenomenological point of view in refs.~\cite{Scimemi:2017etj, Scimemi:2019cmh, Bacchetta:2019sam}. The common conclusion is that, for $q_T/Q<0.25$, the power corrections remain at the level of 1\%, and therefore the data can be safely included in phenomenological extractions. Above this threshold, the deviation between the theory and the measurements grows. However, such a  rule does not work for data with precision of the order of (or better than) 1\%. In this case, the power corrections significantly affect the quality of the description, despite being numerically small. 

In this work, we use the same general strategy for selecting the data as in refs.~\cite{Scimemi:2019cmh, Bertone:2019nxa}. Namely, we only include in our fit a data point if it fulfils the conditions 
\begin{eqnarray}\label{def:data-selection}
\frac{\langle q_T\rangle }{\langle Q \rangle}\equiv\delta<0.25\quad \text{   and      }\quad \(\delta^2<2\sigma\quad \text{or} \quad \langle q_T\rangle<10 \text{ GeV}\),
\end{eqnarray}
where $\langle q_T\rangle$ and $\langle Q\rangle$ are the average values of $q_T$ and $Q$ for the bin, and $\sigma$ is the relative uncorrelated uncertainty. The second condition is actually needed only for the high-energy data, as it is satisfied by all the data from the lower energy experiments with $Q<40$ GeV. The selection rules of eq.~(\ref{def:data-selection}) allow us to keep control of the predictive power of the theory, and still incorporate a large amount of data into the fit procedure. They are slightly softer than the rules used in refs.~\cite{Scimemi:2019cmh, Bertone:2019nxa}, because we plainly include all data with $\langle q_T\rangle<10 \text{ GeV}$. 

The bulk of the data considered here has already been used in previous extractions, such asin refs.~\cite{Scimemi:2017etj, Bertone:2019nxa, Scimemi:2019cmh, Bacchetta:2019sam, Bacchetta:2022awv}. This includes  the fixed-target E288, E605, E772 experiments from FermiLab (263 data points)~\cite{Ito:1980ev, Moreno:1990sf, E772:1994cpf}, the Z-boson production data from the CDF and D0 experiments at Tevatron (107 data points)~\cite{CDF:1999bpw, CDF:2012brb, D0:1999jba, D0:2007lmg, D0:2010dbl}, and the LHC run-1 and run-2 measurements of Z-boson production by the ATLAS, CMS, and LHCb collaborations (75 data points)~\cite{ATLAS:2015iiu, CMS:2011wyd, CMS:2016mwa, LHCb:2015okr, LHCb:2015mad}. Since these datasets are well-known and have been well-studied in the past, we refer the reader to refs.~\cite{Scimemi:2017etj, Scimemi:2019cmh, Bacchetta:2019sam, Bacchetta:2022awv} for a detailed discussion on their properties. In addition to these, we have included the latest measurements done at RHIC~\cite{PHENIX:2018dwt, STAR:SX} and the LHC~\cite{ATLAS:2019zci, CMS:2019raw, CMS:2021oex, CMS:2022ubq, LHCb:2021huf}, and the W-boson production data from Tevatron~\cite{CDF:1991pgi, D0:1998thd}. As we consider these data in the framework of TMD factorization for the first time, we find it worthwhile to highlight the particularities of each set in the following lines. 

\begin{figure}[t]
\centering
\includegraphics[width=0.6\textwidth]{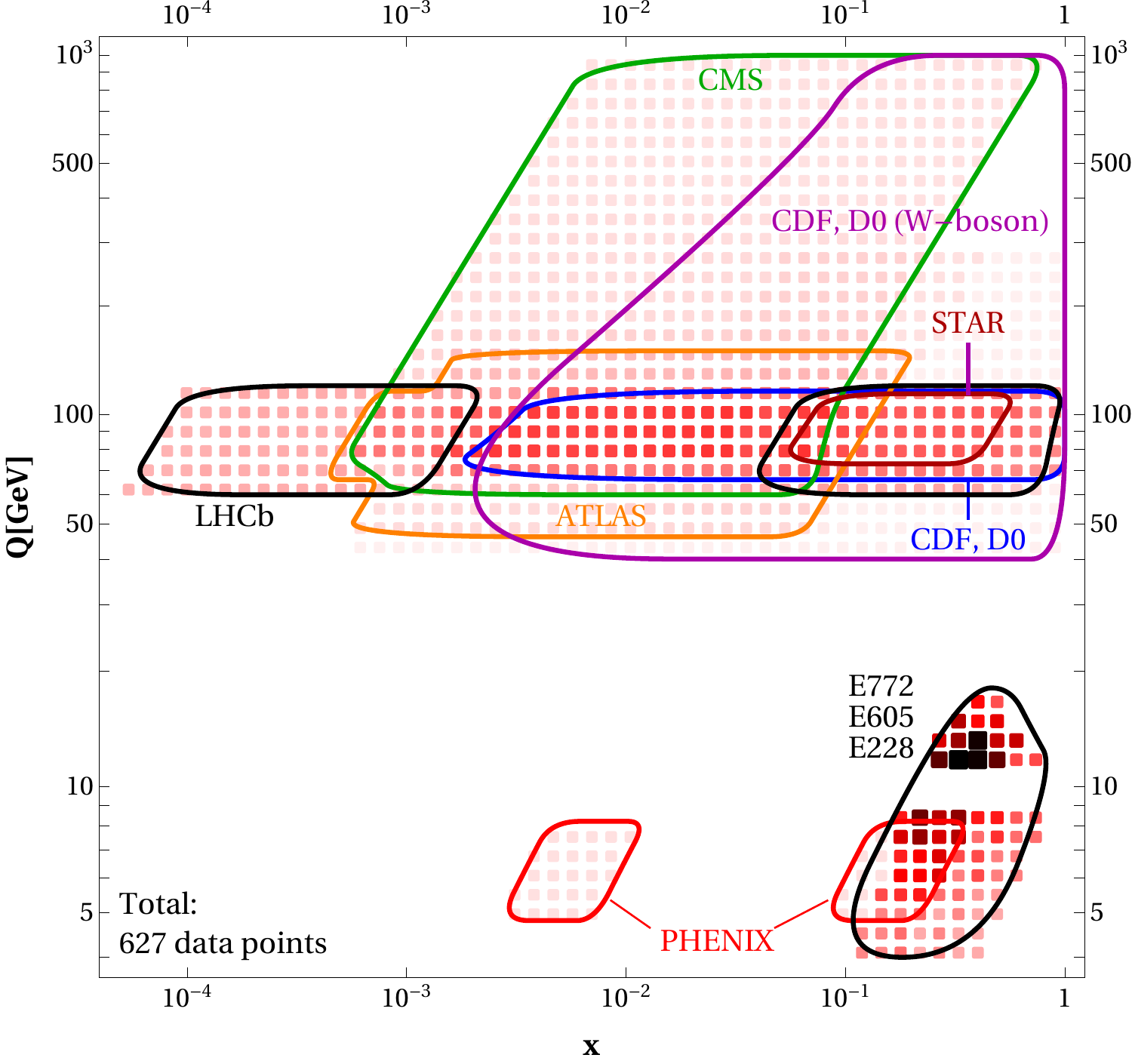}
\caption{Distribution of data in the $(x,Q)$ plane. Each data point can span large regions. The color gradient darkens with an increasing number of data points contributing to a particular $(x,Q)$ point.}
\label{fig:points}
\end{figure}

The PHENIX data~\cite{PHENIX:2018dwt} were taken at $\sqrt{s}=200$ GeV, which restricts the $Q$ range ($\langle Q\rangle=7$ GeV). It is the only modern DY measurement at low energy presently available, and has already been studied within TMD factorization in refs.~\cite{Scimemi:2019cmh, Bacchetta:2019sam, Bacchetta:2022awv}. The $Z/\gamma$-boson production measurement at STAR~\cite{STAR:SX} was made at moderately high energy ($\sqrt{s}=510$ GeV) during the 2018-2020 runs and the final results are currently in preparation for publication. Here we used the preliminary data.

In the present fit we include the recent $y$-differential measurements of Z-boson production at CMS~\cite{CMS:2019raw} and LHCb~\cite{LHCb:2021huf}, at $\sqrt{s}=13$ TeV. These replace the corresponding integrated measurements used in~\cite{Scimemi:2019cmh}. We also include the more precise ($\sim 0.1$\% uncertainty) measurement of the Z-boson differential cross-section by ATLAS~\cite{ATLAS:2019zci}. Finally, we include the high-$Q$ neutral-boson production measurements by the CMS collaboration~\cite{CMS:2022ubq}. This dataset is unique, since it spans up to $Q=1$ TeV (in several bins). For the interpretation of these data one should take into account that the bin-migration effects due to final state radiation are not incorporated in the published tables
\footnote{We thank Louis Moureaux and Buğra Bilin for their help with the interpretation of these data, and especially for sharing with us their code for the computation of the bin-migration effect.}.
Accounting for these effects is critical to confirm the agreement between theory and  measurement. Nevertheless, we are not able to describe the lowest lying bin $50$ GeV$ <Q<76$ GeV, for which we encountered a large difference in the normalization, and therefore, we exclude this bin. We have also excluded the $76$ GeV$ <Q<106$ GeV bin, using instead the $y-$differential measurement from the same run~\cite{CMS:2019raw}.

For the first time in  TMD phenomenology, we include W-boson production data~\cite{CDF:1991pgi, D0:1998thd}. Generally, the description of this observable is problematic within the TMD factorization framework because, usually, the data are integrated over a wide kinematic range, including regions where the TMD factorization conditions are not fulfilled. For a detailed discussion of this issue, see~\cite{Gutierrez-Reyes:2020ouu}. While the measurements~\cite{CDF:1991pgi, D0:1998thd} are fully integrated in $Q$, an explicit restriction on the transverse energy of the electron and neutrino (the missed transverse energy) was also imposed. This permits to find the lowest limit for $Q$. We have restricted the upper limit of integration to $300\,$GeV, since the contribution of higher $Q$ provides a negligible correction. To estimate the cut rules of eq.~(\ref{def:data-selection}) for these data we used $\langle Q \rangle=M_W$. 

In total, the present analysis includes 627 data points, summarized in tab. \ref{tab:data}. The kinematical coverage of the datasets in the $(x,Q)$-plane is shown in fig.~\ref{fig:points}. All the new data (w.r.t.~\cite{Bertone:2019nxa, Scimemi:2019cmh, Bacchetta:2019sam}) are at high energy. In fact, the new dataset totally supersedes the previous ones in both number of points (e.g. the present fit includes 227 points from LHC, vs. 80 in~\cite{Scimemi:2019cmh}) and precision. Therefore, the present selection allows for a more precise determination of the CS kernel (due to increased span in $Q$) and  provides a finer flavor separation due to the W-boson measurements.

\begin{table}[t]
\begin{center}
\scalebox{0.84}{
\begin{tabular}{|c||c|c|c|c|c|c|}
\hline
Experiment & ref. 
&$\sqrt{s}$ [GeV]& $Q$ [GeV] & $y$/$x_F$ & \specialcellcenter{fiducial\\region}  & \specialcellcenter{$N_{\rm pt}$\\after cuts}
\\\hline\hline 
E288 (200) & \cite{Ito:1980ev} 
& 19.4 & \specialcellcenter{4 - 9 in\\ 1~GeV bins$^*$} & $0.1<x_F<0.7$  & -& 43
\\\hline
E288 (300) & \cite{Ito:1980ev} 
& 23.8 & \specialcellcenter{4 - 12 in \\ 1~GeV bins$^*$} & $-0.09<x_F<0.51$ & - & 53
\\\hline
E288 (400) & \cite{Ito:1980ev} 
& 27.4 & \specialcellcenter{5 - 14 in \\ 1~GeV bins$^*$} & $-0.27<x_F<0.33$ & - & 79
\\\hline\hline
E605 & \cite{Moreno:1990sf} 
& 38.8 & \specialcellcenter{7 - 18 in \\ 5 bins$^*$} & $-0.1<x_F<0.2$ & -&  53
\\\hline\hline
E772 & \cite{E772:1994cpf} 
& 38.8 & \specialcellcenter{5 - 15 in \\ 8 bins$^*$} & $0.1<x_F<0.3$ & - & 35
\\\hline\hline
PHENIX & \cite{PHENIX:2018dwt} 
& 200 & 4.8 - 8.2 & $1.2<y<2.2$ & - & 3
\\\hline\hline
STAR & \cite{STAR:SX} 
& 510 & 73 - 114 & $|y|<1$ & \specialcellcenter{$p_T>25$ GeV \\ $|\eta|<1$} & 11
\\\hline\hline
CDF (run1) & \cite{CDF:1999bpw} 
& 1800 & 66 - 116 & - & - & 33
\\\hline
CDF (run2) & \cite{CDF:2012brb} 
& 1960 & 66 - 116 & - & - & 45
\\\hline
CDF (W-boson) & \cite{CDF:1991pgi} 
& 1800 & Q$>$40 & - & $p_{T,e},p_{T,\nu}>20$GeV & 6
\\\hline\hline
D0 (run1) & \cite{D0:1999jba} 
& 1800 & 75 - 105 & - & - & 16
\\\hline
D0 (run2) & \cite{D0:2007lmg} 
& 1960 & 70 - 110 & - & - & 9
\\\hline
D0 (run2)$_\mu$ & \cite{D0:2010dbl} 
& 1960 & 65 - 115 & $|y|<1.7$ & \specialcellcenter{$p_T>15$ GeV \\ $|\eta|<1.7$} & 4
\\\hline
D0 (W-boson) & \cite{D0:1998thd} 
& 1800 & Q$>$50 & - & $p_{T,e},p_{T,\nu}>25$GeV & 7
\\\hline\hline
ATLAS (8TeV) & \cite{ATLAS:2015iiu} 
& 8000 & 66 - 116 & \specialcellcenter{$|y|<2.4$\\ in 6 bins} & \specialcellcenter{$p_T>20$ GeV \\ $|\eta|<2.4$} & 30
\\\hline
ATLAS (8TeV) & \cite{ATLAS:2015iiu} 
& 8000 & 46 - 66 & $|y|<2.4$ & \specialcellcenter{$p_T>20$ GeV \\ $|\eta|<2.4$} & 5
\\\hline
ATLAS (8TeV) & \cite{ATLAS:2015iiu} 
& 8000 & 116 - 150 & $|y|<2.4$ & \specialcellcenter{$p_T>20$ GeV \\ $|\eta|<2.4$} & 9
\\\hline
ATLAS (13TeV) & \cite{ATLAS:2019zci} 
& 13000 & 66 - 116 & \specialcellcenter{$|y|<2.5$} & \specialcellcenter{$p_T>27$ GeV \\ $|\eta|<2.5$} & 5
\\\hline\hline
CMS (7TeV) & \cite{CMS:2011wyd} 
& 7000 & 60 - 120 & $|y|<2.1$ & \specialcellcenter{$p_T>20$ GeV \\ $|\eta|<2.1$} & 8
\\\hline
CMS (8TeV) & \cite{CMS:2016mwa} 
& 8000 & 60 - 120 & $|y|<2.1$ & \specialcellcenter{$p_T>20$ GeV \\ $|\eta|<2.1$} & 8
\\\hline
CMS (13TeV) & \cite{CMS:2019raw} 
& 13000 & 76 - 106 & \specialcellcenter{$|y|<2.4$\\ in 5 bins} & \specialcellcenter{$p_T>25$ GeV \\ $|\eta|<2.4$} & 64
\\\hline
CMS (13TeV) & \cite{CMS:2022ubq} 
& 13000 & \specialcellcenter{106 - 170\\ 170 - 350 \\ 350 - 1000} & $|y|<2.4$ & \specialcellcenter{$p_{1T}>25$ GeV \\ $p_{2T}>20$ GeV \\ $|\eta|<2.4$} & 34
\\\hline\hline
LHCb (7TeV) & \cite{LHCb:2015okr} 
& 7000 & 60 - 120 & $2<y<4.5$ & \specialcellcenter{$p_T>20$ GeV \\ $2<\eta<4.5$} & 8
\\\hline
LHCb (8TeV) & \cite{LHCb:2015mad} 
& 8000 & 60 - 120 & $2<y<4.5$ & \specialcellcenter{$p_T>20$ GeV \\ $2<\eta<4.5$} & 7
\\\hline
LHCb (13TeV) & \cite{LHCb:2021huf} 
& 13000 & 60 - 120 & \specialcellcenter{$2<y<4.5$ \\ in 5 bins} & \specialcellcenter{$p_T>20$ GeV \\ $2<\eta<4.5$} & 49
\\\hline\hline
Total & & 
& & & & 627
\\\hline
\end{tabular}
}
\par
*Bins with $9\lesssim Q \lesssim 11$ are omitted due to the $\Upsilon$ resonance.
\caption{
\label{tab:data}
Summary table of the data included in the fit. For each dataset we report: the reference, the centre-of-mass energy, the coverage in $Q$ and $y/x_F$, the cuts on the fiducial region (if any), and the number of data points that survive after the cut of eq.~(\ref{def:data-selection}).}
\end{center}
\end{table}

\section{Fit procedure}
\label{sec:fit}

Comparing the theoretical predictions with the data, we are able to restrict the free parameters of our ansatz for the TMD distributions, and in this way determine its NP part. This procedure is standard, and the present implementation is generally the same as the ones used in refs.~\cite{Scimemi:2017etj, Bertone:2019nxa, Scimemi:2019cmh, Bacchetta:2019sam, Bacchetta:2022awv}. However, in the present fit we treat the uncertainties more accurately and the PDF uncertainties are taken into account. The details of the procedure are reviewed in this section.

\subsection{Treatment of the experimental data}
\label{sec:treatment_of_data}

In the measurement of the cross-sections in an experiment there are several features that should be treated accurately in order to achieve a better consistency. We point them out one-by-one below.

\textit{Bin integration.} The expression for the cross-section in eq.~(\ref{W:xSec}) is given for a single point in the $(Q^2,y,q_T)$-space. On the experimental side, the measurement is provided for a volume-element of phase space, which can be obtained by averaging the theoretical expression
\begin{eqnarray}
\frac{d\sigma}{dQ^2dyd\vec q_T^2}\Big|_{\text{exp}}
&=&
(Q^2_{\text{max}}-Q^2_{\text{min}})^{-1}
(y_{\text{max}}-y_{\text{min}})^{-1}
(\vec q_{T\text{max}}^2-\vec q^2_{T\text{min}})^{-1}
\\\nn &&\qquad
\times 
\int_{Q_{\text{max}}}^{Q_{\text{min}}} 2QdQ
\int_{y_{\text{max}}}^{y_{\text{min}}} dy
\int_{q_{T\text{max}}}^{q_{T\text{min}}} 2\vec q_T d\vec q_T
\frac{d\sigma}{dQ^2dyd\vec q_T^2}\Big|_{\text{th.}}.
\end{eqnarray}
Here, the $Q_{\text{max}/\text{min}}$, $y_{\text{max}/\text{min}}$,
 and $\vec q_{T\text{max}/\text{min}}$ are the boundaries of the phase-space volume (bin). These integrations could not be simplified analytically, since the expression for the cross-section is too involved (especially in the presence of fiducial cuts). Accounting for the effect of finite bin size is of crucial importance, since for most of the experiments $d\sigma|_{\text{th.}}$ changes significantly and non-linearly within the bin (an explicit study of this  can be found in ref.~\cite{Scimemi:2019cmh}). It should also be taken into account that some experiments provide integrated (rather than averaged) data. For example, the LHCb measurements in refs.~\cite{LHCb:2015okr, LHCb:2015mad} are given for $\Delta \sigma$, that is,  the bin integrated cross-section without any weighting factors.

\textit{Normalization.} In a few cases the measurement is normalized to the total cross-section, i.e.
\begin{equation}
\frac{1}{\sigma}\frac{d\sigma}{dQ^2dyd\vec q_T^2}=
\(\int_0^\infty \frac{d\sigma}{dQ^2dyd\vec q_T^2} d\vec q_T^2\)^{-1}\frac{d\sigma}{dQ^2dyd\vec q_T^2}.
\end{equation}
This practice helps to reduce the normalization uncertainty of the measurement. Our formalism does not allow the computation of the weighting factor, since it includes  values of $q_T$ beyond the factorization range. In these cases we adopt the following procedure~\cite{Scimemi:2017etj, Bertone:2019nxa, Scimemi:2019cmh}. We compute
\begin{eqnarray}\label{def:norm}
\frac{1}{\sigma}\frac{d\sigma}{dQ^2dyd\vec q_T^2}\simeq
\mathcal{N}\frac{d\sigma}{dQ^2dyd\vec q_T^2},
\qquad
\mathcal{N}=
\frac{\sum_{\substack{\text{included}\\\text{bins}}}
\sigma^{-1}\frac{d\sigma}{dQ^2dyd\vec q_T^2}\Big|_{\text{exp.}}}{\sum_{\substack{\text{included}\\\text{bins}}}
\frac{d\sigma}{dQ^2dyd\vec q_T^2}\Big|_{\text{th.}}},
\end{eqnarray}
i.e. we normalize to the area of included data points. Note, that the normalization is done after the bin-integration. The numerator of $\mathcal{N}$ tells which part of the cross-section is included into the fit, while the denominator gives the theoretical estimate of normalization. If the shape of data is perfectly described by the theory, the factor $\mathcal{N}$ is independent on the number of included bins. Due to it, we expect that the deviation of the normalization factor computed with the full data set (in a framework that describes also large-$q_T$ data) from eq.~(\ref{def:norm}) is minimal and it cannot significantly impact the results of the fit.

The normalization procedure reduces the amount of information which can be gained from the data and it also potentially introduces an unknown uncertainty due to eq.~(\ref{def:norm}); therefore, we use unnormalized data whenever available. Only the following datasets require a normalization factor: D0 (run2)~\cite{D0:2007lmg,D0:2010dbl}, ATLAS (13 TeV)~\cite{ATLAS:2019zci}, and W-boson measurements~\cite{CDF:1991pgi, D0:1998thd}. The normalization factor is computed independently for each new set of values of the NP parameters (including the PDF replicas), and for each replica of the data, which provides the proper propagation of the fitting and data uncertainties.

\textit{Nuclear effects.} The fixed target experiments are done on nuclear targets (\textit{Cu} for E288 and E605~\cite{Ito:1980ev, Moreno:1990sf}, and $\,^2\!H$ for E772~\cite{E772:1994cpf}). To simulate the nuclear environment we perform the iso-spin rotation of the TMD distribution. Namely, we set
\begin{eqnarray}
f_{1,u\ot A}(x,b)&=&
\frac{Z}{A}f_{1,u\ot p}(x,b)
+
\frac{A-Z}{A}f_{1,d\ot p}(x,b),
\\\nn
f_{1,d\ot A}(x,b)&=&
\frac{Z}{A}f_{1,d\ot p}(x,b)
+
\frac{A-Z}{A}f_{1,u\ot p}(x,b),
\end{eqnarray}
where $A$ is the atomic number and $Z$ is the charge of a nuclear target. The effects of heavier mass in the TMD distribution can be ignored since they are compensated by the scaling of the momentum fraction, as it has been shown in ref.~\cite{Moos:2020wvd}. We do not include any finer modifications, since the fixed target data are not precise enough to distinguish them. 

\textit{Artemide.} The computation of the theory prediction, as well as all integrals and factors required for comparison with the data, are done in \texttt{artemide}, a multi-purpose code for the phenomenology of TMD factorization. It is based on the $\zeta$-prescription, which allows for many simplifications of the code and improves the speed of the computation. In particular, the computation of the theory prediction for the full dataset (627 points) with all integrations required for comparison with experiment consumes 20-30 seconds on the average desktop (12 cores processor) depending on the NP input. The code of \texttt{artemide} is written in FORTRAN95. It is open-source and available at~\cite{artemide}. The values of $\alpha_s$ and the collinear PDFs are obtained from the LHAPDF interface~\cite{Buckley:2014ana}.

The legacy of \texttt{artemide} is established in many previous global fits, including fits of various unpolarized~\cite{Scimemi:2017etj, Bertone:2019nxa, Vladimirov:2019bfa, Scimemi:2019cmh, Hautmann:2020cyp, Gutierrez-Reyes:2020ouu, Bury:2022czx}, and polarized~\cite{Bury:2020vhj, Bury:2021sue, Horstmann:2022xkk} observables. For the present work we have updated it with the expressions for N$^3$LO coefficient functions and N$^3$LO anomalous dimensions, without further modifications of the internal structure of the code.

\subsection{Definition of the $\chi^2$-test function and related quantities}

The agreement of the theory prediction and the data is quantified by the $\chi^2$-test function. We use the standard definition adopted from the fits of collinear PDFs in refs.~\cite{Ball:2008by, Ball:2012wy}, to which we refer the reader for a detailed discussion. The $\chi^2$-test function is defined as
\begin{eqnarray}
\chi^2=\sum_{i,j\in \text{data}}(m_i-t_i)V^{-1}_{ij}(m_j-t_j),
\end{eqnarray}
where $i$ and $j$ run over all data points included into the fit, $m_i$ and $t_i$ are the experimental value and theoretical prediction for point $i$, respectively, and $V^{-1}_{ij}$ is the inverse of the covariance matrix. The covariance matrix is defined as
\begin{eqnarray}\label{def:V}
V_{ij}=\delta_{ij}\Delta_{i,\text{uncorr}.}^2+\sum_{l}\Delta_{i,\text{corr}.}^{(l)}\Delta_{j,\text{corr}.}^{(l)},
\end{eqnarray}
where $\Delta_{i,\text{uncorr}.}$ is the uncorrelated uncertainty of the measurement (if there are more than one, they are summed in squares), and $\Delta_{i,\text{corr}.}^{(l)}$ is the $l$-th correlated uncertainty. The normalization uncertainty (due to the luminosity) is included in the $\chi^2$ as one of the correlated uncertainties. This definition of the covariance matrix and the $\chi^2$-test function takes into account the nature of the experimental uncertainties and also provides a faithful estimate of the agreement between
data and theoretical predictions.

Additionally, this definition of $\chi^2$ allows for an one-to-one separation of correlated and uncorrelated parts~\cite{Ball:2012wy} for each correlated dataset. Therefore,
\begin{eqnarray}
\chi^2=\chi^2_D+\chi^2_\lambda,
\end{eqnarray}
where $\chi^2_D$ incorporates the contribution due to the uncorrelated uncertainties of the measurement, while $\chi^2_\lambda$ is the rest. The definitions of $\chi_D^2$ and $\chi_\lambda^2$ are
\begin{eqnarray}
\chi_D^2=\sum_{i}\frac{(m_i-\bar t_i)^2}{\Delta_{i,\text{uncorr.}}^2},
\qquad
\chi_\lambda^2=\sum_{l}\lambda_l^2,
\end{eqnarray}
with $\lambda$ and $\bar t$ defined below. Loosely speaking, $\chi^2_D$ ($\chi^2_\lambda$) shows the agreement in the shape (normalization) between the theory prediction and the experimental measurement. 

To perform this decomposition, one should compute the ``nuisance parameters'' $\lambda_l$ ($l$ enumerates the number of correlated uncertainties for the given dataset)~\cite{Ball:2012wy, Bertone:2019nxa} for a given theory prediction. Then the theory prediction is decomposed as
\begin{eqnarray}\label{t=t-d}
t_i=\bar t_i-d_i,\qquad d_i=\sum_l \lambda_l \Delta^{(l)}_{\text{corr.}}~.
\end{eqnarray}
The terms $d_i$ are interpreted as correlated shifts in the predictions which generate the $\chi^2_\lambda$ contribution. The terms $\bar t_i$ represent the part of the theory curve that contributes solely to $\chi^2_D$, and thus has a ``perfect'' normalization. 

This decomposition is often useful for analysis and visualization because the correlated (normalization) uncertainty is generally much larger than the uncorrelated one. Hence, in plots comparing with data, we show the $\bar t_i$ part of the prediction.  This allows us to visually confirm the ``good'' values of $\chi^2$. The average value of correlated shifts relative to the cross-section $\langle d_i/\sigma\rangle$ is presented in tab.~\ref{tab:chi2}. It exhibits the general disagreement in the normalization between the predictions of TMD factorization  and the measurements.

The computation of the $\chi^2$ value and further manipulations are performed with the \texttt{DataProcessor} library, which is written in PYTHON and interfaced to \texttt{artemide} via the standard \texttt{f2py} library. The code of \texttt{DataProcessor}, together with the collection of the experimental data points, and all programs used for the present work, can be found in~\cite{DataProcessor}.

\subsection{Minimisation procedure and uncertainty estimation}
\label{sec:uncert}

The ansatz for the TMD distributions contains in total 13 parameters, which we denote as $\overrightarrow{\lambda}$
\begin{eqnarray}
\overrightarrow{\lambda}=\{B_{\text{NP}}, c_0, c_1, \lambda_1^u, \lambda_2^{u}, \lambda_1^d, \lambda_2^{d}, \lambda_1^{\bar u}, \lambda_2^{\bar u}, \lambda_1^{\bar d}, \lambda_2^{\bar d}, \lambda_1^{sea}, \lambda_2^{sea}\}.
\end{eqnarray}
To find the optimal values of $\overrightarrow{\lambda}$ we minimize the $\chi^2$ using the library \texttt{iMinuit}~\cite{iminuit}. The resulting value $\overrightarrow{\lambda}_{\text{center}}$ is called \textit{central value} fit. The central value fit is used  only as an initial assumption for all further minimization procedures described below.

 The propagation of initial uncertainties to extracted values of the TMD distributions is both the central component and the most time-consuming step in the computation. We employ the resampling method to perform this task, generating samples of setups distributed according to the initial uncertainties.
 We distinguish two sources of uncertainties:
\begin{itemize}
\item \textit{Experimental uncertainties}. These are uncertainties due to 
 experimental measurements (statistical, systematic, etc.). To propagate these uncertainties we generate replicas of pseudo-data. A replica consisting of pseudo-data is obtained by adding Gaussian noise to the values of the data points. The parameters of the noise are dictated by the correlated
and uncorrelated experimental uncertainties. Here, one should account for the nature of the correlated uncertainty in order to avoid the so-called D'Agostini bias~\cite{DAgostini:2003syq}. The detailed algorithm for the generation of pseudo-data is given in ref.~\cite{Ball:2008by}.
\item \textit{Uncertainty in collinear PDFs}. To propagate these uncertainties we use the Monte-Carlo sampling of the PDF distributions. The MSHT PDF set is given with (asymmetric) Hessian uncertainties, and the Monte-Carlo samples are generated according to the prescription given in ref.~\cite{Hou:2016sho}. 
\end{itemize}
Other sources of uncertainties (such as uncertainties in $M_Z$ or $\alpha_s$, uncertainties due to missed higher perturbative orders, etc) are considered negligible.

The initial uncertainties included in the analysis originate from different sources, and it is not always clear how they should be combined. This is because the propagation mechanism for each uncertainty differs. The experimental uncertainties modify the expression of $\chi^2$ (by changing $V$ and $m$), while the PDF uncertainty changes the theory expression (by changing the boundary value of the TMD distribution). In this work, we consider them on the same foot and generate samples by varying \textit{data and PDF simultaneously}. Here, the PDF replica is randomly selected from the pre-generated sample
\footnote{We use a distribution with 1000 replicas. The preservation of this ensemble is important for the future use of extracted TMD distributions. We will provide the sample used in this work in the LHAPDF format upon request.}, and thus could be present in the final ensemble several times. This method of uncertainty propagation is more accurate than any used before. For example, in the method of ref.~\cite{Bury:2022czx} both types of uncertainties were sampled independently and then combined into a single uncertainty band; in refs.~\cite{Bacchetta:2019sam, Bacchetta:2022awv} the PDF uncertainty was introduced into the definition of $\chi^2$.

For each setup sample, we minimize the $\chi^2$-function and find a set of parameters $\overrightarrow{\lambda}$, which defines the minimum. This procedure is repeated 1000 times, giving us an ensemble $\{\Lambda_i=(\overrightarrow{\lambda}_i,n_i)\}$, where $i=1,...,1000$ and $n_i$ is the serial number of the PDF replica used in the sample setup. The PDF replica's serial numbers must be preserved since the values of  $\overrightarrow{\lambda}$ are essentially correlated with it. The ensemble of $\Lambda$'s entirely describes the TMD distributions and the CS kernel. This ensemble is used in all further manipulations. The list of values of $\Lambda_i$ can be found in the \texttt{artemide} repository~\cite{artemide}, in the (human-readable) format suitable for automatic processing by the \texttt{DataProcessor}.

Using the ensemble $\Lambda$, we can find the mean values of the parameters $\overrightarrow{\lambda}_0=\langle \overrightarrow{\lambda}_i \rangle$ associated with the central PDF replica (since the averaging of PDF replicas produces the central one by definition). The distribution is not entirely Gaussian, and this uncertainty band is associated with the $68\%$-confidence interval (68\%CI). The boundaries of the 68\%CI are computed using the resampling method by computing the 16\% and 84\% quantiles.

The central values and uncertainty bands for secondary values, such as TMD distributions, cross-sections, $\chi^2$, etc., are computed starting from the ensemble $\Lambda$. For example, to obtain the TMD distribution we compute the ensemble $F_i=F[\Lambda_i]$. The central value is then the mean $\langle F_i\rangle$ and the uncertainty band is 68\%CI of $F_i$. Note, that $\langle F[\Lambda_i]\rangle \neq F[\langle \Lambda_i\rangle]$, due to the correlations in-between members of $\Lambda_i$. The procedure described above allows to propagate all  correlations correctly.

This work presents a comprehensive analysis of error propagation in TMD phenomenology, which is the first of its kind. The proposed procedure is expected to reduce the dependence on PDFs as input parameters. However, the approach comes at the cost of increased computational complexity.
In the present work, we use the MSHT20 PDF set~\cite{Bailey:2020ooq}, which we present as the main result. To cross check we also made an independent run (with 300 replicas) with the NNPDF3.1 PDF set~\cite{NNPDF:2017mvq}. The results of this run are given in  appendix \ref{app:NNPDF}.

\section{Results}
\label{sec:finalresults}
In this section we present the results of the fitting procedure, starting with the quality of the data description, and finishing with the presentation of the extracted TMD distributions and CS kernel.

\subsection{Quality of data description}

\begin{figure}[t]
\centering
\includegraphics[width=0.4\textwidth]{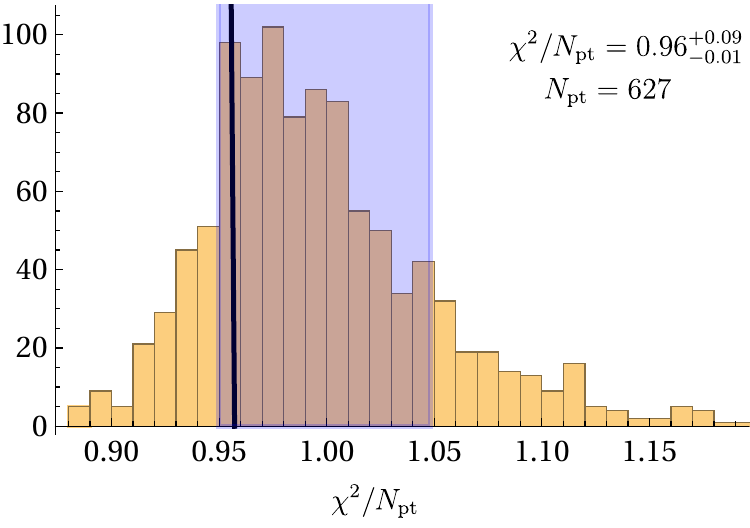}
\caption{The histogram of the $\chi^2$ values for the full dataset. The black line marks the position of the mean prediction, and the blue band shows the 68\%CI of the $\chi^2$ distribution.}
\label{fig:chi2_total}
\end{figure}

We find that the current setup perfectly describes the data. The central value fit results in $\chi^2/N_{\text{pt}}= 0.93$. For the mean prediction (i.e. $\langle d\sigma[\Lambda_i] \rangle$), $\chi^2/N_{\text{pt}}= 0.957$, with the 68\%CI (0.950, 1.048). The histogram of $\chi^2/N_{\text{pt}}$ is given in fig.~\ref{fig:chi2_total}. The complete list of the $\chi^2$-values for all datasets is presented in tab. \ref{tab:chi2}.

\begin{table}[t]
\def\arraystretch{1.2}
\centering
\begin{tabular}{|l|c||c|c|c||c|}
\hline
dataset & $N_{\text{pt}}$ & $\chi^2_D/ N_{\text{pt}}$& $\chi^2_\lambda/ N_{\text{pt}}$
& $\chi^2/ N_{\text{pt}}$ & $\langle d/\sigma \rangle$
\\\hline
CDF (run1)      & 33 & 0.51 & 0.16 & $0.67_{-0.03}^{+0.05}$ & 9.1\%
\\\hline
CDF (run2)      & 45 & 1.58 & 0.11 & $1.59_{-0.14}^{+0.26}$ & 4.0\%
\\\hline
CDF (W-boson)   & 6 & 0.33 & 0.00 & $0.33_{-0.01}^{+0.01}$ & --
\\\hline\hline
D0 (run1)   & 16 & 0.69 & 0.00 & $0.69_{-0.03}^{+0.08}$ & 7.1\%
\\\hline
D0 (run2)   & 13 & 2.16 & 0.16 & $2.32_{-0.32}^{+0.40}$ & --
\\\hline
D0 (W-boson)   & 7 & 2.39 & 0.00 & $2.39_{-0.18}^{+0.20}$ & --
\\\hline\hline
ATLAS (8TeV, $Q\sim M_Z$)   & 30 & 1.60 & 0.49 & $2.09_{-0.35}^{+1.09}$ & 4.1\%
\\\hline
ATLAS (8TeV)   & 14 & 1.11 & 0.11 & $1.22_{-0.21}^{+0.47}$ & 2.3\%
\\\hline
ATLAS (13 TeV)   & 5 & 1.94 & 1.75 & $3.70_{-2.24}^{+16.5}$ & --
\\\hline\hline
CMS (7TeV)       & 8 & 1.30 & 0.00 & $1.30_{-0.01}^{+0.03}$ & --
\\\hline
CMS (8TeV)   & 8 & 0.79 & 0.00 & $0.78_{-0.01}^{+0.02}$ & --
\\\hline
CMS (13 TeV, $Q\sim M_Z$)   & 64 & 0.63 & 0.24 & $0.86_{-0.11}^{+0.23}$ & 4.3\%
\\\hline
CMS (13 TeV, $Q>M_Z$)   & 33 & 0.73 & 0.12 & $0.92_{-0.15}^{+0.40}$ & 1.0\%
\\\hline\hline
LHCb (7 TeV)   & 10 & 1.21 & 0.56 & $1.77_{-0.31}^{+0.53}$ & 5.0\%
\\\hline
LHCb (8 TeV)   & 9 & 0.77 & 0.78 & $1.55_{-0.50}^{+0.94}$ & 4.3\%
\\\hline
LHCb (13 TeV)   & 49 & 1.07 & 0.10 & $1.18_{-0.01}^{+0.25}$ & 4.5\%
\\\hline\hline
PHENIX   & 3 & 0.29 & 0.12 & $0.42_{-0.10}^{+0.15}$ & 10.\%
\\\hline
STAR   & 11 & 1.91 & 0.28 & $2.19_{-0.31}^{+0.51}$ & 15.\%
\\\hline\hline
E288 (200)   & 43 & 0.31 & 0.07 & $0.38_{-0.05}^{+0.12}$ & 44.\%
\\\hline
E288 (300)   & 53 & 0.36 & 0.07 & $0.43_{-0.04}^{+0.08}$ & 48.\%
\\\hline
E288 (400)   & 79 & 0.37 & 0.05 & $0.48_{-0.03}^{+0.11}$ & 48.\%
\\\hline
E772   & 35 & 0.87 & 0.21 & $1.08_{-0.05}^{+0.08}$ & 27.\%
\\\hline
E605   & 53 & 0.18 & 0.21 & $0.39_{-0.00}^{+0.03}$ & 49.\%
\\\hline\hline
\textbf{Total}   & 627 & 0.79 & 0.17 & $0.96_{-0.01}^{+0.09}$ & \\
\hline
\end{tabular}
\caption{The values of $\chi^2$ for the individual datasets. The last column shows the average relative shifts computed by eq.~(\ref{t=t-d}), which indicate the size of normalization  discrepancy.}
\label{tab:chi2}
\end{table}

In comparison to the SV19 fit~\cite{Scimemi:2019cmh} we observe an overall improvement in the $\chi^2$, which is especially significant for the description of the LHC data ($\chi^2_{\text{LHC}}/N_{\text{pt}}=1.26_{-0.15}^{+0.76}$ with $N_{\text{pt}}=230$), and the low-energy DY data ($\chi^2_{\text{low}}/N_{\text{pt}}=0.50_{-0.03}^{+0.09}$ with $N_{\text{pt}}=266$). Similarly to SV19, we observe that the low-energy DY data suffer of deficits in the normalization. This is a known feature of TMD factorization (see e.g. the extended discussions in refs.~\cite{Vladimirov:2019bfa, Scimemi:2019cmh}). Given that the data have very large normalization uncertainties, these deficits do not significantly impact the value of $\chi^2$; therefore it is not clear at the moment whether the problem arises from a shortcoming of the theory or of the measurements. Let us also mention that the PHENIX measurement ($\langle Q\rangle=7$ GeV) does not show any problem with the normalization.

\begin{figure}[t]
\centering
\includegraphics[width=0.95\textwidth]{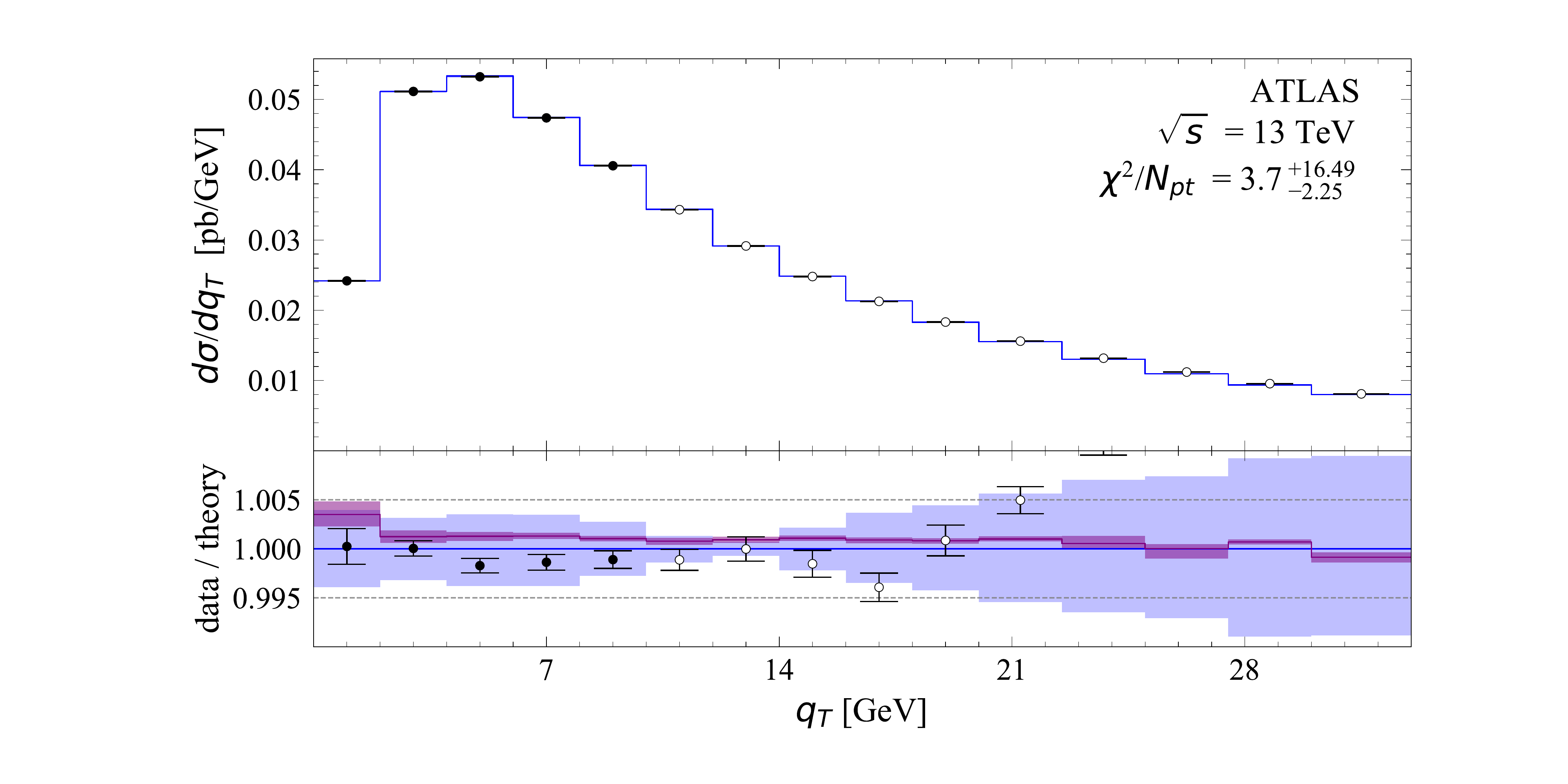}
\caption{Comparison of data (here, the Z-Boson production by ATLAS $\sqrt{s}=13$ TeV measurement in ref.~\cite{ATLAS:2019zci}) with the theoretical prediction (blue). The blue band is the 68\%CI. The filled points are included into the fit and the displayed $\chi^2$ corresponds to those only. The purple line (band) shows the prediction (uncertainty) obtained without taking into account the PDF uncertainties.}
\label{fig:example-data}
\end{figure}

In fig.~\ref{fig:example-data} we present the comparison of  theory vs.  ATLAS 13 TeV measurement, which is the most precise measurement at our disposal (with uncorrelated uncertainties $<0.5\%$). In this plot one can see that TMD factorization works up to $q_T\simeq 0.2\ Q$ (even if in this particular case only data up to $q_T=10$ GeV are included into the fit). At larger $q_T$, the theory prediction is systematically lower than the measurement: this is a signal of the necessity for power corrections. The full collection of data plots is given in  appendix \ref{app:plots}.

We emphasize that the uncertainty band obtained in this fit is larger than in  previous analyses with  \texttt{artemide}~\cite{Bertone:2019nxa, Scimemi:2019cmh} because of the PDF uncertainty evaluated in the present work. Also, since we cannot control the PDF uncertainty, the band is often larger than the uncertainty of the measurement. It indicates that a simultaneous extraction of PDF and TMD distributions will reduce the uncertainties of both. We also have observed that most part of our uncertainty band is correlated. To determine the size of the correlation we computed the covariance matrix for the full set of replicas,  $\text{cov}(d\sigma(\Lambda_i),d\sigma(\Lambda_j))$ and fitted it with the form of eq.~(\ref{def:V}), and determined $\Delta_{\text{th.uncorr}}$ and $\Delta_{\text{th.corr}}$ for the theory prediction. We found that the portions of correlated and uncorrelated parts depend on $x$. Generally, the higher the value of $x$, the larger the uncorrelated part of the band. For example, for the Z-boson production at $\sqrt{s}=13$ TeV 
$\langle \Delta_{\text{th.uncorr}}\rangle\simeq 8.5\%$ at $|y|<0.4$,
$\langle \Delta_{\text{th.uncorr}}\rangle\simeq 15\%$ at $2.0<y<2.4$,
and
$\langle \Delta_{\text{th.uncorr}}\rangle\simeq 48\%$ at $4<|y|<4.5$.

The values of the cross-sections predicted by the TMD factorization are somewhat lower than the experimental ones, see the last column in table~\ref{tab:chi2}. This is a known feature and has been observed in many analyses (see e.g.  refs.~\cite{Scimemi:2019cmh, Bertone:2019nxa, Vladimirov:2019bfa, Bacchetta:2019sam, Bacchetta:2022awv}). The discrepancy is of the order of a few percent for LHC energies (increasing for larger rapidity bins), and of the order of $\sim 40\%$ for fixed-target experiments. These discrepancies are within the published correlated uncertainties, and thus result in a minor increase of the $\chi^2$. The corresponding contribution is provided in table~\ref{tab:chi2} (column labeled  $\chi^2_\lambda/N_{\text{pt}}$). Theoretically, there could be several sources of disagreement in normalization, f.i. power corrections, see for instance ref.~\cite{Vladimirov:2023aot}.

Few experiments provide the normalized cross-section, which we match using the procedure described in sec.~\ref{sec:treatment_of_data}. The values of $\mathcal{N}$ provide us estimates for the integrated cross-sections for the experiments with normalized data
\begin{eqnarray}\label{norm_th}
\sigma_{\text{th.}}=\mathcal{N}^{-1}.
\end{eqnarray}
We present these estimations in table~\ref{tab:norms}, together with the experimental measurement (if available in the publication), and the theoretical prediction by DYNNLO~\cite{Catani:2007vq, Catani:2009sm} (done with MSHT20 collinear PDFs). Clearly, the results of DYNNLO agree within errors to the ones determined in our simplified procedure, demonstrating the consistency of the approach.

\begin{table}[]
\centering
\begin{tabular}{|l|c|c|c|}
\hline
dataset & $\sigma_{\text{exp.}}$ [pb]   & $\sigma_{\text{DYNNLO}}$[pb] & $\sigma_{\text{th.}}$[pb] 
\\\hline
ATLAS (13 TeV)     & $736.2\pm 16.7$ & $717.7\pm 6.8$ & $711.3\pm14.5$
\\\hline
CMS (7 TeV)     & - & $397.6 \pm 10.4$ & $386.4\pm8.2$
\\\hline
CMS (8 TeV)     & $440\pm 14$ & $453.4\pm 9.0$ & $441.0\pm 8.2$
\\\hline
D0 (run2)     & - & $252.0\pm7.3$ & $248.1 \pm 4.9$
\\\hline
\end{tabular}
\caption{\label{tab:norms} Comparison of the total cross-section values for the normalized datasets. The column $\sigma_{\text{exp.}}$ shows the measured value (if available in the original publications) with uncertainties from all sources. The column $\sigma_{\text{DYNNLO}}$ shows the prediction by DYNNLO, with PDF uncertainty. The column $\sigma_{\text{th.}}$ shows the value computed by eq.~(\ref{norm_th}), with uncertainties from TMD extraction only.}
\end{table}

\subsection{Collins-Soper kernel}

\begin{figure}[t]
\centering
\includegraphics[width=0.55\textwidth]{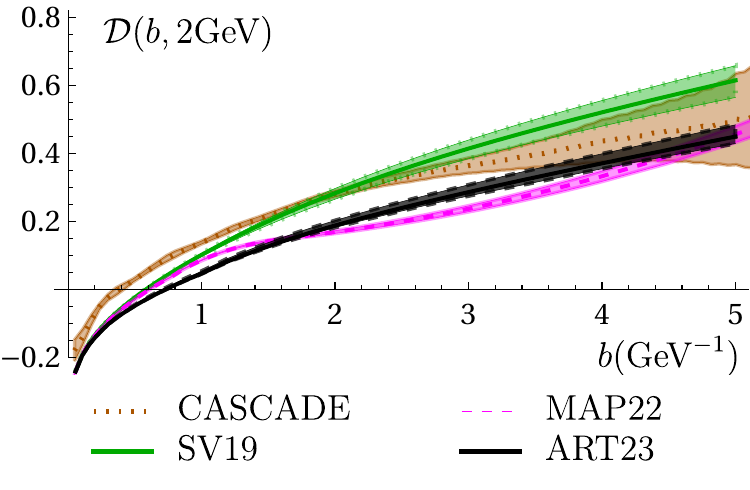}
\caption{Plot of the Collins-Soper kernel at $\mu=2$ GeV. Different lines correspond to the independent extractions CASCADE~\cite{BermudezMartinez:2022ctj}, SV19~\cite{Scimemi:2019cmh}, MAP22~\cite{Bacchetta:2022awv}, and ART23 (this work).}
\label{fig:CS}
\end{figure}

The plot of the CS kernel is presented in fig.~\ref{fig:CS}. The values of parameters that we obtain are
\begin{eqnarray}\label{param:CS}
B_{\text{NP}}=1.56^{+0.13}_{-0.09}\text{GeV},
\qquad
c_0=3.69^{+0.65}_{-0.61}\cdot 10^{-2},
\qquad
c_1=5.82^{+0.64}_{-0.88}\cdot 10^{-2}.
\end{eqnarray}
The parameters $B_{\text{NP}}$ and $c_0$ are compatible with the ones extracted in SV19. In particular, for the MMHT14 PDF (predecessor of MSHT20) SV19 found $B_{\text{NP}}=1.55\pm0.29$ and $c_0=(4.7\pm1.47) \cdot 10^{-2}$. Nonetheless, the shape of the distributions changes significantly due to the new logarithmic term $\sim c_1$ in the ansatz of eq.~(\ref{CS:NP}). This term modifies the shape of the distribution at $b\sim 1-3$ GeV, leaving the large-$b$ asymptotic behaviour untouched. Removing this parameter, the fit shows worse values of $\chi^2$, $\chi^2/N_{pt}=1.51$ (for central replicas of PDF and the data). The general shape and value of the CS kernel are in good agreement with the MAP22 determination, as can be seen in fig.~\ref{fig:CS}, despite the differences in the two codes.

A significative feature of the current fit is that the size of the uncertainty band for the CS kernel is reduced, in contrast to that of the TMD distribution itself. This is correct, since in the $\zeta$-prescription the CS kernel is exactly decorrelated from the TMDPDF (on the theory side), and the quality of the data has increased. We also notice that the parameter $c_1$ is clearly non-zero, which indicates the presence of a non-negligible logarithmic behaviour in the next-to-leading power term of the small-$b$ expansion of the CS kernel. 

\subsection{Unpolarized TMD distribution}

The values of the TMDPDF parameters extracted in the fit are
\begin{eqnarray}\label{param:lambdas}
\lambda_1^{u}=0.87_{-0.10}^{+0.10},
&\qquad&
\lambda_2^{u}=0.91_{-0.29}^{+0.33},
\\\nn
\lambda_1^{d}=0.99_{-0.12}^{+0.09},
&\qquad&
\lambda_2^{d}=6.06_{-1.34}^{+1.36},
\\\nn
\lambda_1^{\bar u}=0.35_{-0.22}^{+0.23},
&\qquad&
\lambda_2^{\bar u}=46.6_{-8.1}^{+7.9},
\\\nn
\lambda_1^{\bar d}=0.12_{-0.11}^{+0.13},
&\qquad&
\lambda_2^{\bar d}=1.53_{-0.17}^{+0.54},
\\\nn
\lambda_1^{sea}=1.32_{-0.24}^{+0.23},
&\qquad&
\lambda_2^{sea}=0.46_{-0.45}^{+0.13},
\end{eqnarray}
Most of these parameters have reasonable sizes, and they agree (within uncertainty) with similar ones found in ref.~\cite{Bury:2022czx}. However, the parameters $\lambda_1^{\bar d}$ and $\lambda_2^{sea}$ show some problematic behaviour. Namely, they almost vanish at their lower boundary. For negligible values of $\lambda$'s the $b-$profile of the corresponding TMDPDF flattens. This is a clearly non-physical behavior, which results in disturbed shapes of the uncertainty bands for $\bar d$ and $sea$ flavors at large-$b$. Simultaneously, it does not produce any problem in the prediction for the cross-section, since the TMDPDFs contributes in products with the evolution factors. It merely indicates that the present observables/data are not restrictive enough for these flavor combinations.

\begin{figure}
\centering
\includegraphics[width=0.49\textwidth]{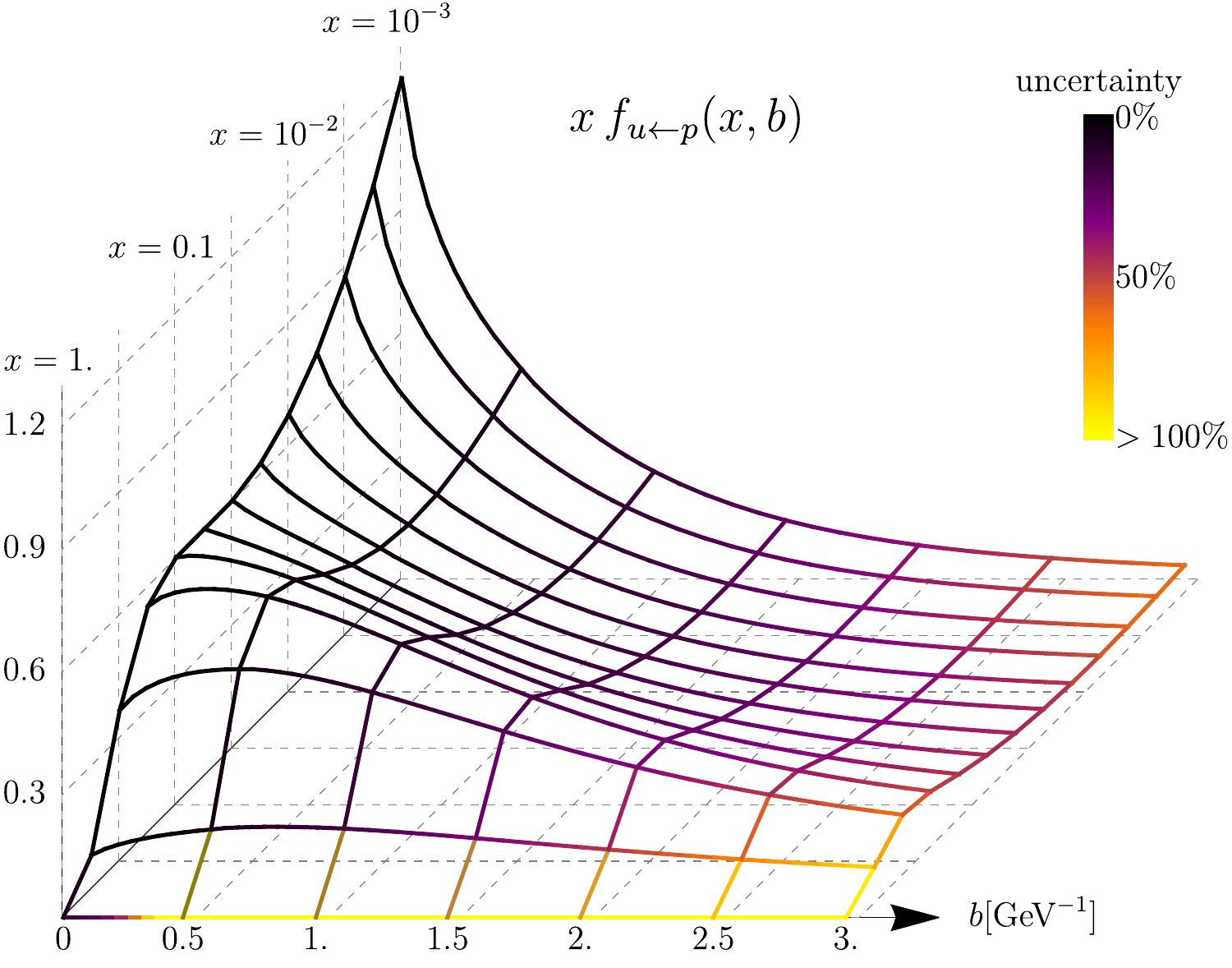}
\includegraphics[width=0.49\textwidth]{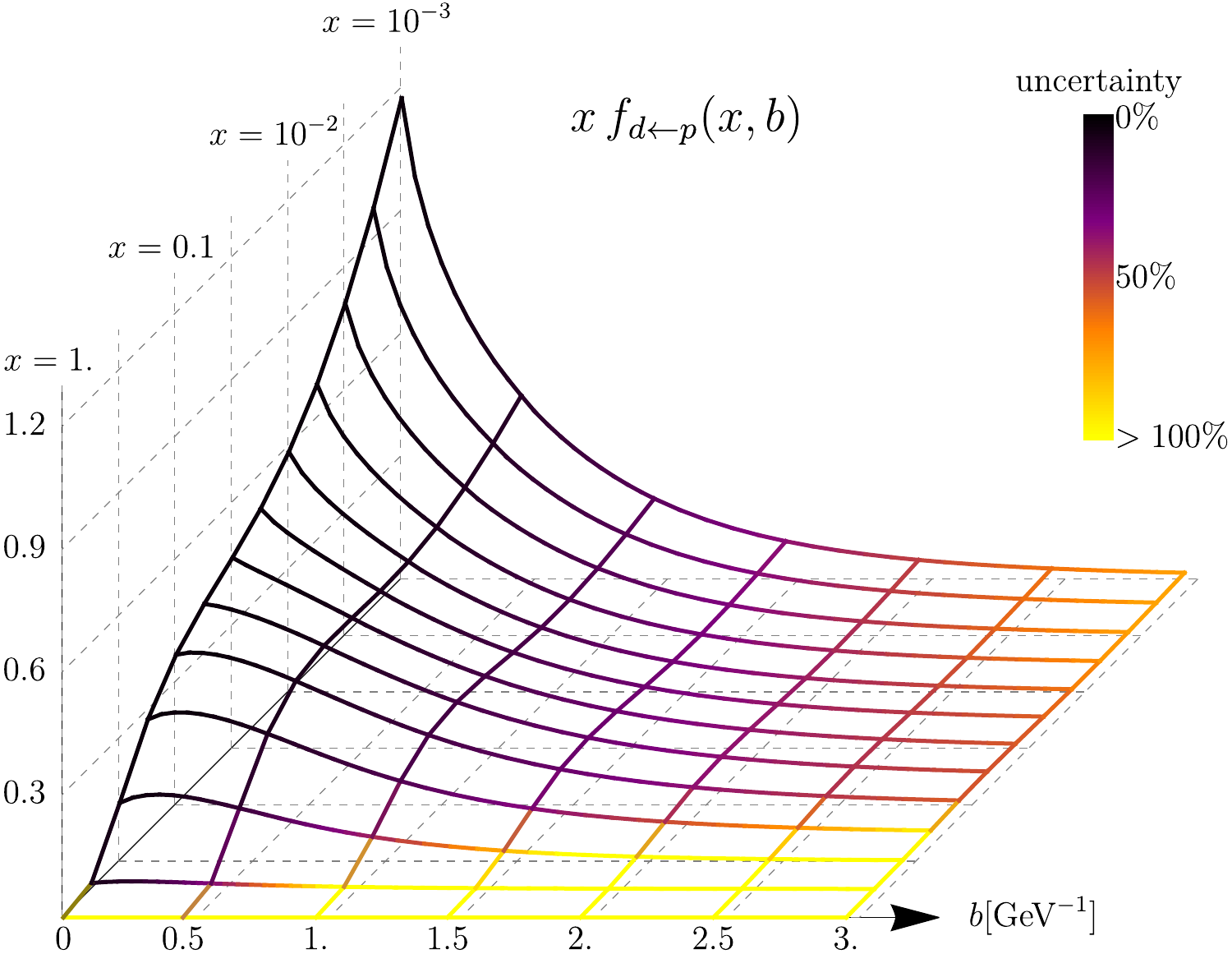}
\caption{Shape of TMDs in the (x,b)-space. The color indicates the uncertainty.}
\label{fig:sofa}
\end{figure}

\begin{figure}
\centering
\includegraphics[width=0.89\textwidth]{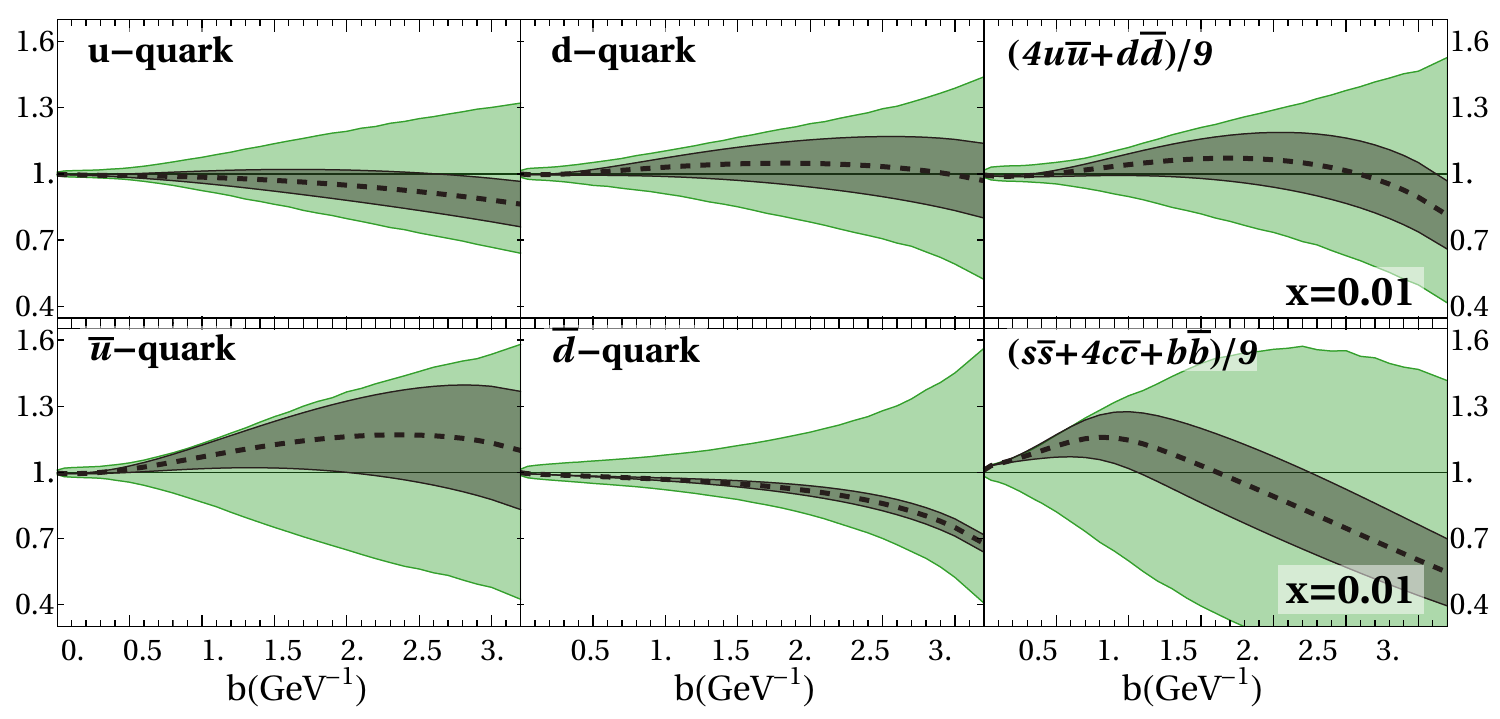}\\
\includegraphics[width=0.9\textwidth]{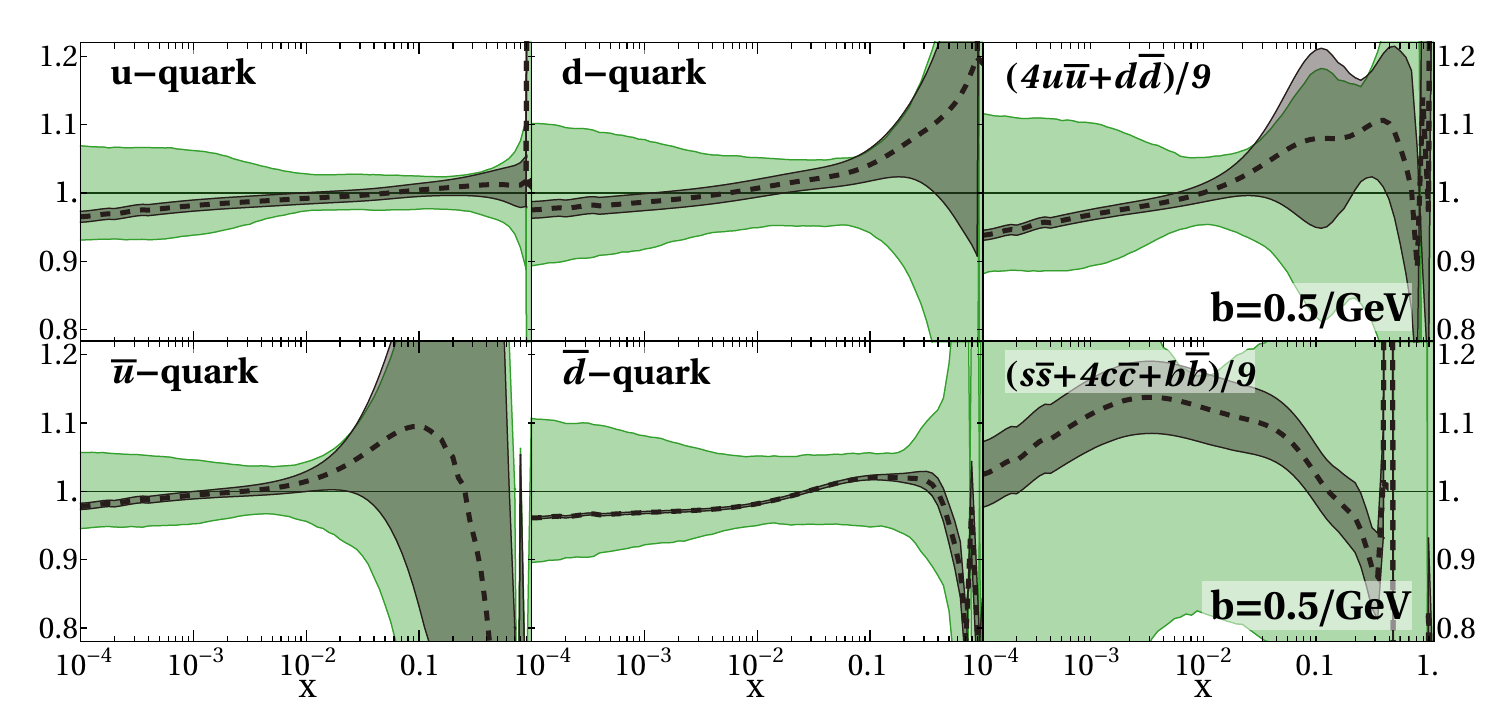}
\caption{Sizes of the full uncertainty bands in ART23 (green) in comparison to the extraction at the central PDF replica (grey). The upper panel shows the $b$-dependence at $x=0.01$. The lower panel shows the $x-$dependence at $b=0.5$ GeV$^{-1}$.}
\label{fig:uncertainties}
\end{figure}

The shapes of the TMDPDFs are shown in fig.~\ref{fig:sofa} for $u$ and $d$ quarks (other flavors show similar behaviour). The sizes of the uncertainty bands are shown in fig.~\ref{fig:uncertainties}. Generally, the uncertainty bands are increased by an order of magnitude in comparison to the SV19 fit, and grow faster with the increase of $b$. This is the result of incorporating the PDF uncertainties, which helps to account for the PDF-bias and allows for a more realistic uncertainty estimation. The $x$-shape of the uncertainties has become more involved. Their minimum is at $x\sim 10^{-2}$, where  more precise data are located. The sizes of quark- and anti-quark uncertainties are compatible, because most part of the data depends on the product $f_{1q} f_{1\bar q}$ that does not distinguish between quarks and anti-quarks.

\subsection{Impact of inclusion the PDF uncertainty}

The inclusion of PDF uncertainties in an analysis of TMDPDFs inflates the uncertainty on TMDPDFs themselves by almost an order of magnitude. This feature has been already analysed in ref.~\cite{Bury:2022czx} using the SV19 setup as  baseline. In this subsection we present the impact of PDF uncertainty to our analysis. 

To determine this impact we have performed an independent fit of the data including into the fit procedure only \textit{experimental uncertainties} (see sec.~\ref{sec:uncert}). We have observed that the central values of parameters are largely unaffected, while the uncertainty band is essentially reduced. The quality of the data description remains high as $\chi^2/N_{\text{pt}}=0.95^{+0.008}_{-0.006}$. Note that the uncertainty band here is reduced by an order of magnitude, which already tells that the largest part of the replica distribution is due to the PDF uncertainty rather than due to the data.

The values of the NP parameters obtained in this fit are
\begin{gather*}
\boxed{\textbf{No PDF uncertainty}}
\\
B_{\text{NP}}=1.51^{+0.03}_{-0.01}\text{GeV},
\qquad
c_0=3.79^{+0.37}_{-0.39}\cdot 10^{-2},
\qquad
c_1=5.27^{+0.15}_{-0.02}\cdot 10^{-2},
\\
\lambda_1^{u}=0.87_{-0.03}^{+0.04},
\qquad
\lambda_2^{u}=0.80_{-0.11}^{+0.10},
\\
\lambda_1^{d}=0.92_{-0.03}^{+0.02},
\qquad
\lambda_2^{d}=5.59_{-0.55}^{+0.45},
\\\nn
\lambda_1^{\bar u}=0.22_{-0.06}^{+0.05},
\qquad
\lambda_2^{\bar u}=44.8_{-5.1}^{+5.0},
\\\nn
\lambda_1^{\bar d}=0.10_{-0.04}^{+0.03},
\qquad
\lambda_2^{\bar d}=1.77_{-0.17}^{+0.18},
\\\nn
\lambda_1^{sea}=1.25_{-0.04}^{+0.03},
\qquad
\lambda_2^{sea}=0.43_{-0.41}^{+0.42}.
\end{gather*}
Comparing these parameters to eq.~(\ref{param:CS}, \ref{param:lambdas}) we see that they stay within the uncertainty band of the full analysis, and in most cases have practically the same central value. The uncertainty band is reduced by a factor 2-10. There are two parameters ($\lambda_{2}^u$ and $\lambda_1^{\bar u}$) whose central value shifted more in comparison to the complete fit. It possibly indicates that these parts of PDFs have some tension with TMD data. The comparison of shapes is presented in fig.~\ref{fig:uncertainties}.

The theoretical uncertainty in the cross-section also shrinks by an order of magnitude. The central value  almost coincides with the central line for the full fit, although in some cases there are deviations in the low-$q_T$ bins. An example is shown in fig.~\ref{fig:example-data}.

\section{Conclusions}
\label{sec:conclusions}

The  present extraction of unpolarized TMDPDF from the global fit of Drell-Yan data (refereed as ART23) represents a significant step forward in comparison to previous analyses. The main improvements  are a higher order perturbative input (which reaches N$^4$LL with NNLO evolution for the collinear PDFs), a consistent treatment of PDF uncertainties in the error analysis, and the inclusion of additional data. We also use the flavor dependent form of the fitting ansatz for TMDPDF. The introduction of a flavor dependence reduces the sensitivity to the choice of PDF sets, as observed in~\cite{Bury:2022czx}. Furthermore, the newly incorporated non-perturbative logarithmic dependence of the CS evolution kernel, eq.~(\ref{CS:NP}), plays a crucial role in achieving a successful fit.

We also find it particularly interesting that several groups find a reasonable agreement on the CS kernel (see fig.~\ref{fig:CS}) despite somewhat different functional forms and uncertainty-estimation procedure. The agreement with the new Z-boson data at LHC and W-boson mediated data at Tevatron is particularly encouraging.

It is important to stress that the PDF uncertainties dominate the TMDPDF extraction, see fig.~\ref{fig:uncertainties}. For that reason, the uncertainties on TMDPDFs that we find are larger by almost an order of magnitude in comparison to other global extractions~\cite{Scimemi:2019cmh, Bertone:2019nxa, Bacchetta:2019sam, Bacchetta:2022awv}, even though the present dataset is about 30-40\% larger and more precise than those considered earlier. We argue that this increased uncertainty is more realistic, and previous studies are biased in several aspects. The $\chi^2$ value that we obtain is very stable, see fig.~\ref{fig:chi2_total}, and it shows a very good agreement between the theory and the data. This observation is noteworthy and suggests that future improvements to the TMDPDFs determination could be achieved by a joint fit of PDFs and TMDPDFs. These possibilities will be explored in due time along with the inclusion of new data. 

The values of ART23 unpolarized TMDPDFs, as well as, the  \texttt{artemide} code release used for their extraction is published in the artemide-repository~\cite{artemide}. The accompanying code (for minimisation, generation of plots, etc) is presented in ref.~\cite{DataProcessor}. We also release ART23 TMDPFs in the unified format of TMDlib2~\cite{Abdulov:2021ivr}.

Our current understanding of the TMDPDFs is summarized in fig.~\ref{fig:sofa}, which illustrates the shape and the uncertainty of the unpolarized TMD distributions in the $(x,b)$ plane for up and down quarks. An important point for future consideration is the impact of power corrections, for which preliminary theoretical results have been obtained but are not yet directly applicable to our present analysis~\cite{Moos:2020wvd, Ebert:2021jhy, Rodini:2022wki}. The present extraction opens the path to the N$^4$LL analysis of SIDIS data, where, in addition to the unpolarized TMDPDFs and CS kernel, one can also determine TMD fragmentation functions.

\acknowledgments
We thank the STAR collaboration, explicitly Salvatore Fazio and Xiaoxuan Chu, for sharing their preliminary results with us, such that these data could be included in the fit. We also thank Louis Moureaux and Buğra Bilin for their help with the interpretation of the CMS Q-differential data.
A.V. is funded by the \textit{Atracci\'on de Talento Investigador} program of the Comunidad de Madrid (Spain) No. 2020-T1/TIC-20204 and Europa Excelencia EUR2023-143460 from Spanish Ministerio de Ciencias y Innovaci\'on. P.Z. is funded by the \textit{Atracci\'on de Talento Investigador} program of the Comunidad de Madrid (Spain) No. 2022-T1/TIC-24024.  This work was partially supported by DFG FOR 2926 ``Next Generation pQCD for  Hadron  Structure:  Preparing  for  the  EIC'',  project number 430824754.
This project is supported by the Spanish Ministerio de Ciencias y Innovaci\'on grant PID2019-106080GB-C21 and PID2022-136510NB-C31.
This project has received funding from the European Union Horizon 2020 research and innovation program under grant agreement Num. 824093 (STRONG-2020).

\appendix

\section{Fit using NNPDF3.1}
\label{app:NNPDF}

In the present approach the TMDPDF suffers the PDF-bias~\cite{Bury:2022czx}, i.e. is strongly dependent on the collinear PDF. To explore this dependence, we have performed an additional fit using the NNPDF3.1 collinear PDF~\cite{NNPDF:2017mvq} with identically the same technique as presented in the main text. In this appendix we discuss the outcomes of the fit.

\begin{figure}[t]
\begin{center}
\includegraphics[width=0.5\textwidth]{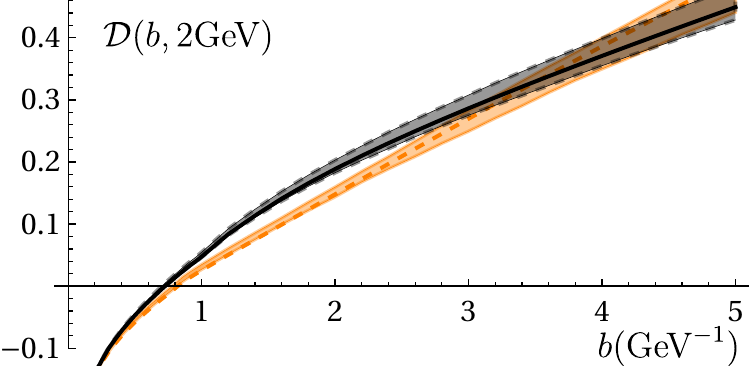}
\end{center}
\caption{Comparison of the CS kernels extracted with the MSHT20 (solid grey curve) and NNPDF3.1 (dashed orange curve) collinear PDFs.}
\label{fig:CS-nnpdf}
\end{figure}

The fitting procedure is described in sec. \ref{sec:uncert}. In this fit we utilize the 1000-replica distribution of NNPDF3.1, and computed 300 replicas in total. The quality of the fit with NNPDF3.1 is worse, we obtain $\chi^2/N_{\text{pt}}=1.18^{+0.11}_{-0.05}$. The resulting values of parameters for the CS kernel are:
\begin{eqnarray}
B_{\text{NP}}=1.82^{+0.37}_{-0.29}\text{ GeV},
\qquad
c_0=5.03^{+1.02}_{-0.81}\cdot 10^{-2},
\qquad
c_1=1.04^{+0.09}_{-0.15}\cdot 10^{-1}.
\end{eqnarray}
These parameters are quite different from the MSHT values eq.~(\ref{param:CS}). The comparison of the resulting CS kernels is given in fig.~\ref{fig:CS-nnpdf}. In particular, we observe that the value of $B_{\text{NP}}$ has very large uncertanties, in contrast to the MSHT case.

\begin{figure}[t]
\begin{center}
\includegraphics[width=0.85\textwidth]{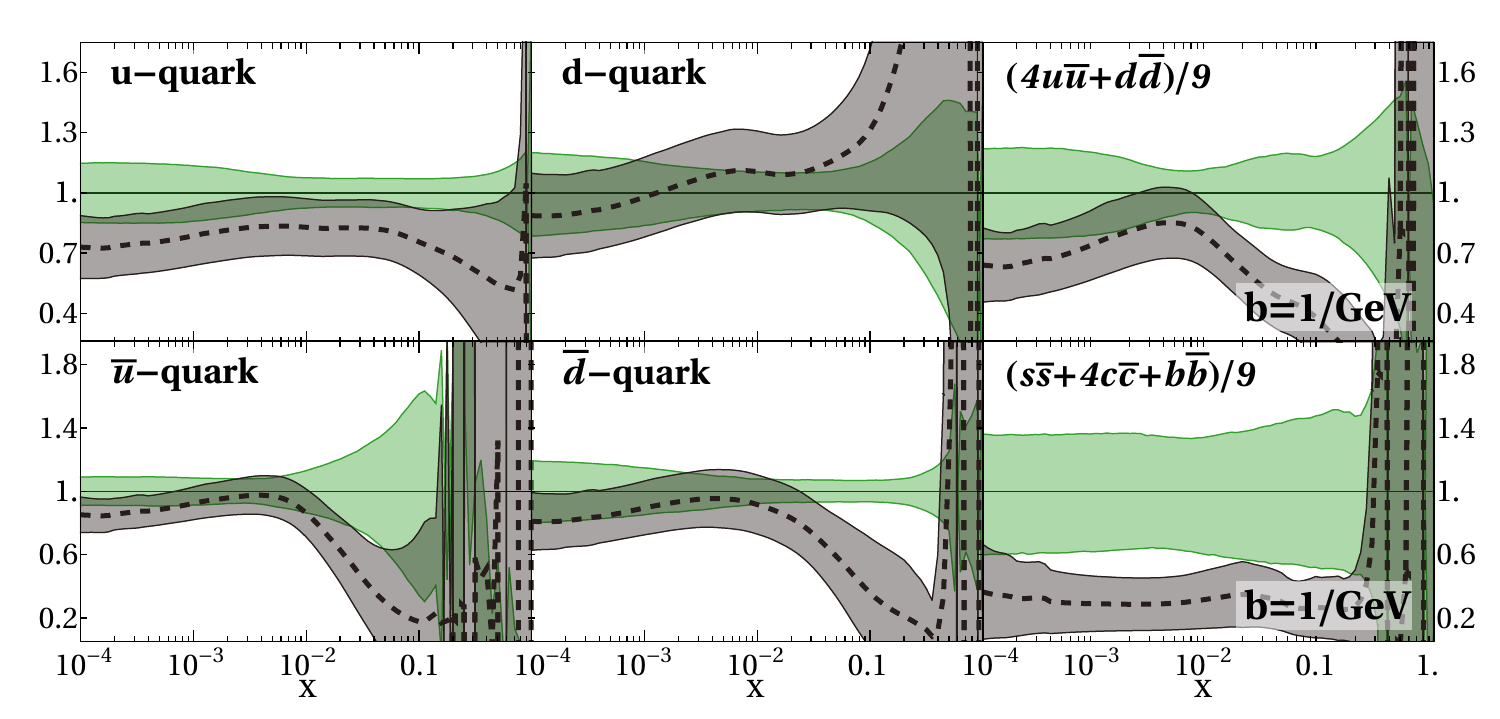}
\includegraphics[width=0.85\textwidth]{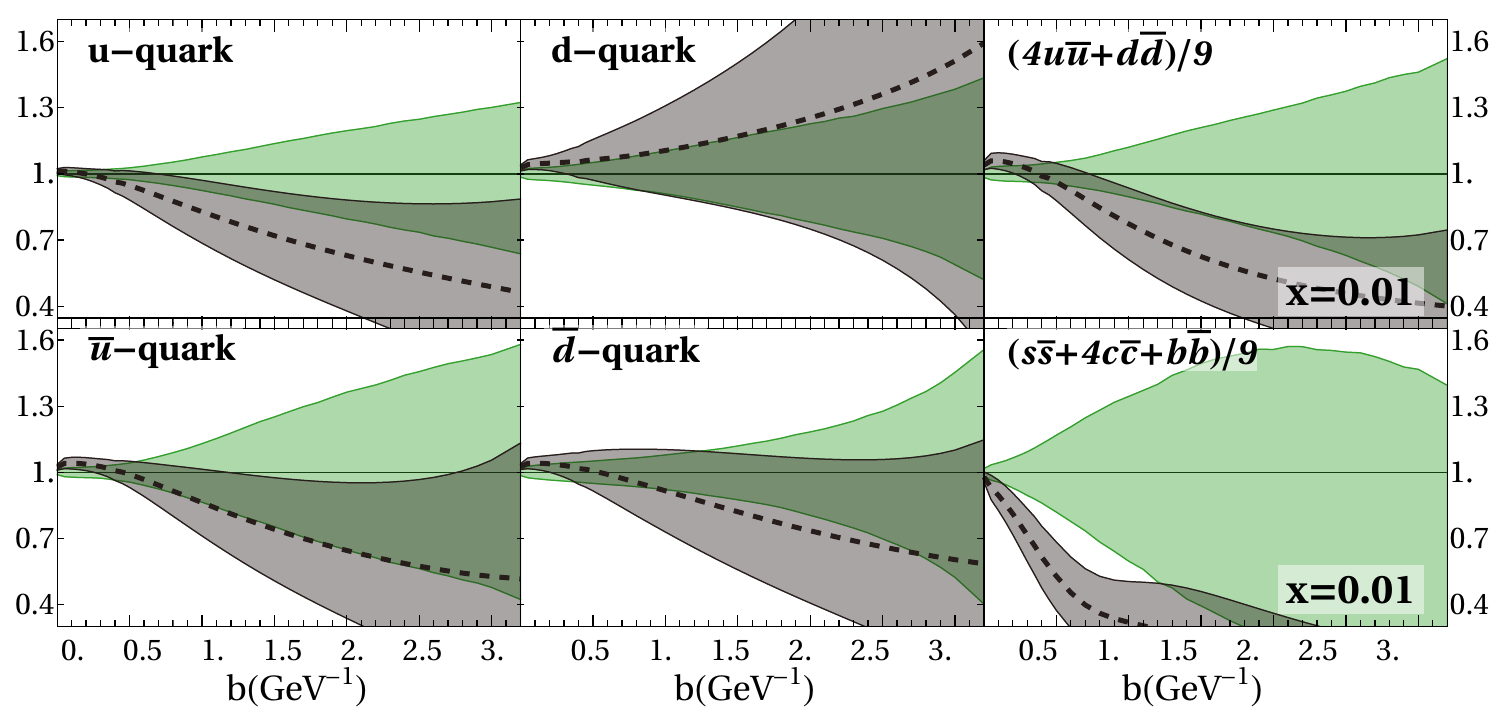}
\end{center}
\caption{Comparison of uncertainty bands for unpolarized TMDPDFs extracted with MSHT20 (green) and NNPDF3.1 (gray) collinear distributions. The plots are normalized to the MSHT20 case.}
\label{fig:MSHTvsNNPDF}
\end{figure}

The parameters of the TMD distribution's ansatz are
\begin{eqnarray}
\lambda_1^{u}=1.11_{-0.17}^{+0.15},
&\qquad&
\lambda_2^{u}=2.69_{-2.53}^{+1.79},
\\\nn
\lambda_1^{d}=0.99_{-0.17}^{+0.14},
&\qquad&
\lambda_2^{d}=3.60_{-3.57}^{+4.36},
\\\nn
\lambda_1^{\bar u}=0.37_{-0.15}^{+0.16},
&\qquad&
\lambda_2^{\bar u}=75.0_{-13.2}^{+14.8},
\\\nn
\lambda_1^{\bar d}=0.37_{-0.32}^{+0.30},
&\qquad&
\lambda_2^{\bar d}=17.7_{-14.2}^{+6.91},
\\\nn
\lambda_1^{sea}=1.88_{-0.29}^{+0.34},
&\qquad&
\lambda_2^{sea}=2.71_{-2.69}^{+2.44}.
\end{eqnarray}
Generally, we observe that the NNPDF3.1-fit yields a different spread in the parameters with a larger uncertainty. Most of the parameters are in  relative agreement; however, the parameters $\lambda_2^u$, $\lambda_2^{\bar u}$, $\lambda_2^{\bar d}$ and $\lambda_2^{sea}$ disagree, and do not overlap within the uncertainty bands. All these parameters are responsible for the behaviour of the TMDPDF at large-$x$, i.e. exactly in the region where collinear distributions differ. However, for the middle-$x$ range ($x\sim 10^{-2}-10^{-3}$) the TMDPDFs are in general agreement, except for the sea-quark (see fig.~\ref{fig:MSHTvsNNPDF}).

\begin{figure}[]
\begin{center}
\includegraphics[width=0.95\textwidth]{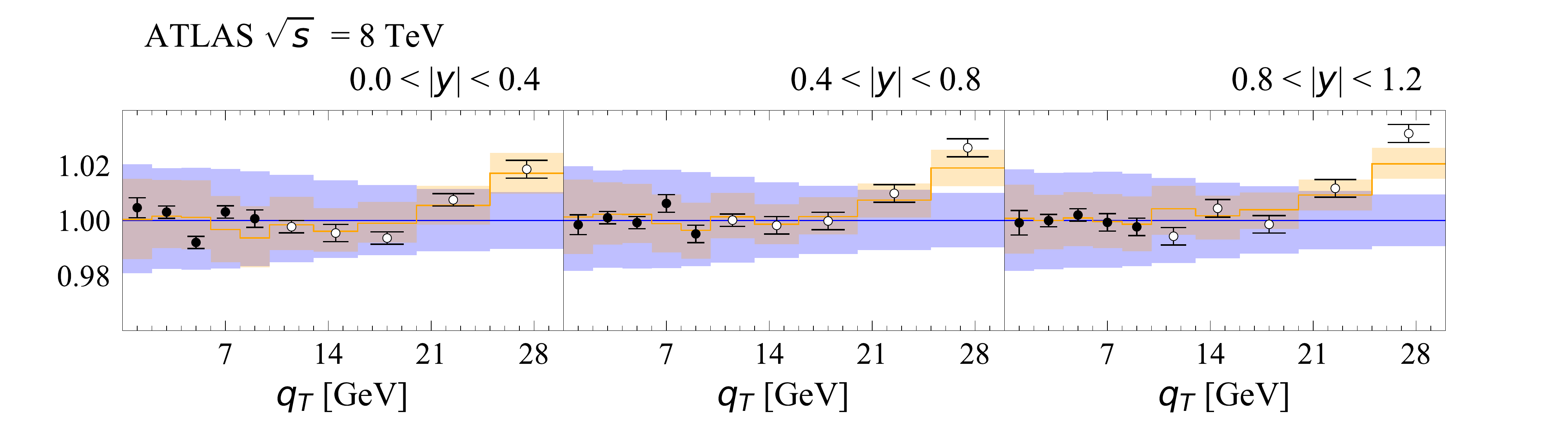}
\includegraphics[width=0.95\textwidth]{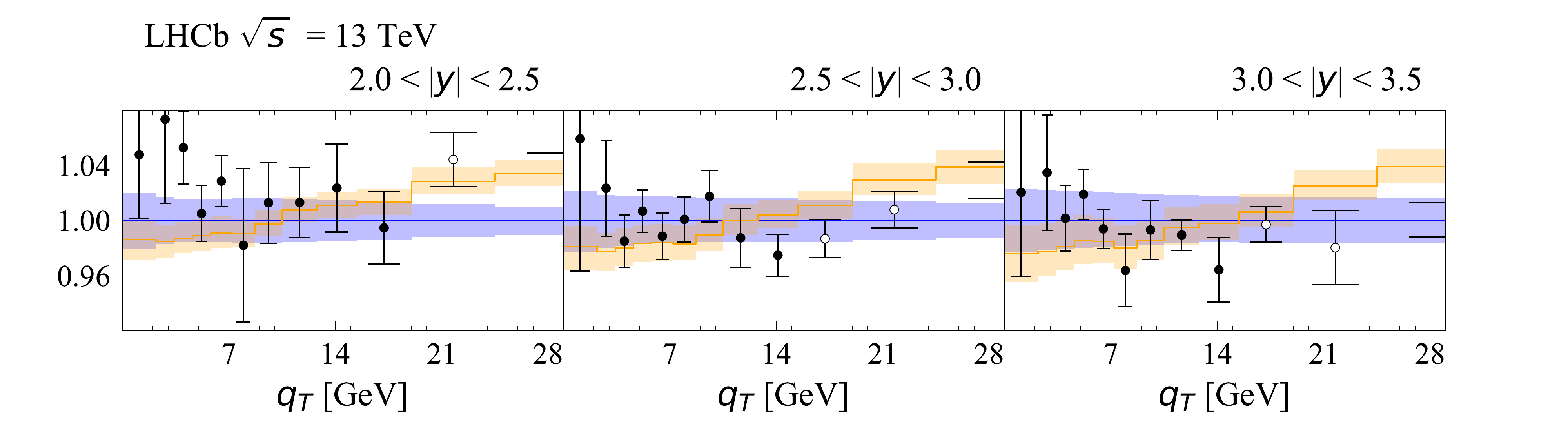}
\end{center}
\caption{Comparison of theory predictions based on MHST20 (blue) or NNPDF3.1 (orange) collinear PDFs. The rapidity range is indicated above the plot. All data and uncertainties are normalized to the blue line (MSHT20 result).}
\label{fig:LHCb_uncer_vs_NNPDF}
\end{figure}

The resulting theory predictions are given in fig.~\ref{fig:LHCb_uncer_vs_NNPDF}, for several bins of ATLAS at $\sqrt{s}=8$ TeV, and LHCb at $\sqrt{s}=13$ TeV. Other experiments demonstrate a similar picture. Comparing these two experiments one can see that both PDFs produce very similar results for ATLAS (which has $x\sim 10^{-2}-10^{-3}$), while for LHCb (which has $x\sim 10^{-1}$) the curves are very different. It confirms our previous conclusion that the extractions of TMDPDFs are very sensitive to the large-$x$ range.

Let us stress that without the inclusion of the flavor-dependence the extractions with MSHT20 and NNPDF3.1 diagree drastically~\cite{Bury:2022czx}. The inclusion of the flavor-dependence reduces the problem of the PDF-bias, but does not resolve it completely.

\clearpage
\section{Comparison with data}
\label{app:plots}

In this appendix, we present all the data used for the fit, along with the resulting theory prediction of our main fit (with MSHT20 PDF input). The depicted theory prediction is the distributions (of replicas) average. The 68\%CI of the theory prediction (see the discussion in sec.~\ref{sec:uncert}) is shown as a blue band. For a better visual comparison of data and theory predictions, the theory curves are shifted by a factor $d$ (see eq.~(\ref{t=t-d})) computed for the central line of the prediction. In all plots, we demonstrate more data than those described by the TMD factorization theorem. The data points used in the fit are shown by filled points, while the rest are shown by empty points.

\begin{figure}[htb]
\begin{center}
\includegraphics[height=0.28\textheight]{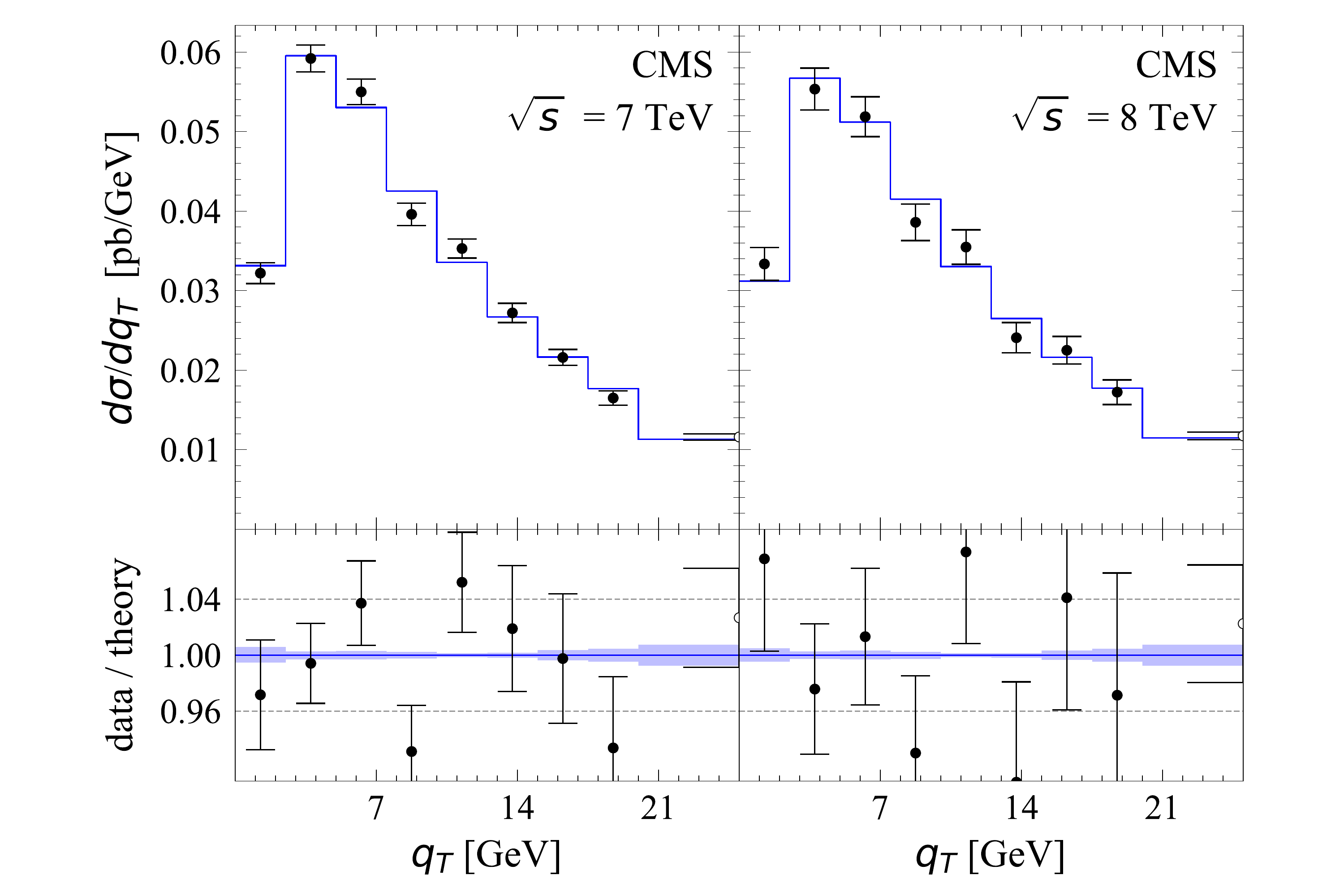}
\includegraphics[height=0.28\textheight]{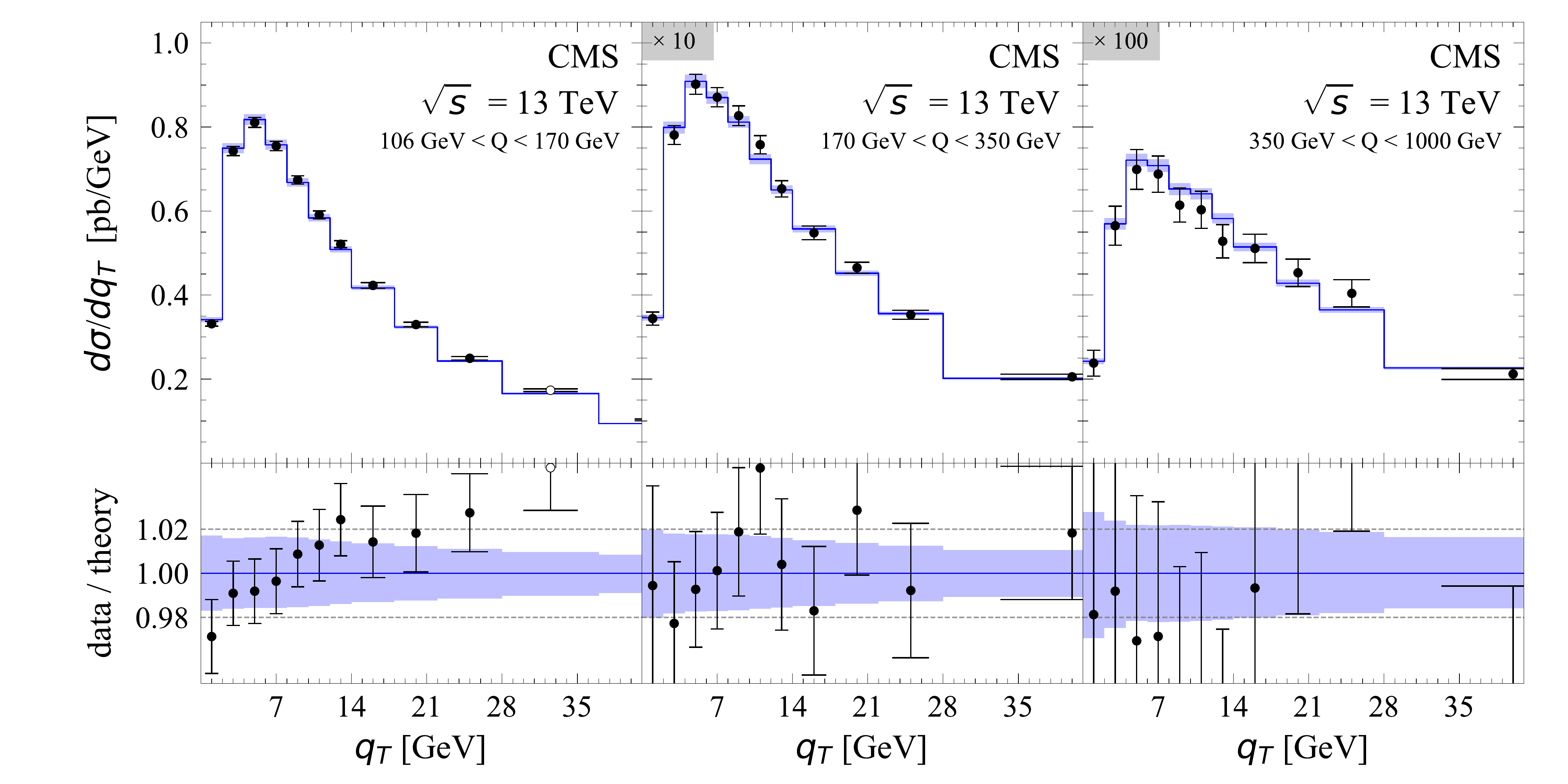}
\end{center}
\caption{Differential cross-section for the $Z/\gamma^*$ boson production measured by CMS, at different values of $\sqrt{s}$ and $Q$, according to the legends in the plots. }
\label{fig:CMS1}
\end{figure}

\begin{figure}[htb]
\begin{center}
\includegraphics[height=0.28\textheight]{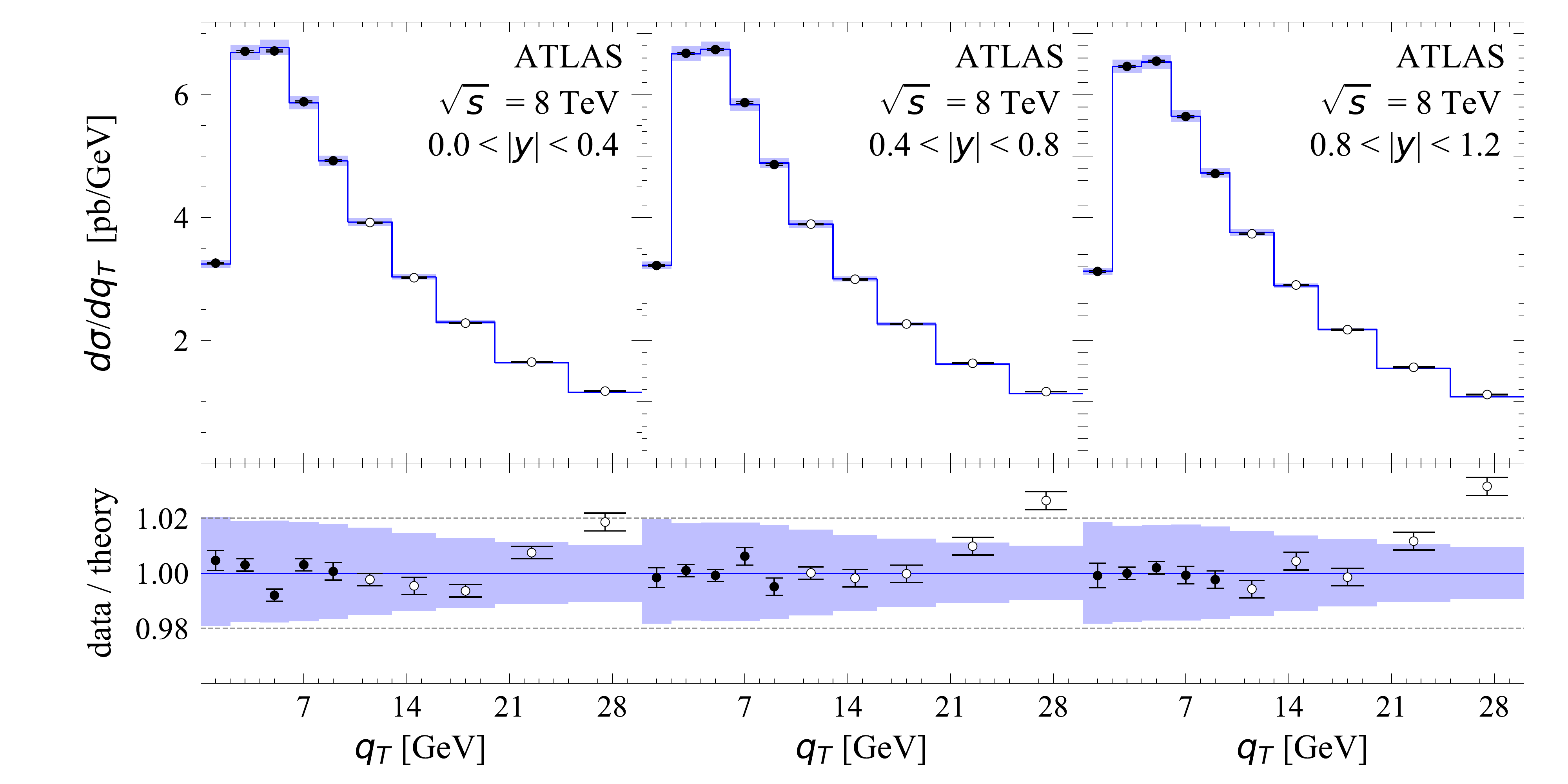}
\includegraphics[height=0.28\textheight]{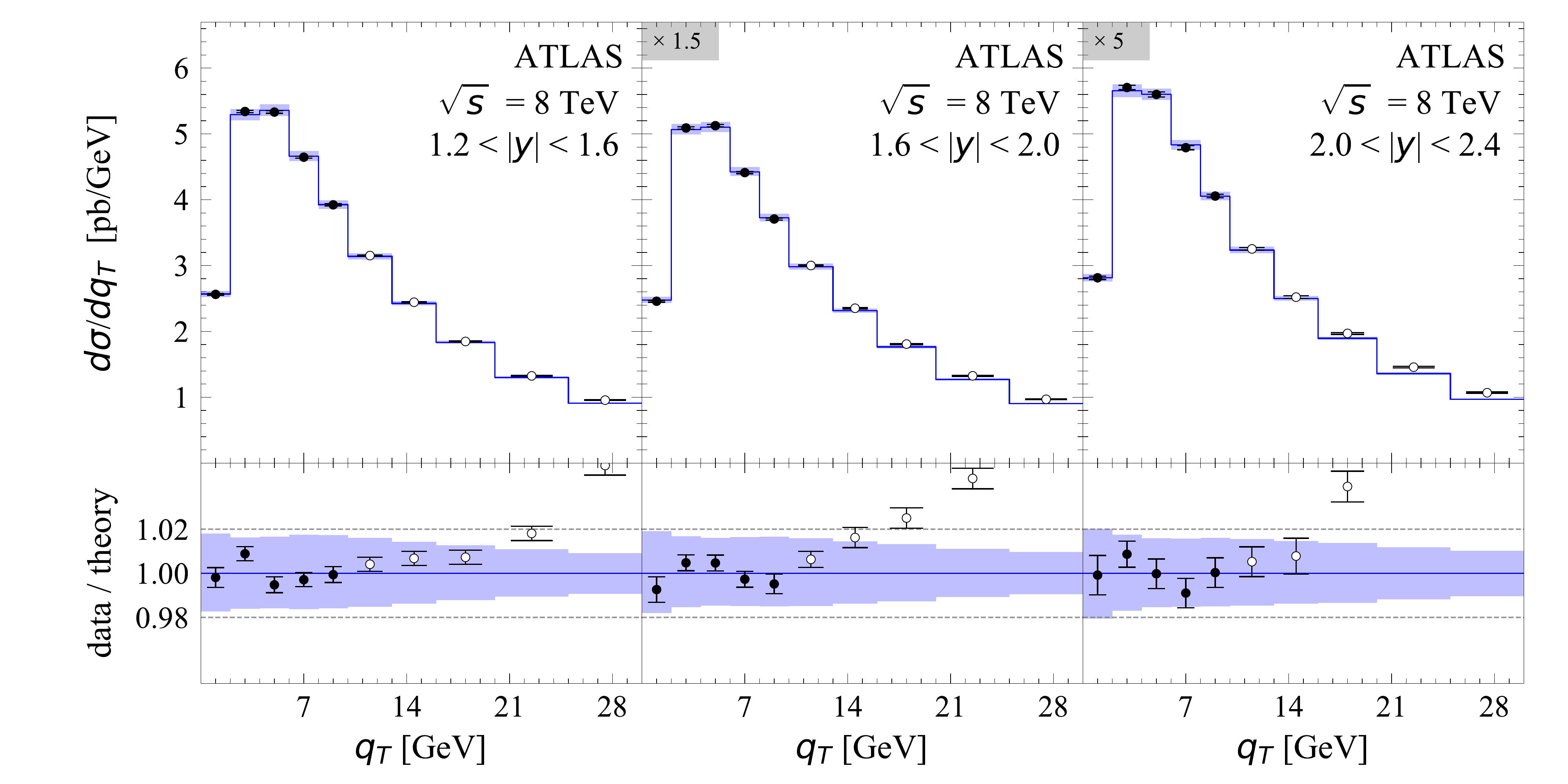}
\includegraphics[height=0.28\textheight]{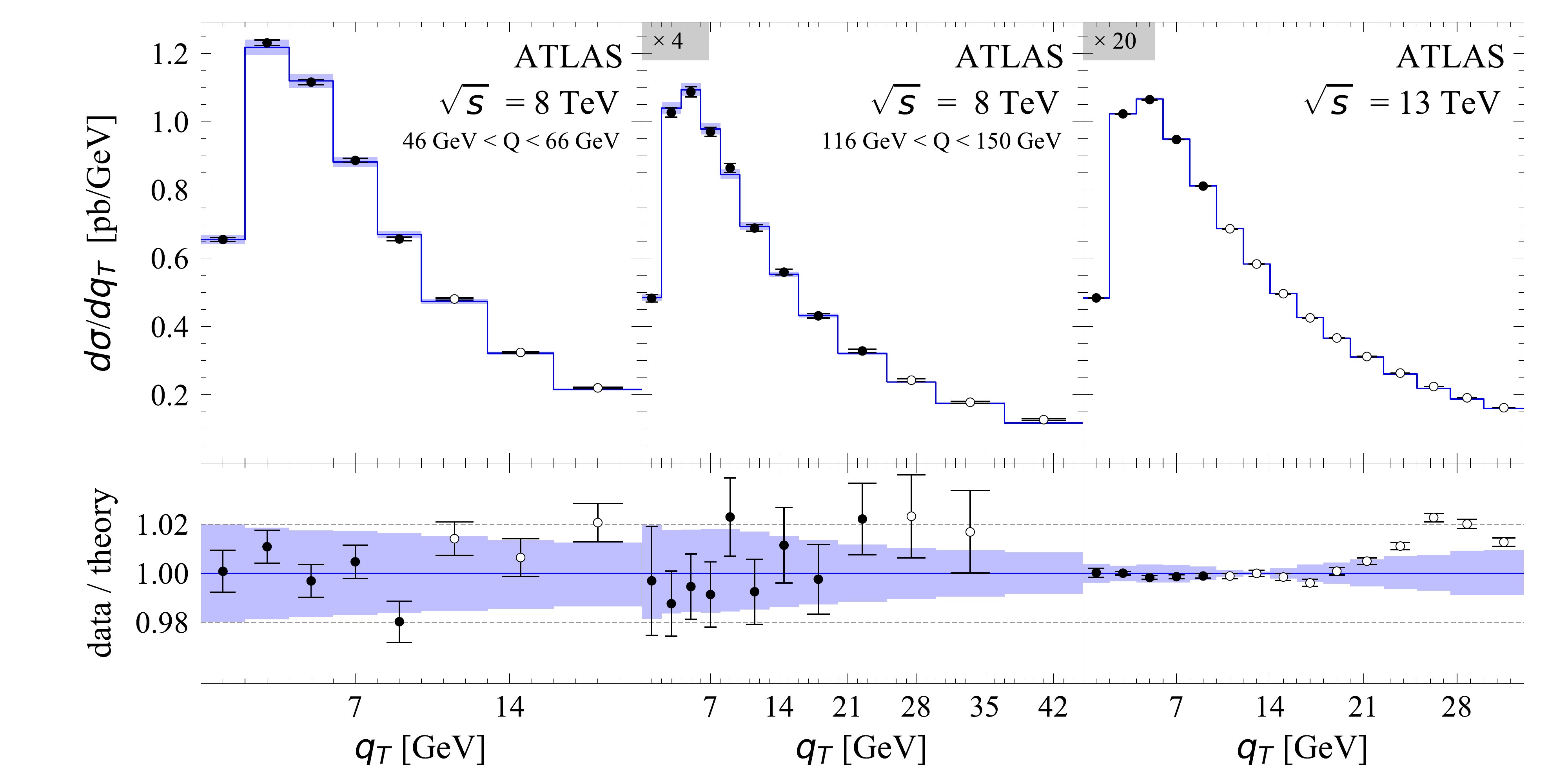}    
\end{center}
\caption{Differential cross-section for the $Z/\gamma^*$ boson production measured by ATLAS, at different values of $\sqrt{s}$, $y$, and $Q$. All details pertaining the values of the kinematic variables and their cuts can be seen in the plots. The plot for comparison with ATLAS at $\sqrt{s}=13$ TeV can be found in fig.~\ref{fig:example-data} in a larger scale.}
\label{fig:ATLAS}
\end{figure}

\begin{figure}[htb]
\begin{center}
\includegraphics[height=0.29\textheight]{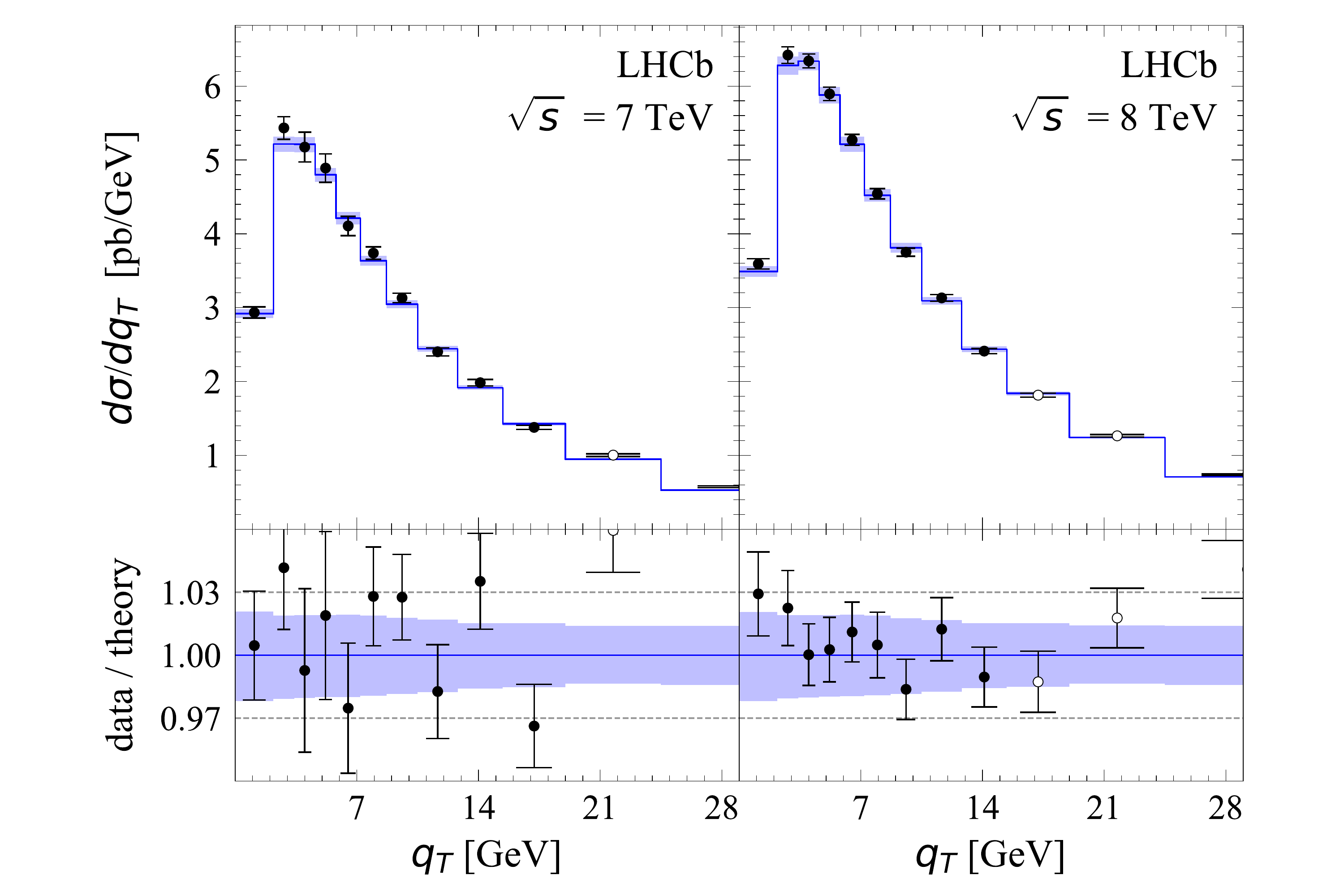}
\includegraphics[height=0.29\textheight]{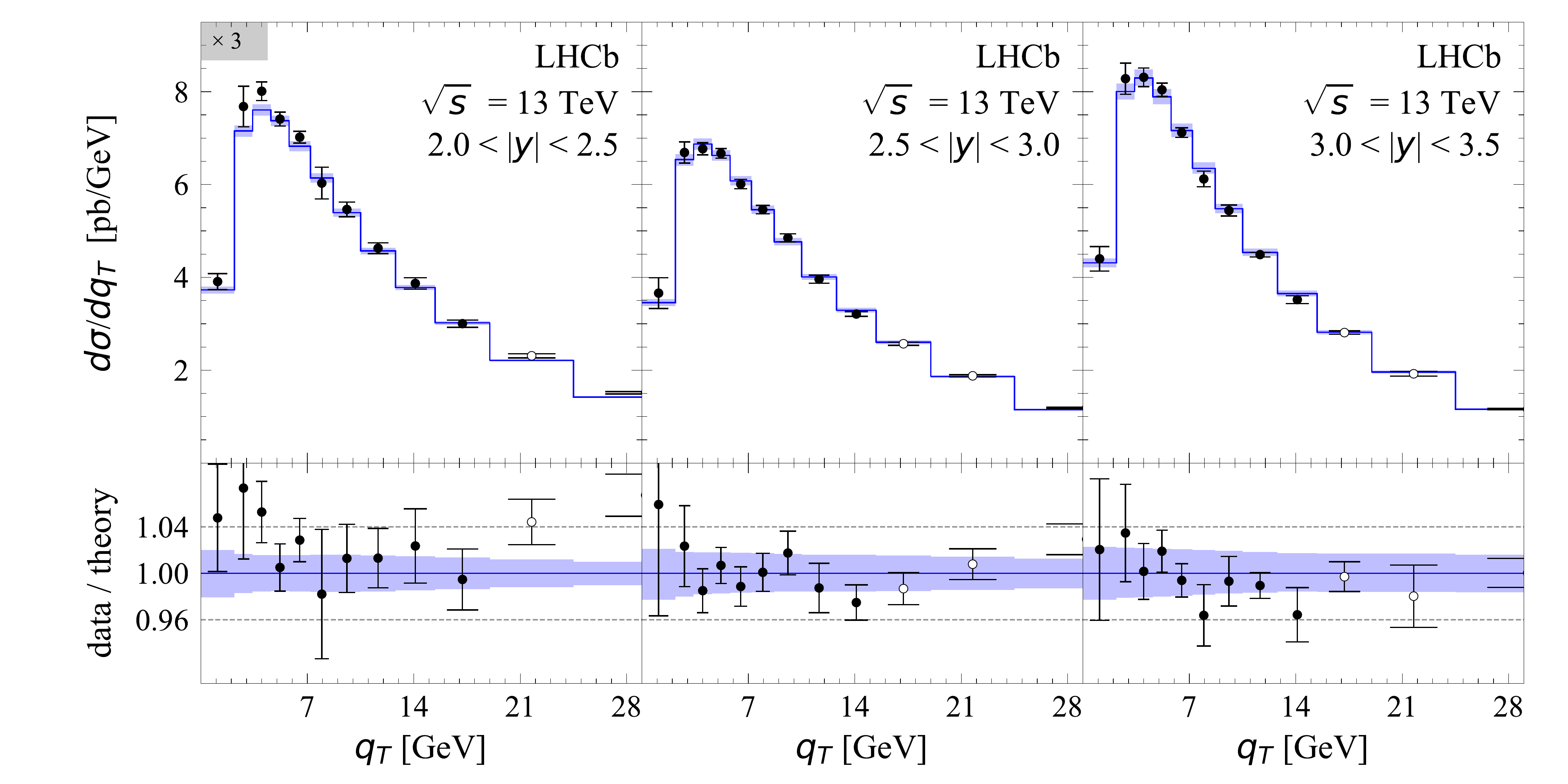}
\includegraphics[height=0.29\textheight]{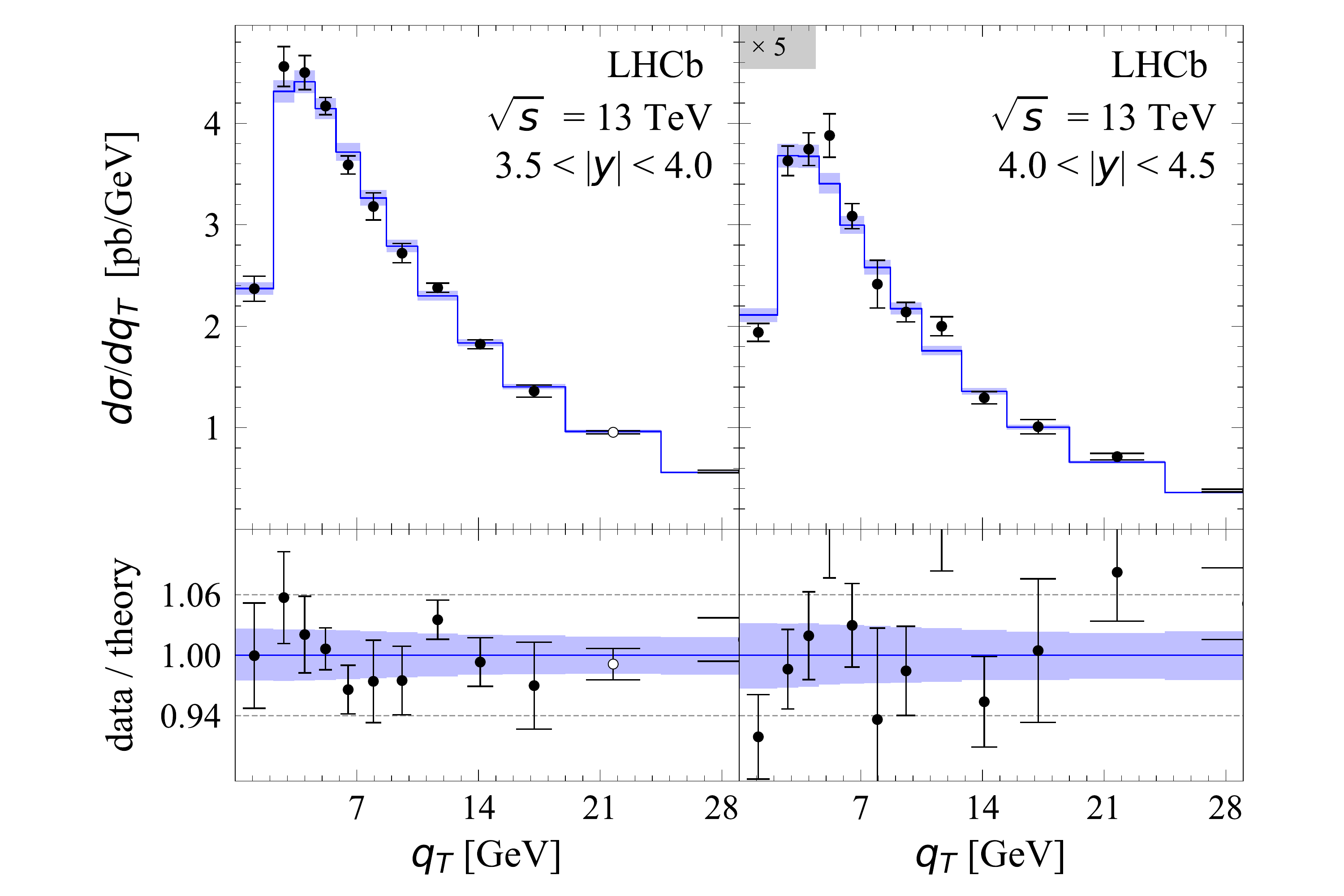}
\end{center}
\caption{Differential cross-section for the $Z/\gamma^*$ boson production measured by LHCb, at different values of $\sqrt{s}$ and $y$, according to the legends in the plots.}
\label{fig:LHCb}
\end{figure}

\begin{figure}[htb]
\begin{center}
\includegraphics[height=0.28\textheight]{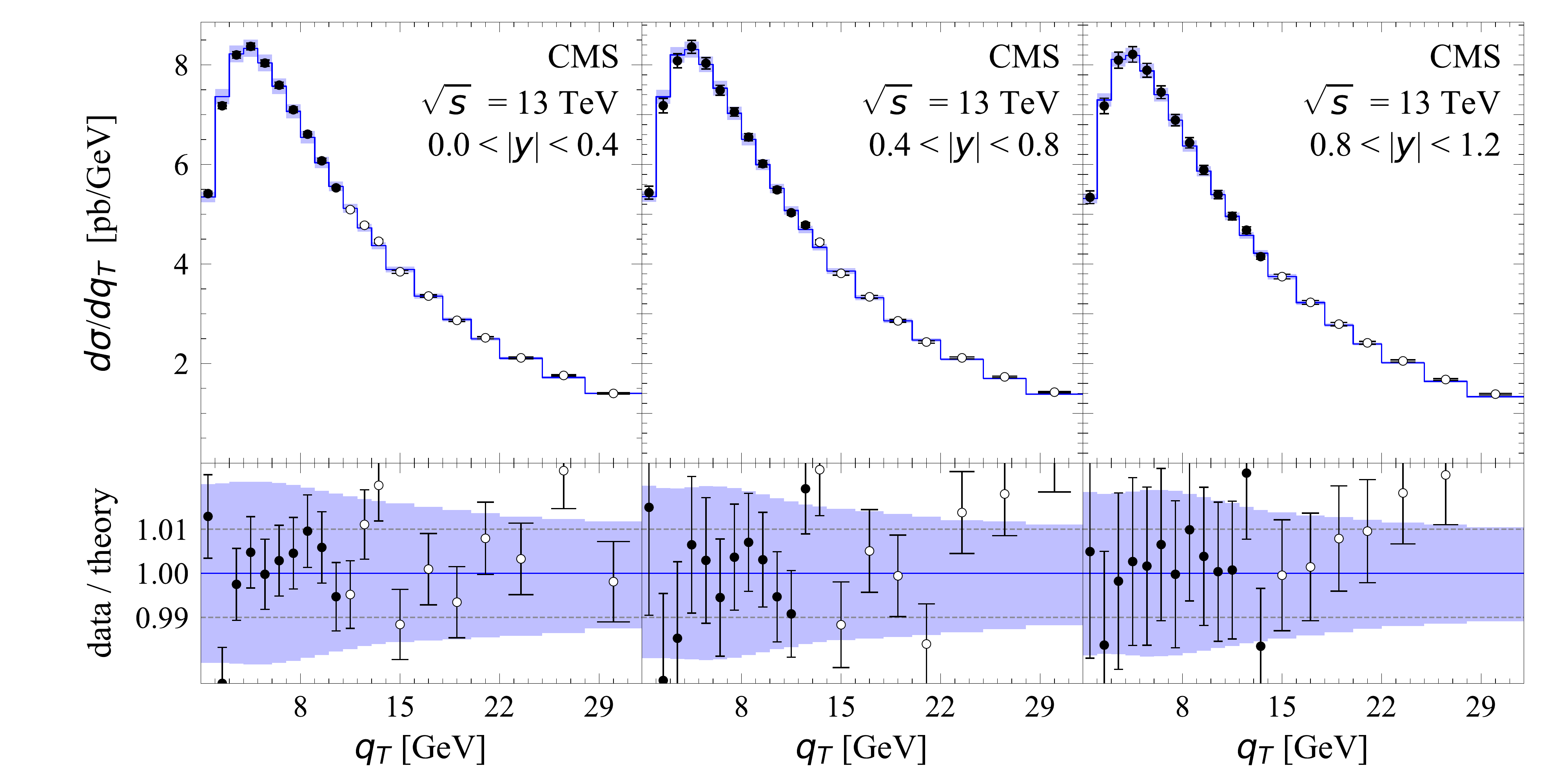}
\includegraphics[height=0.28\textheight]{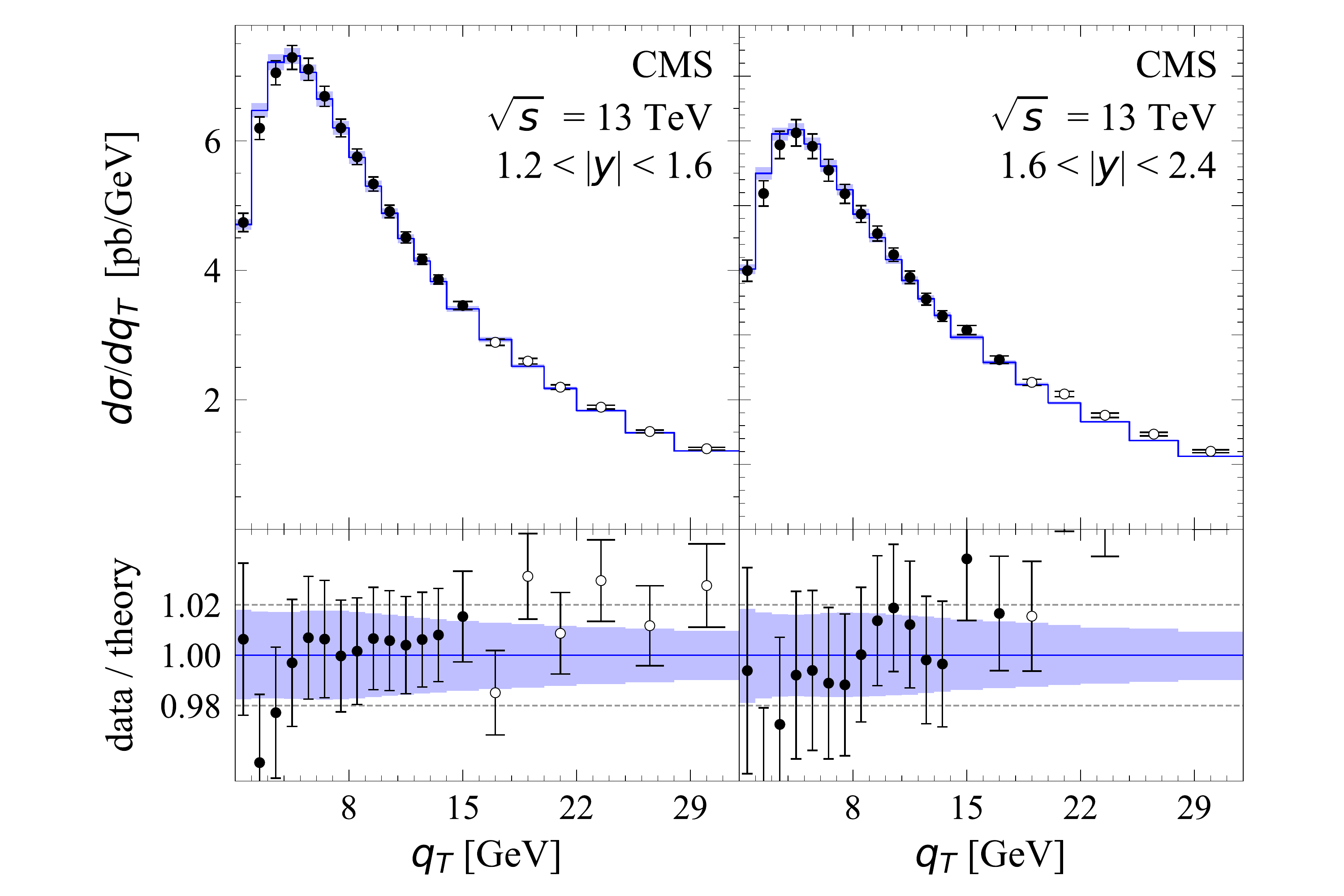}
\includegraphics[height=0.28\textheight]{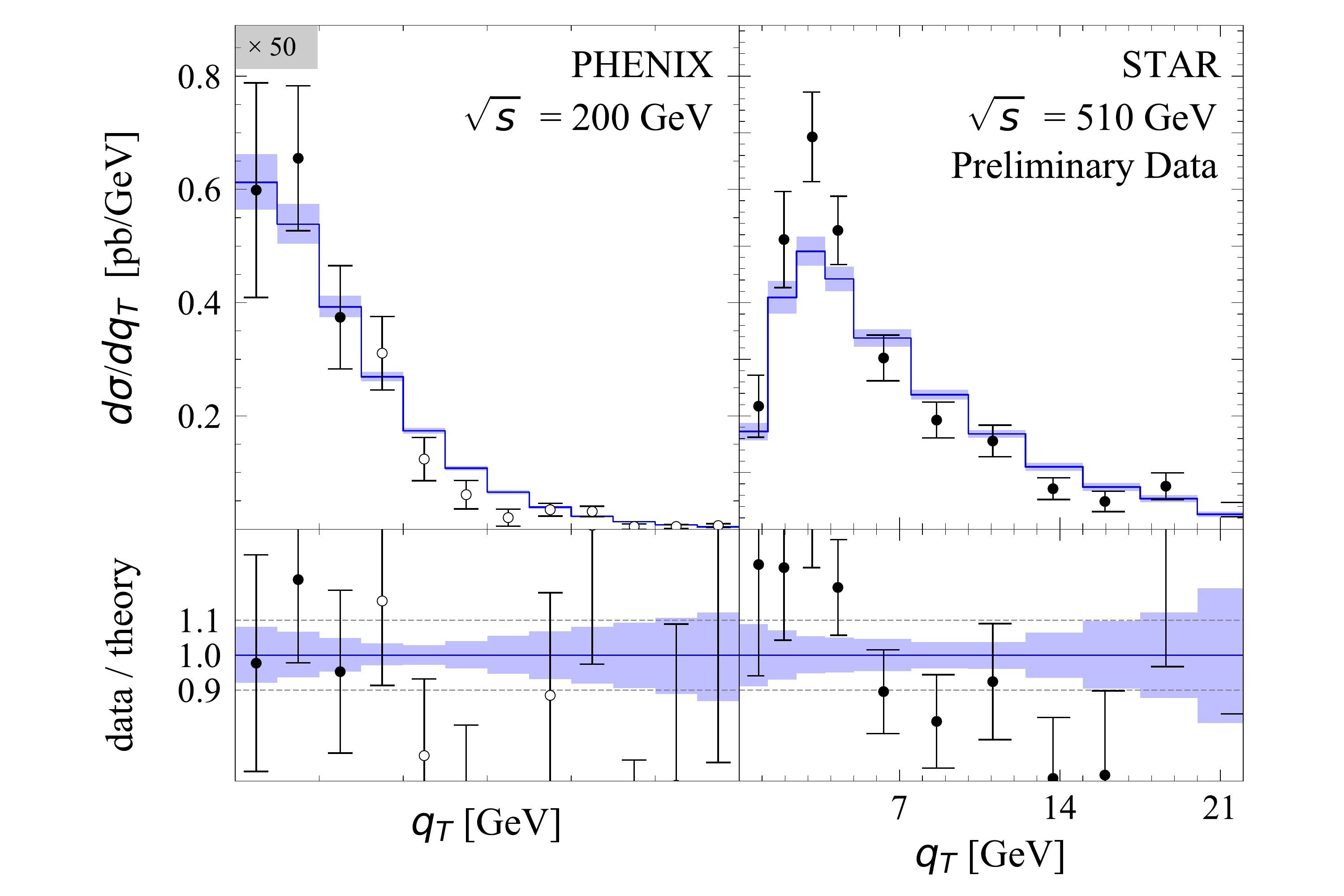}
\end{center}
\caption{Upper and central rows: differential cross-section of the $Z/\gamma^*$ boson production measured by CMS at different values of $y$ (see the text on the plots for details). Lower row: differential cross-section for the DY process measured at PHENIX (left) and STAR (right).}
\label{fig:CMS2}
\end{figure}

\begin{figure}[h]
\begin{center}
\includegraphics[height=0.28\textheight]{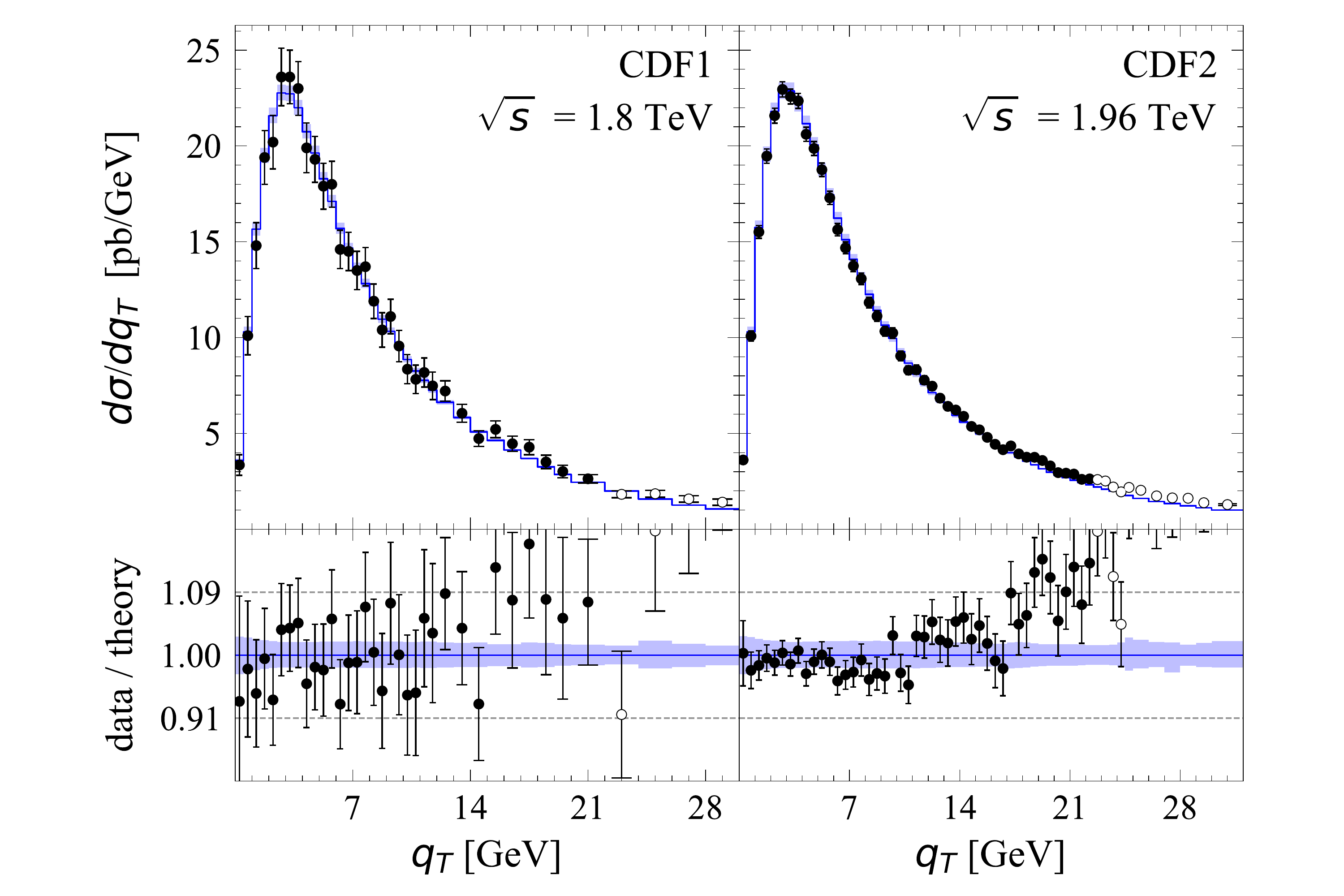}
\includegraphics[height=0.28\textheight]{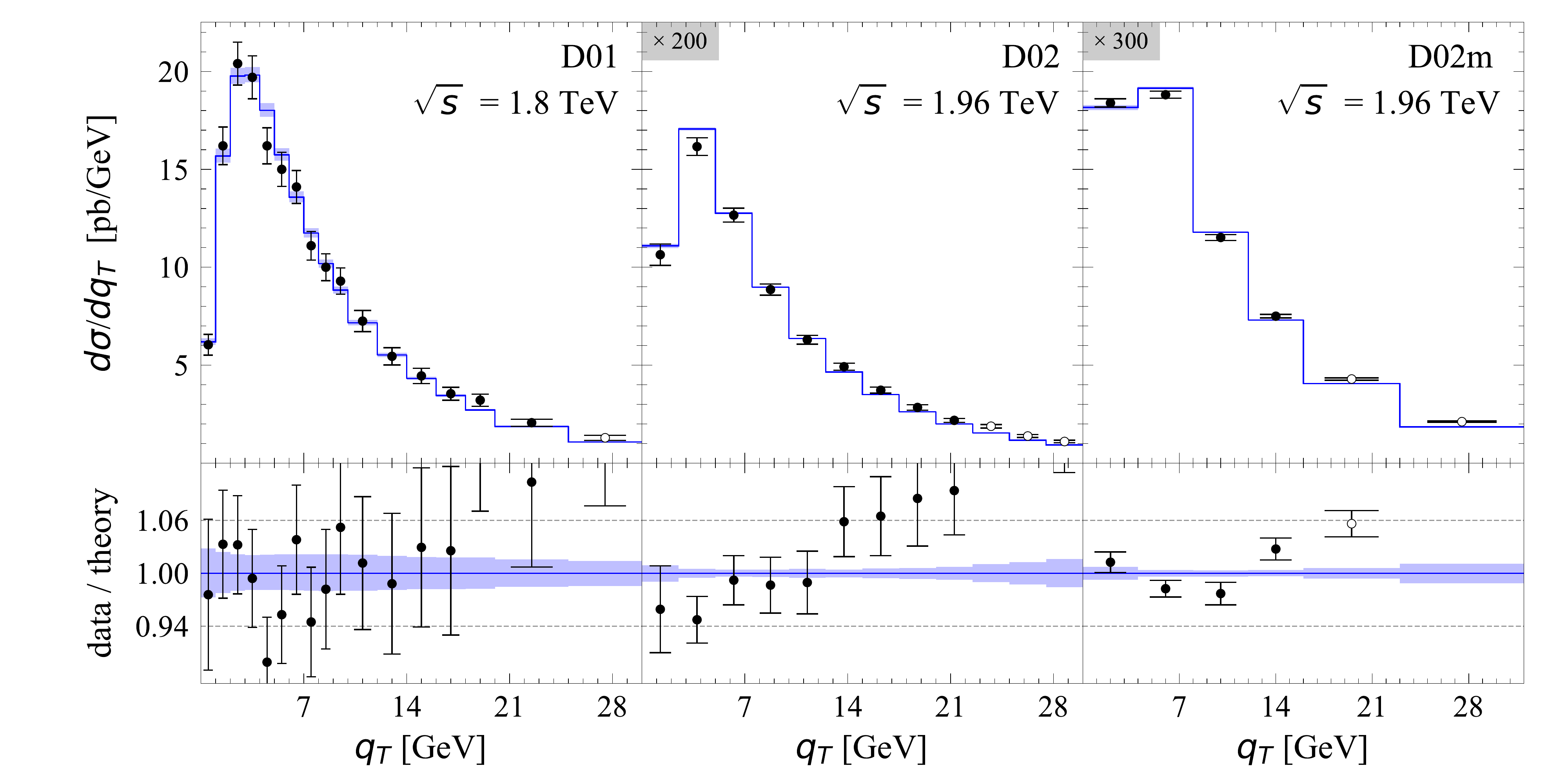}
\includegraphics[height=0.28\textheight]{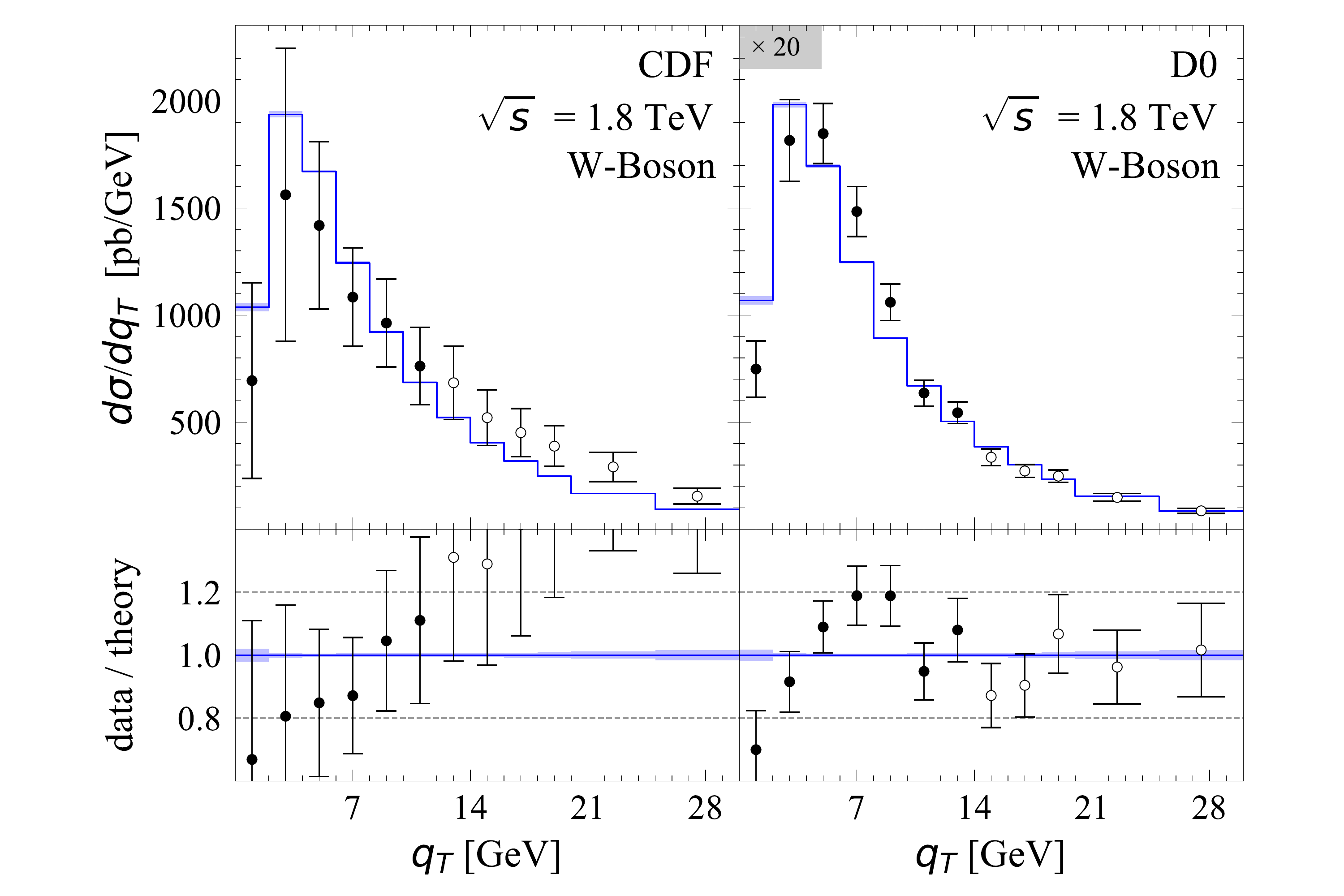}
\end{center}
\caption{Differential cross-section for the $Z/\gamma^*$ (upper and center rows) and $W$ (lower row) boson production in the DY-process measured by CDF and D0, at different values of $\sqrt{s}$.}
\label{fig:Tevatron}
\end{figure}

\begin{figure}[htb]
\begin{center}
\includegraphics[width=0.95\textwidth]{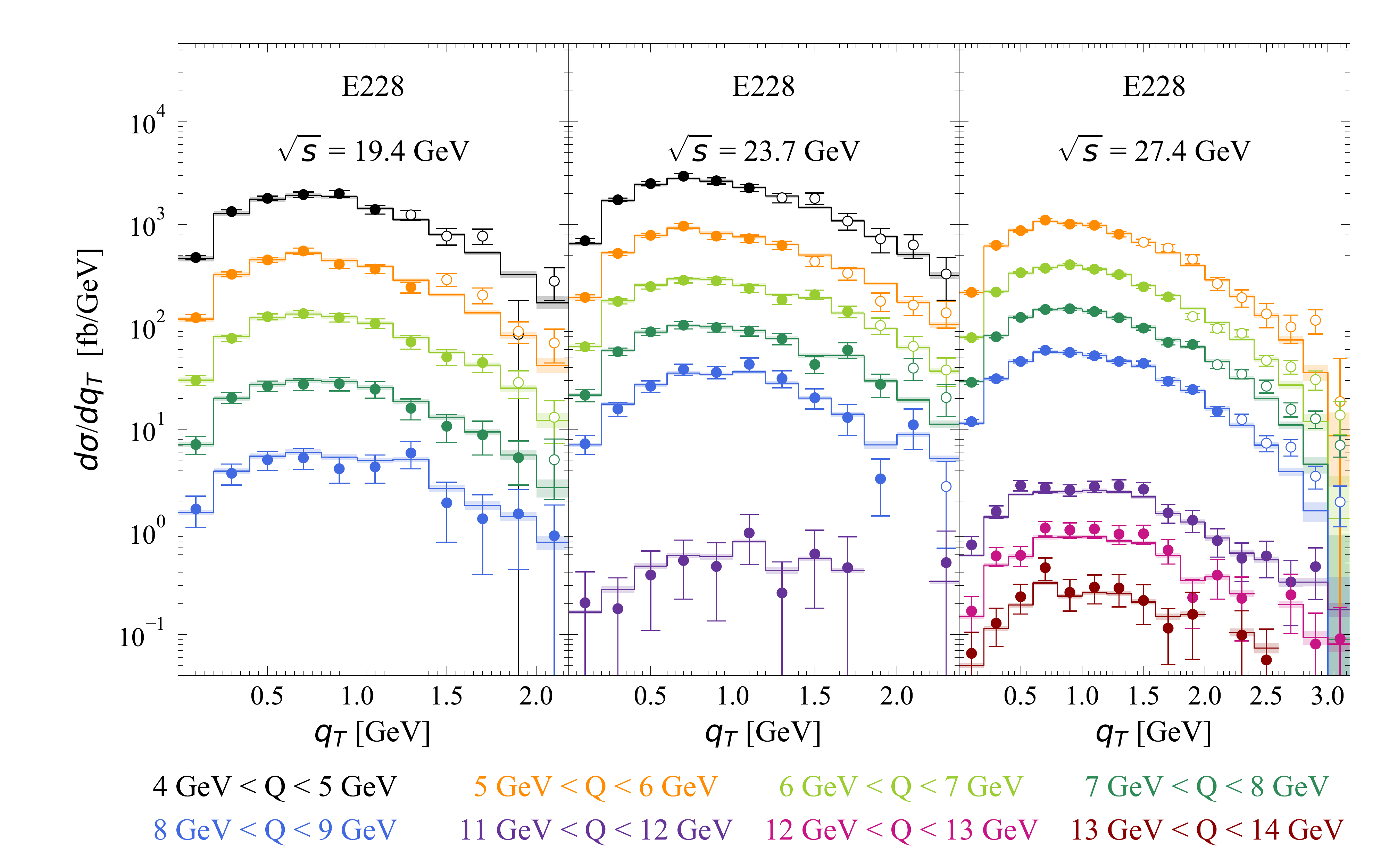}
\includegraphics[width=0.95\textwidth]{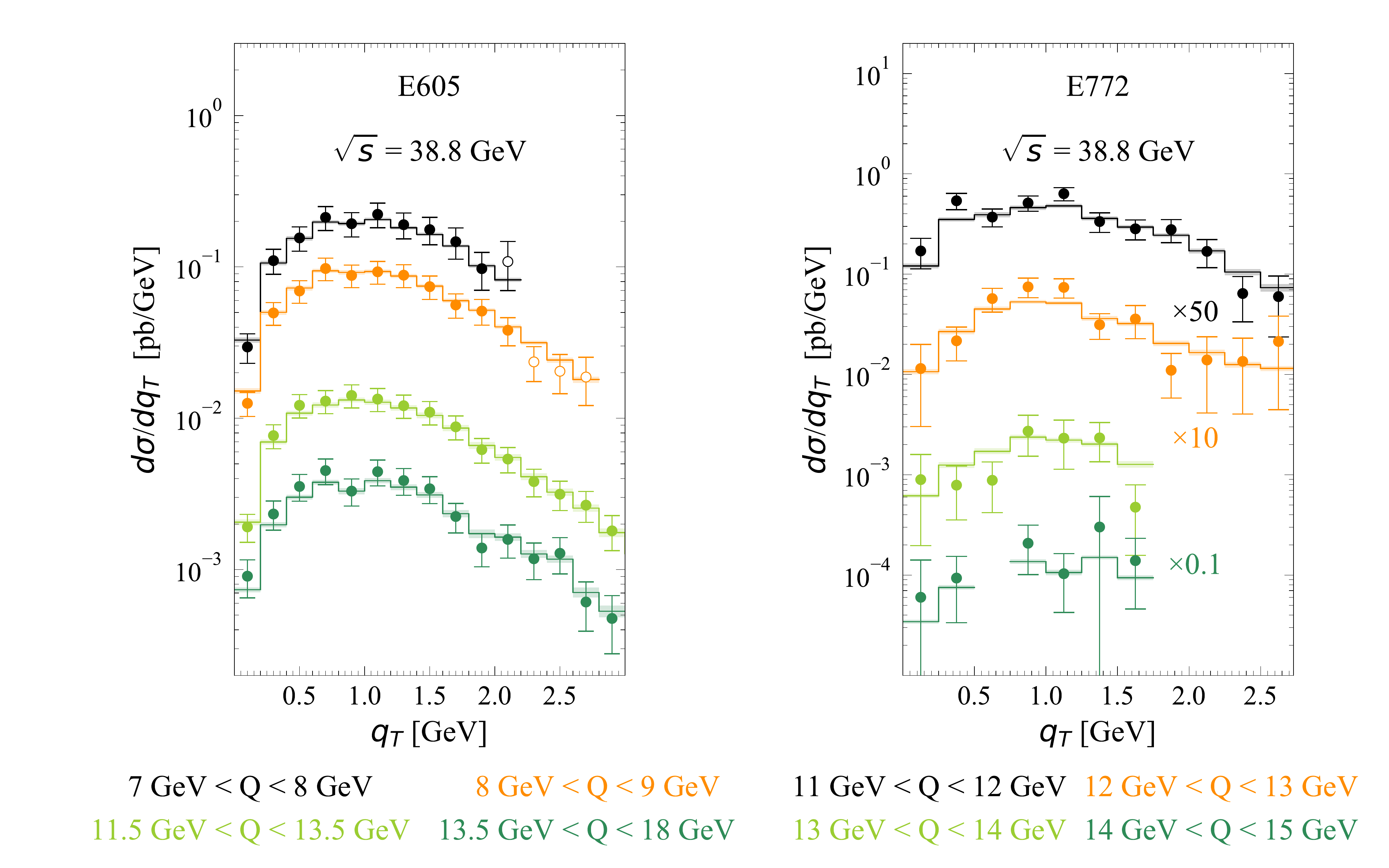}
\end{center}
\caption{Differential cross section of the DY process measured by the E228, E605 and E772 experiments at different values of $\sqrt{s}$ and $Q$. To increase visibility, the data in different $Q-$bins of the E772 experiment are multiplied by the factors indicated in the plots. }
\label{fig:E-series}
\end{figure}

\clearpage

\bibliography{bibFILE}

\providecommand{\href}[2]{#2}\begingroup\raggedright\begin{thebibliography}{10}

\bibitem{Becher:2010tm}
T.~Becher and M.~Neubert, \emph{{Drell-Yan Production at Small $q_T$,
  Transverse Parton Distributions and the Collinear Anomaly}},
  \href{https://doi.org/10.1140/epjc/s10052-011-1665-7}{\emph{Eur. Phys. J. C}
  {\bfseries 71} (2011) 1665}
  [\href{https://arxiv.org/abs/1007.4005}{{\ttfamily 1007.4005}}].

\bibitem{Collins:2011zzd}
J.~Collins, \emph{{Foundations of perturbative QCD}}, vol.~32, Cambridge
  University Press (11, 2013).

\bibitem{Echevarria:2011epo}
M.G.~Echevarria, A.~Idilbi and I.~Scimemi, \emph{{Factorization Theorem For
  Drell-Yan At Low $q_T$ And Transverse Momentum Distributions
  On-The-Light-Cone}},
  \href{https://doi.org/10.1007/JHEP07(2012)002}{\emph{JHEP} {\bfseries 07}
  (2012) 002} [\href{https://arxiv.org/abs/1111.4996}{{\ttfamily 1111.4996}}].

\bibitem{Echevarria:2012js}
M.G.~Echevarr\'\i{}a, A.~Idilbi and I.~Scimemi, \emph{{Soft and Collinear
  Factorization and Transverse Momentum Dependent Parton Distribution
  Functions}},
  \href{https://doi.org/10.1016/j.physletb.2013.09.003}{\emph{Phys. Lett. B}
  {\bfseries 726} (2013) 795}
  [\href{https://arxiv.org/abs/1211.1947}{{\ttfamily 1211.1947}}].

\bibitem{Scimemi:2017etj}
I.~Scimemi and A.~Vladimirov, \emph{{Analysis of vector boson production within
  TMD factorization}},
  \href{https://doi.org/10.1140/epjc/s10052-018-5557-y}{\emph{Eur. Phys. J. C}
  {\bfseries 78} (2018) 89} [\href{https://arxiv.org/abs/1706.01473}{{\ttfamily
  1706.01473}}].

\bibitem{Bacchetta:2019sam}
A.~Bacchetta, V.~Bertone, C.~Bissolotti, G.~Bozzi, F.~Delcarro, F.~Piacenza
  et~al., \emph{{Transverse-momentum-dependent parton distributions up to
  N$^{3}$LL from Drell-Yan data}},
  \href{https://doi.org/10.1007/JHEP07(2020)117}{\emph{JHEP} {\bfseries 07}
  (2020) 117} [\href{https://arxiv.org/abs/1912.07550}{{\ttfamily
  1912.07550}}].

\bibitem{Bertone:2019nxa}
V.~Bertone, I.~Scimemi and A.~Vladimirov, \emph{{Extraction of unpolarized
  quark transverse momentum dependent parton distributions from
  Drell-Yan/Z-boson production}},
  \href{https://doi.org/10.1007/JHEP06(2019)028}{\emph{JHEP} {\bfseries 06}
  (2019) 028} [\href{https://arxiv.org/abs/1902.08474}{{\ttfamily
  1902.08474}}].

\bibitem{Scimemi:2019cmh}
I.~Scimemi and A.~Vladimirov, \emph{{Non-perturbative structure of
  semi-inclusive deep-inelastic and Drell-Yan scattering at small transverse
  momentum}}, \href{https://doi.org/10.1007/JHEP06(2020)137}{\emph{JHEP}
  {\bfseries 06} (2020) 137}
  [\href{https://arxiv.org/abs/1912.06532}{{\ttfamily 1912.06532}}].

\bibitem{Bacchetta:2022awv}
A.~Bacchetta, V.~Bertone, C.~Bissolotti, G.~Bozzi, M.~Cerutti, F.~Piacenza
  et~al., \emph{{Unpolarized Transverse Momentum Distributions from a global
  fit of Drell-Yan and Semi-Inclusive Deep-Inelastic Scattering data}},
  \href{https://arxiv.org/abs/2206.07598}{{\ttfamily 2206.07598}}.

\bibitem{Gehrmann:2012ze}
T.~Gehrmann, T.~Lubbert and L.L.~Yang, \emph{{Transverse parton distribution
  functions at next-to-next-to-leading order: the quark-to-quark case}},
  \href{https://doi.org/10.1103/PhysRevLett.109.242003}{\emph{Phys. Rev. Lett.}
  {\bfseries 109} (2012) 242003}
  [\href{https://arxiv.org/abs/1209.0682}{{\ttfamily 1209.0682}}].

\bibitem{Gehrmann:2014yya}
T.~Gehrmann, T.~Luebbert and L.L.~Yang, \emph{{Calculation of the transverse
  parton distribution functions at next-to-next-to-leading order}},
  \href{https://doi.org/10.1007/JHEP06(2014)155}{\emph{JHEP} {\bfseries 06}
  (2014) 155} [\href{https://arxiv.org/abs/1403.6451}{{\ttfamily 1403.6451}}].

\bibitem{Echevarria:2015byo}
M.G.~Echevarria, I.~Scimemi and A.~Vladimirov, \emph{{Universal transverse
  momentum dependent soft function at NNLO}},
  \href{https://doi.org/10.1103/PhysRevD.93.054004}{\emph{Phys. Rev. D}
  {\bfseries 93} (2016) 054004}
  [\href{https://arxiv.org/abs/1511.05590}{{\ttfamily 1511.05590}}].

\bibitem{Echevarria:2016scs}
M.G.~Echevarria, I.~Scimemi and A.~Vladimirov, \emph{{Unpolarized Transverse
  Momentum Dependent Parton Distribution and Fragmentation Functions at
  next-to-next-to-leading order}},
  \href{https://doi.org/10.1007/JHEP09(2016)004}{\emph{JHEP} {\bfseries 09}
  (2016) 004} [\href{https://arxiv.org/abs/1604.07869}{{\ttfamily
  1604.07869}}].

\bibitem{Vladimirov:2016dll}
A.A.~Vladimirov, \emph{{Correspondence between Soft and Rapidity Anomalous
  Dimensions}},
  \href{https://doi.org/10.1103/PhysRevLett.118.062001}{\emph{Phys. Rev. Lett.}
  {\bfseries 118} (2017) 062001}
  [\href{https://arxiv.org/abs/1610.05791}{{\ttfamily 1610.05791}}].

\bibitem{Li:2016ctv}
Y.~Li and H.X.~Zhu, \emph{{Bootstrapping Rapidity Anomalous Dimensions for
  Transverse-Momentum Resummation}},
  \href{https://doi.org/10.1103/PhysRevLett.118.022004}{\emph{Phys. Rev. Lett.}
  {\bfseries 118} (2017) 022004}
  [\href{https://arxiv.org/abs/1604.01404}{{\ttfamily 1604.01404}}].

\bibitem{ATLAS:2019zci}
{\scshape ATLAS} collaboration, \emph{{Measurement of the transverse momentum
  distribution of Drell\textendash{}Yan lepton pairs in
  proton\textendash{}proton collisions at $\sqrt{s}=13$ TeV with the ATLAS
  detector}}, \href{https://doi.org/10.1140/epjc/s10052-020-8001-z}{\emph{Eur.
  Phys. J. C} {\bfseries 80} (2020) 616}
  [\href{https://arxiv.org/abs/1912.02844}{{\ttfamily 1912.02844}}].

\bibitem{Das:2019btv}
G.~Das, S.-O.~Moch and A.~Vogt, \emph{{Soft corrections to inclusive
  deep-inelastic scattering at four loops and beyond}},
  \href{https://doi.org/10.1007/JHEP03(2020)116}{\emph{JHEP} {\bfseries 03}
  (2020) 116} [\href{https://arxiv.org/abs/1912.12920}{{\ttfamily
  1912.12920}}].

\bibitem{Luo:2019szz}
M.-x.~Luo, T.-Z.~Yang, H.X.~Zhu and Y.J.~Zhu, \emph{{Quark Transverse Parton
  Distribution at the Next-to-Next-to-Next-to-Leading Order}},
  \href{https://doi.org/10.1103/PhysRevLett.124.092001}{\emph{Phys. Rev. Lett.}
  {\bfseries 124} (2020) 092001}
  [\href{https://arxiv.org/abs/1912.05778}{{\ttfamily 1912.05778}}].

\bibitem{vonManteuffel:2020vjv}
A.~von Manteuffel, E.~Panzer and R.M.~Schabinger, \emph{{Cusp and collinear
  anomalous dimensions in four-loop QCD from form factors}},
  \href{https://doi.org/10.1103/PhysRevLett.124.162001}{\emph{Phys. Rev. Lett.}
  {\bfseries 124} (2020) 162001}
  [\href{https://arxiv.org/abs/2002.04617}{{\ttfamily 2002.04617}}].

\bibitem{Duhr:2020seh}
C.~Duhr, F.~Dulat and B.~Mistlberger, \emph{{Drell-Yan Cross Section to Third
  Order in the Strong Coupling Constant}},
  \href{https://doi.org/10.1103/PhysRevLett.125.172001}{\emph{Phys. Rev. Lett.}
  {\bfseries 125} (2020) 172001}
  [\href{https://arxiv.org/abs/2001.07717}{{\ttfamily 2001.07717}}].

\bibitem{Ebert:2020yqt}
M.A.~Ebert, B.~Mistlberger and G.~Vita, \emph{{Transverse momentum dependent
  PDFs at N$^3$LO}}, \href{https://doi.org/10.1007/JHEP09(2020)146}{\emph{JHEP}
  {\bfseries 09} (2020) 146}
  [\href{https://arxiv.org/abs/2006.05329}{{\ttfamily 2006.05329}}].

\bibitem{Moch:2021qrk}
S.~Moch, B.~Ruijl, T.~Ueda, J.A.M.~Vermaseren and A.~Vogt, \emph{{Low moments
  of the four-loop splitting functions in QCD}},
  \href{https://doi.org/10.1016/j.physletb.2021.136853}{\emph{Phys. Lett. B}
  {\bfseries 825} (2022) 136853}
  [\href{https://arxiv.org/abs/2111.15561}{{\ttfamily 2111.15561}}].

\bibitem{Chen:2021rft}
L.~Chen, M.~Czakon and M.~Niggetiedt, \emph{{The complete singlet contribution
  to the massless quark form factor at three loops in QCD}},
  \href{https://doi.org/10.1007/JHEP12(2021)095}{\emph{JHEP} {\bfseries 12}
  (2021) 095} [\href{https://arxiv.org/abs/2109.01917}{{\ttfamily
  2109.01917}}].

\bibitem{Duhr:2022yyp}
C.~Duhr, B.~Mistlberger and G.~Vita, \emph{{Four-Loop Rapidity Anomalous
  Dimension and Event Shapes to Fourth Logarithmic Order}},
  \href{https://doi.org/10.1103/PhysRevLett.129.162001}{\emph{Phys. Rev. Lett.}
  {\bfseries 129} (2022) 162001}
  [\href{https://arxiv.org/abs/2205.02242}{{\ttfamily 2205.02242}}].

\bibitem{Moult:2022xzt}
I.~Moult, H.X.~Zhu and Y.J.~Zhu, \emph{{The four loop QCD rapidity anomalous
  dimension}}, \href{https://doi.org/10.1007/JHEP08(2022)280}{\emph{JHEP}
  {\bfseries 08} (2022) 280}
  [\href{https://arxiv.org/abs/2205.02249}{{\ttfamily 2205.02249}}].

\bibitem{Bury:2022czx}
M.~Bury, F.~Hautmann, S.~Leal-Gomez, I.~Scimemi, A.~Vladimirov and P.~Zurita,
  \emph{{PDF bias and flavor dependence in TMD distributions}},
  \href{https://doi.org/10.1007/JHEP10(2022)118}{\emph{JHEP} {\bfseries 10}
  (2022) 118} [\href{https://arxiv.org/abs/2201.07114}{{\ttfamily
  2201.07114}}].

\bibitem{Vladimirov:2019bfa}
A.~Vladimirov, \emph{{Pion-induced Drell-Yan processes within TMD
  factorization}}, \href{https://doi.org/10.1007/JHEP10(2019)090}{\emph{JHEP}
  {\bfseries 10} (2019) 090}
  [\href{https://arxiv.org/abs/1907.10356}{{\ttfamily 1907.10356}}].

\bibitem{STAR:SX}
S.~Fazio and X.~Chu. private communication.

\bibitem{CMS:2019raw}
{\scshape CMS} collaboration, \emph{{Measurements of differential Z boson
  production cross sections in proton-proton collisions at $ \sqrt{s} $ = 13
  TeV}}, \href{https://doi.org/10.1007/JHEP12(2019)061}{\emph{JHEP} {\bfseries
  12} (2019) 061} [\href{https://arxiv.org/abs/1909.04133}{{\ttfamily
  1909.04133}}].

\bibitem{CMS:2021oex}
{\scshape CMS} collaboration, \emph{{Measurement of mass dependence of the
  transverse momentum of Drell Yan lepton pairs in proton-proton collisions at
  $\sqrt{s} = 13~\mathrm{TeV}$}}, .

\bibitem{CMS:2022ubq}
{\scshape CMS} collaboration, \emph{{Measurement of the mass dependence of the
  transverse momentum of lepton pairs in Drell-Yan production in proton-proton
  collisions at $\sqrt{s}$ = 13 TeV}},
  \href{https://doi.org/10.1140/epjc/s10052-023-11631-7}{\emph{Eur. Phys. J. C}
  {\bfseries 83} (2023) 628}
  [\href{https://arxiv.org/abs/2205.04897}{{\ttfamily 2205.04897}}].

\bibitem{LHCb:2021huf}
{\scshape LHCb} collaboration, \emph{{Precision measurement of forward $Z$
  boson production in proton-proton collisions at $\sqrt{s} = 13$ TeV}},
  \href{https://doi.org/10.1007/JHEP07(2022)026}{\emph{JHEP} {\bfseries 07}
  (2022) 026} [\href{https://arxiv.org/abs/2112.07458}{{\ttfamily
  2112.07458}}].

\bibitem{CDF:1991pgi}
{\scshape CDF} collaboration, \emph{{Measurement of the W P(T) distribution in
  $\bar{p}p$ collisions at $\sqrt{s} = 1.8$ TeV}},
  \href{https://doi.org/10.1103/PhysRevLett.66.2951}{\emph{Phys. Rev. Lett.}
  {\bfseries 66} (1991) 2951}.

\bibitem{D0:1998thd}
{\scshape D0} collaboration, \emph{{Measurement of the shape of the transverse
  momentum distribution of $W$ bosons produced in $p\bar{p}$ collisions at
  $\sqrt{s} = 1.8$ TeV}},
  \href{https://doi.org/10.1103/PhysRevLett.80.5498}{\emph{Phys. Rev. Lett.}
  {\bfseries 80} (1998) 5498}
  [\href{https://arxiv.org/abs/hep-ex/9803003}{{\ttfamily hep-ex/9803003}}].

\bibitem{Bailey:2020ooq}
S.~Bailey, T.~Cridge, L.A.~Harland-Lang, A.D.~Martin and R.S.~Thorne,
  \emph{{Parton distributions from LHC, HERA, Tevatron and fixed target data:
  MSHT20 PDFs}},
  \href{https://doi.org/10.1140/epjc/s10052-021-09057-0}{\emph{Eur. Phys. J. C}
  {\bfseries 81} (2021) 341}
  [\href{https://arxiv.org/abs/2012.04684}{{\ttfamily 2012.04684}}].

\bibitem{artemide}
``\texttt{artemide}.'' stable version:
  \url{https://github.com/VladimirovAlexey/artemide-public} \\ in-production
  version: \url{https://github.com/VladimirovAlexey/artemide-development}.

\bibitem{NNPDF:2017mvq}
{\scshape NNPDF} collaboration, \emph{{Parton distributions from high-precision
  collider data}},
  \href{https://doi.org/10.1140/epjc/s10052-017-5199-5}{\emph{Eur. Phys. J. C}
  {\bfseries 77} (2017) 663}
  [\href{https://arxiv.org/abs/1706.00428}{{\ttfamily 1706.00428}}].

\bibitem{Chiu:2012ir}
J.-Y.~Chiu, A.~Jain, D.~Neill and I.Z.~Rothstein, \emph{{A Formalism for the
  Systematic Treatment of Rapidity Logarithms in Quantum Field Theory}},
  \href{https://doi.org/10.1007/JHEP05(2012)084}{\emph{JHEP} {\bfseries 05}
  (2012) 084} [\href{https://arxiv.org/abs/1202.0814}{{\ttfamily 1202.0814}}].

\bibitem{Scimemi:2018xaf}
I.~Scimemi and A.~Vladimirov, \emph{{Systematic analysis of double-scale
  evolution}}, \href{https://doi.org/10.1007/JHEP08(2018)003}{\emph{JHEP}
  {\bfseries 08} (2018) 003}
  [\href{https://arxiv.org/abs/1803.11089}{{\ttfamily 1803.11089}}].

\bibitem{Collins:1989gx}
J.C.~Collins, D.E.~Soper and G.F.~Sterman, \emph{{Factorization of Hard
  Processes in QCD}},
  \href{https://doi.org/10.1142/9789814503266_0001}{\emph{Adv. Ser. Direct.
  High Energy Phys.} {\bfseries 5} (1989) 1}
  [\href{https://arxiv.org/abs/hep-ph/0409313}{{\ttfamily hep-ph/0409313}}].

\bibitem{Vladimirov:2017ksc}
A.~Vladimirov, \emph{{Structure of rapidity divergences in multi-parton
  scattering soft factors}},
  \href{https://doi.org/10.1007/JHEP04(2018)045}{\emph{JHEP} {\bfseries 04}
  (2018) 045} [\href{https://arxiv.org/abs/1707.07606}{{\ttfamily
  1707.07606}}].

\bibitem{Vladimirov:2021hdn}
A.~Vladimirov, V.~Moos and I.~Scimemi, \emph{{Transverse momentum dependent
  operator expansion at next-to-leading power}},
  \href{https://arxiv.org/abs/2109.09771}{{\ttfamily 2109.09771}}.

\bibitem{Gutierrez-Reyes:2020ouu}
D.~Gutierrez-Reyes, S.~Leal-Gomez and I.~Scimemi, \emph{{W-boson production in
  TMD factorization}},
  \href{https://doi.org/10.1140/epjc/s10052-021-09202-9}{\emph{Eur. Phys. J. C}
  {\bfseries 81} (2021) 418}
  [\href{https://arxiv.org/abs/2011.05351}{{\ttfamily 2011.05351}}].

\bibitem{ParticleDataGroup:2022pth}
{\scshape Particle Data Group} collaboration, \emph{{Review of Particle
  Physics}}, \href{https://doi.org/10.1093/ptep/ptac097}{\emph{PTEP} {\bfseries
  2022} (2022) 083C01}.

\bibitem{Balitsky:2017gis}
I.~Balitsky and A.~Tarasov, \emph{{Power corrections to TMD factorization for
  Z-boson production}},
  \href{https://doi.org/10.1007/JHEP05(2018)150}{\emph{JHEP} {\bfseries 05}
  (2018) 150} [\href{https://arxiv.org/abs/1712.09389}{{\ttfamily
  1712.09389}}].

\bibitem{Collins:1981va}
J.C.~Collins and D.E.~Soper, \emph{{Back-To-Back Jets: Fourier Transform from B
  to K-Transverse}},
  \href{https://doi.org/10.1016/0550-3213(82)90453-9}{\emph{Nucl. Phys. B}
  {\bfseries 197} (1982) 446}.

\bibitem{Herzog:2018kwj}
F.~Herzog, S.~Moch, B.~Ruijl, T.~Ueda, J.A.M.~Vermaseren and A.~Vogt,
  \emph{{Five-loop contributions to low-N non-singlet anomalous dimensions in
  QCD}}, \href{https://doi.org/10.1016/j.physletb.2019.01.060}{\emph{Phys.
  Lett. B} {\bfseries 790} (2019) 436}
  [\href{https://arxiv.org/abs/1812.11818}{{\ttfamily 1812.11818}}].

\bibitem{Agarwal:2021zft}
B.~Agarwal, A.~von Manteuffel, E.~Panzer and R.M.~Schabinger, \emph{{Four-loop
  collinear anomalous dimensions in QCD and N=4 super Yang-Mills}},
  \href{https://doi.org/10.1016/j.physletb.2021.136503}{\emph{Phys. Lett. B}
  {\bfseries 820} (2021) 136503}
  [\href{https://arxiv.org/abs/2102.09725}{{\ttfamily 2102.09725}}].

\bibitem{Luo:2020epw}
M.-x.~Luo, T.-Z.~Yang, H.X.~Zhu and Y.J.~Zhu, \emph{{Unpolarized quark and
  gluon TMD PDFs and FFs at N$^{3}$LO}},
  \href{https://doi.org/10.1007/JHEP06(2021)115}{\emph{JHEP} {\bfseries 06}
  (2021) 115} [\href{https://arxiv.org/abs/2012.03256}{{\ttfamily
  2012.03256}}].

\bibitem{Moch:2004pa}
S.~Moch, J.A.M.~Vermaseren and A.~Vogt, \emph{{The Three loop splitting
  functions in QCD: The Nonsinglet case}},
  \href{https://doi.org/10.1016/j.nuclphysb.2004.03.030}{\emph{Nucl. Phys. B}
  {\bfseries 688} (2004) 101}
  [\href{https://arxiv.org/abs/hep-ph/0403192}{{\ttfamily hep-ph/0403192}}].

\bibitem{Vladimirov:2020umg}
A.A.~Vladimirov, \emph{{Self-contained definition of the Collins-Soper
  kernel}}, \href{https://doi.org/10.1103/PhysRevLett.125.192002}{\emph{Phys.
  Rev. Lett.} {\bfseries 125} (2020) 192002}
  [\href{https://arxiv.org/abs/2003.02288}{{\ttfamily 2003.02288}}].

\bibitem{Moos:2020wvd}
V.~Moos and A.~Vladimirov, \emph{{Calculation of transverse momentum dependent
  distributions beyond the leading power}},
  \href{https://doi.org/10.1007/JHEP12(2020)145}{\emph{JHEP} {\bfseries 12}
  (2020) 145} [\href{https://arxiv.org/abs/2008.01744}{{\ttfamily
  2008.01744}}].

\bibitem{Signori:2013gra}
A.~Signori, A.~Bacchetta and M.~Radici, \emph{{Flavor dependence of unpolarized
  TMDs from semi-inclusive pion production}},
  \href{https://doi.org/10.1142/S2010194514600209}{\emph{Int. J. Mod. Phys.
  Conf. Ser.} {\bfseries 25} (2014) 1460020}
  [\href{https://arxiv.org/abs/1309.5929}{{\ttfamily 1309.5929}}].

\bibitem{Bacchetta:2018lna}
A.~Bacchetta, G.~Bozzi, M.~Radici, M.~Ritzmann and A.~Signori, \emph{{Effect of
  Flavor-Dependent Partonic Transverse Momentum on the Determination of the $W$
  Boson Mass in Hadronic Collisions}},
  \href{https://doi.org/10.1016/j.physletb.2018.11.002}{\emph{Phys. Lett. B}
  {\bfseries 788} (2019) 542}
  [\href{https://arxiv.org/abs/1807.02101}{{\ttfamily 1807.02101}}].

\bibitem{Echevarria:2012pw}
M.G.~Echevarria, A.~Idilbi, A.~Sch\"afer and I.~Scimemi,
  \emph{{Model-Independent Evolution of Transverse Momentum Dependent
  Distribution Functions (TMDs) at NNLL}},
  \href{https://doi.org/10.1140/epjc/s10052-013-2636-y}{\emph{Eur. Phys. J. C}
  {\bfseries 73} (2013) 2636}
  [\href{https://arxiv.org/abs/1208.1281}{{\ttfamily 1208.1281}}].

\bibitem{Neumann:2022lft}
T.~Neumann and J.~Campbell, \emph{{Fiducial Drell-Yan production at the LHC
  improved by transverse-momentum resummation at N4LLp+N3LO}},
  \href{https://doi.org/10.1103/PhysRevD.107.L011506}{\emph{Phys. Rev. D}
  {\bfseries 107} (2023) L011506}
  [\href{https://arxiv.org/abs/2207.07056}{{\ttfamily 2207.07056}}].

\bibitem{Ebert:2020dfc}
M.A.~Ebert, J.K.L.~Michel, I.W.~Stewart and F.J.~Tackmann, \emph{{Drell-Yan
  $q_{T}$ resummation of fiducial power corrections at N$^{3}$LL}},
  \href{https://doi.org/10.1007/JHEP04(2021)102}{\emph{JHEP} {\bfseries 04}
  (2021) 102} [\href{https://arxiv.org/abs/2006.11382}{{\ttfamily
  2006.11382}}].

\bibitem{Lee:2022nhh}
R.N.~Lee, A.~von Manteuffel, R.M.~Schabinger, A.V.~Smirnov, V.A.~Smirnov and
  M.~Steinhauser, \emph{{Quark and Gluon Form Factors in Four-Loop QCD}},
  \href{https://doi.org/10.1103/PhysRevLett.128.212002}{\emph{Phys. Rev. Lett.}
  {\bfseries 128} (2022) 212002}
  [\href{https://arxiv.org/abs/2202.04660}{{\ttfamily 2202.04660}}].

\bibitem{Landry:2002ix}
F.~Landry, R.~Brock, P.M.~Nadolsky and C.P.~Yuan, \emph{{Tevatron Run-1 $Z$
  boson data and Collins-Soper-Sterman resummation formalism}},
  \href{https://doi.org/10.1103/PhysRevD.67.073016}{\emph{Phys. Rev. D}
  {\bfseries 67} (2003) 073016}
  [\href{https://arxiv.org/abs/hep-ph/0212159}{{\ttfamily hep-ph/0212159}}].

\bibitem{Ito:1980ev}
A.S.~Ito et~al., \emph{{Measurement of the Continuum of Dimuons Produced in
  High-Energy Proton - Nucleus Collisions}},
  \href{https://doi.org/10.1103/PhysRevD.23.604}{\emph{Phys. Rev. D} {\bfseries
  23} (1981) 604}.

\bibitem{Moreno:1990sf}
G.~Moreno et~al., \emph{{Dimuon production in proton - copper collisions at
  $\sqrt{s}$ = 38.8-GeV}},
  \href{https://doi.org/10.1103/PhysRevD.43.2815}{\emph{Phys. Rev. D}
  {\bfseries 43} (1991) 2815}.

\bibitem{E772:1994cpf}
{\scshape E772} collaboration, \emph{{Cross-sections for the production of high
  mass muon pairs from 800-GeV proton bombardment of H-2}},
  \href{https://doi.org/10.1103/PhysRevD.50.3038}{\emph{Phys. Rev. D}
  {\bfseries 50} (1994) 3038}.

\bibitem{CDF:1999bpw}
{\scshape CDF} collaboration, \emph{{The transverse momentum and total cross
  section of $e^+e^-$ pairs in the $Z$ boson region from $p\bar{p}$ collisions
  at $\sqrt{s} = 1.8$ TeV}},
  \href{https://doi.org/10.1103/PhysRevLett.84.845}{\emph{Phys. Rev. Lett.}
  {\bfseries 84} (2000) 845}
  [\href{https://arxiv.org/abs/hep-ex/0001021}{{\ttfamily hep-ex/0001021}}].

\bibitem{CDF:2012brb}
{\scshape CDF} collaboration, \emph{{Transverse momentum cross section of
  $e^+e^-$ pairs in the $Z$-boson region from $p\bar{p}$ collisions at
  $\sqrt{s}=1.96$ TeV}},
  \href{https://doi.org/10.1103/PhysRevD.86.052010}{\emph{Phys. Rev. D}
  {\bfseries 86} (2012) 052010}
  [\href{https://arxiv.org/abs/1207.7138}{{\ttfamily 1207.7138}}].

\bibitem{D0:1999jba}
{\scshape D0} collaboration, \emph{{Measurement of the inclusive differential
  cross section for $Z$ bosons as a function of transverse momentum in
  $\bar{p}p$ collisions at $\sqrt{s} = 1.8$ TeV}},
  \href{https://doi.org/10.1103/PhysRevD.61.032004}{\emph{Phys. Rev. D}
  {\bfseries 61} (2000) 032004}
  [\href{https://arxiv.org/abs/hep-ex/9907009}{{\ttfamily hep-ex/9907009}}].

\bibitem{D0:2007lmg}
{\scshape D0} collaboration, \emph{{Measurement of the shape of the boson
  transverse momentum distribution in $p \bar{p} \to Z / \gamma^{*} \to e^+ e^-
  + X$ events produced at $\sqrt{s}$=1.96-TeV}},
  \href{https://doi.org/10.1103/PhysRevLett.100.102002}{\emph{Phys. Rev. Lett.}
  {\bfseries 100} (2008) 102002}
  [\href{https://arxiv.org/abs/0712.0803}{{\ttfamily 0712.0803}}].

\bibitem{D0:2010dbl}
{\scshape D0} collaboration, \emph{{Measurement of the Normalized $Z/\gamma^*
  -> \mu^+\mu^-$ Transverse Momentum Distribution in $p\bar{p}$ Collisions at
  $\sqrt{s}=1.96$ TeV}},
  \href{https://doi.org/10.1016/j.physletb.2010.09.012}{\emph{Phys. Lett. B}
  {\bfseries 693} (2010) 522}
  [\href{https://arxiv.org/abs/1006.0618}{{\ttfamily 1006.0618}}].

\bibitem{ATLAS:2015iiu}
{\scshape ATLAS} collaboration, \emph{{Measurement of the transverse momentum
  and $\phi ^*_{\eta }$ distributions of Drell\textendash{}Yan lepton pairs in
  proton\textendash{}proton collisions at $\sqrt{s}=8$ TeV with the ATLAS
  detector}}, \href{https://doi.org/10.1140/epjc/s10052-016-4070-4}{\emph{Eur.
  Phys. J. C} {\bfseries 76} (2016) 291}
  [\href{https://arxiv.org/abs/1512.02192}{{\ttfamily 1512.02192}}].

\bibitem{CMS:2011wyd}
{\scshape CMS} collaboration, \emph{{Measurement of the Rapidity and Transverse
  Momentum Distributions of $Z$ Bosons in $pp$ Collisions at $\sqrt{s}=7$
  TeV}}, \href{https://doi.org/10.1103/PhysRevD.85.032002}{\emph{Phys. Rev. D}
  {\bfseries 85} (2012) 032002}
  [\href{https://arxiv.org/abs/1110.4973}{{\ttfamily 1110.4973}}].

\bibitem{CMS:2016mwa}
{\scshape CMS} collaboration, \emph{{Measurement of the transverse momentum
  spectra of weak vector bosons produced in proton-proton collisions at $
  \sqrt{s}=8 $ TeV}},
  \href{https://doi.org/10.1007/JHEP02(2017)096}{\emph{JHEP} {\bfseries 02}
  (2017) 096} [\href{https://arxiv.org/abs/1606.05864}{{\ttfamily
  1606.05864}}].

\bibitem{LHCb:2015okr}
{\scshape LHCb} collaboration, \emph{{Measurement of the forward $Z$ boson
  production cross-section in $pp$ collisions at $\sqrt{s}=7$ TeV}},
  \href{https://doi.org/10.1007/JHEP08(2015)039}{\emph{JHEP} {\bfseries 08}
  (2015) 039} [\href{https://arxiv.org/abs/1505.07024}{{\ttfamily
  1505.07024}}].

\bibitem{LHCb:2015mad}
{\scshape LHCb} collaboration, \emph{{Measurement of forward W and Z boson
  production in $pp$ collisions at $ \sqrt{s}=8 $ TeV}},
  \href{https://doi.org/10.1007/JHEP01(2016)155}{\emph{JHEP} {\bfseries 01}
  (2016) 155} [\href{https://arxiv.org/abs/1511.08039}{{\ttfamily
  1511.08039}}].

\bibitem{PHENIX:2018dwt}
{\scshape PHENIX} collaboration, \emph{{Measurements of $\mu\mu$ pairs from
  open heavy flavor and Drell-Yan in $p+p$ collisions at $\sqrt{s}=200$ GeV}},
  \href{https://doi.org/10.1103/PhysRevD.99.072003}{\emph{Phys. Rev. D}
  {\bfseries 99} (2019) 072003}
  [\href{https://arxiv.org/abs/1805.02448}{{\ttfamily 1805.02448}}].

\bibitem{Buckley:2014ana}
A.~Buckley, J.~Ferrando, S.~Lloyd, K.~Nordstr\"om, B.~Page, M.~R\"ufenacht
  et~al., \emph{{LHAPDF6: parton density access in the LHC precision era}},
  \href{https://doi.org/10.1140/epjc/s10052-015-3318-8}{\emph{Eur. Phys. J. C}
  {\bfseries 75} (2015) 132} [\href{https://arxiv.org/abs/1412.7420}{{\ttfamily
  1412.7420}}].

\bibitem{Hautmann:2020cyp}
F.~Hautmann, I.~Scimemi and A.~Vladimirov, \emph{{Non-perturbative
  contributions to vector-boson transverse momentum spectra in hadronic
  collisions}},
  \href{https://doi.org/10.1016/j.physletb.2020.135478}{\emph{Phys. Lett. B}
  {\bfseries 806} (2020) 135478}
  [\href{https://arxiv.org/abs/2002.12810}{{\ttfamily 2002.12810}}].

\bibitem{Bury:2020vhj}
M.~Bury, A.~Prokudin and A.~Vladimirov, \emph{{Extraction of the Sivers
  Function from SIDIS, Drell-Yan, and $W^{\pm}/Z$ Data at
  Next-to-Next-to-Next-to Leading Order}},
  \href{https://doi.org/10.1103/PhysRevLett.126.112002}{\emph{Phys. Rev. Lett.}
  {\bfseries 126} (2021) 112002}
  [\href{https://arxiv.org/abs/2012.05135}{{\ttfamily 2012.05135}}].

\bibitem{Bury:2021sue}
M.~Bury, A.~Prokudin and A.~Vladimirov, \emph{{Extraction of the Sivers
  function from SIDIS, Drell-Yan, and $W^\pm/Z$ boson production data with TMD
  evolution}}, \href{https://doi.org/10.1007/JHEP05(2021)151}{\emph{JHEP}
  {\bfseries 05} (2021) 151}
  [\href{https://arxiv.org/abs/2103.03270}{{\ttfamily 2103.03270}}].

\bibitem{Horstmann:2022xkk}
M.~Horstmann, A.~Schafer and A.~Vladimirov, \emph{{Study of the worm-gear-T
  function g1T with semi-inclusive DIS data}},
  \href{https://doi.org/10.1103/PhysRevD.107.034016}{\emph{Phys. Rev. D}
  {\bfseries 107} (2023) 034016}
  [\href{https://arxiv.org/abs/2210.07268}{{\ttfamily 2210.07268}}].

\bibitem{Ball:2008by}
{\scshape NNPDF} collaboration, \emph{{A Determination of parton distributions
  with faithful uncertainty estimation}},
  \href{https://doi.org/10.1016/j.nuclphysb.2008.09.037}{\emph{Nucl. Phys. B}
  {\bfseries 809} (2009) 1} [\href{https://arxiv.org/abs/0808.1231}{{\ttfamily
  0808.1231}}].

\bibitem{Ball:2012wy}
R.D.~Ball et~al., \emph{{Parton Distribution Benchmarking with LHC Data}},
  \href{https://doi.org/10.1007/JHEP04(2013)125}{\emph{JHEP} {\bfseries 04}
  (2013) 125} [\href{https://arxiv.org/abs/1211.5142}{{\ttfamily 1211.5142}}].

\bibitem{DataProcessor}
``\texttt{artemide-DataProcessor}.''
  https://github.com/VladimirovAlexey/artemide-DataProcessor.

\bibitem{iminuit}
H.~Dembinski and P.O.~et~al., \emph{scikit-hep/iminuit}, .

\bibitem{DAgostini:2003syq}
G.~D'Agostini, \emph{{Bayesian reasoning in data analysis: A critical
  introduction}} (2003).

\bibitem{Hou:2016sho}
T.-J.~Hou et~al., \emph{{Reconstruction of Monte Carlo replicas from Hessian
  parton distributions}},
  \href{https://doi.org/10.1007/JHEP03(2017)099}{\emph{JHEP} {\bfseries 03}
  (2017) 099} [\href{https://arxiv.org/abs/1607.06066}{{\ttfamily
  1607.06066}}].

\bibitem{Vladimirov:2023aot}
A.~Vladimirov, \emph{{Kinematic power corrections in TMD factorization
  theorem}}, \href{https://doi.org/10.1007/JHEP12(2023)008}{\emph{JHEP}
  {\bfseries 12} (2023) 008}
  [\href{https://arxiv.org/abs/2307.13054}{{\ttfamily 2307.13054}}].

\bibitem{Catani:2007vq}
S.~Catani and M.~Grazzini, \emph{{An NNLO subtraction formalism in hadron
  collisions and its application to Higgs boson production at the LHC}},
  \href{https://doi.org/10.1103/PhysRevLett.98.222002}{\emph{Phys. Rev. Lett.}
  {\bfseries 98} (2007) 222002}
  [\href{https://arxiv.org/abs/hep-ph/0703012}{{\ttfamily hep-ph/0703012}}].

\bibitem{Catani:2009sm}
S.~Catani, L.~Cieri, G.~Ferrera, D.~de~Florian and M.~Grazzini, \emph{{Vector
  boson production at hadron colliders: a fully exclusive QCD calculation at
  NNLO}}, \href{https://doi.org/10.1103/PhysRevLett.103.082001}{\emph{Phys.
  Rev. Lett.} {\bfseries 103} (2009) 082001}
  [\href{https://arxiv.org/abs/0903.2120}{{\ttfamily 0903.2120}}].

\bibitem{BermudezMartinez:2022ctj}
A.~Bermudez~Martinez and A.~Vladimirov, \emph{{Determination of the
  Collins-Soper kernel from cross-sections ratios}},
  \href{https://doi.org/10.1103/PhysRevD.106.L091501}{\emph{Phys. Rev. D}
  {\bfseries 106} (2022) L091501}
  [\href{https://arxiv.org/abs/2206.01105}{{\ttfamily 2206.01105}}].

\bibitem{Abdulov:2021ivr}
N.A.~Abdulov et~al., \emph{{TMDlib2 and TMDplotter: a platform for 3D hadron
  structure studies}},
  \href{https://doi.org/10.1140/epjc/s10052-021-09508-8}{\emph{Eur. Phys. J. C}
  {\bfseries 81} (2021) 752}
  [\href{https://arxiv.org/abs/2103.09741}{{\ttfamily 2103.09741}}].

\bibitem{Ebert:2021jhy}
M.A.~Ebert, A.~Gao and I.W.~Stewart, \emph{{Factorization for Azimuthal
  Asymmetries in SIDIS at Next-to-Leading Power}},
  \href{https://arxiv.org/abs/2112.07680}{{\ttfamily 2112.07680}}.

\bibitem{Rodini:2022wki}
S.~Rodini and A.~Vladimirov, \emph{{Definition and evolution of transverse
  momentum dependent distribution of twist-three}},
  \href{https://arxiv.org/abs/2204.03856}{{\ttfamily 2204.03856}}.

\end{thebibliography}\endgroup

\end{document}